%% file: main.tex
\pdfoutput=1
\documentclass[12pt,letterpaper]{article}
\providecommand\JTjournal{generic}           
\providecommand\JTmode{submission}      
\def\JTpaperoptions{thmcounter=separate, thmwithin=section}
\input{jtload}             
\input{macros}

\JTtitle{Testing Density Discontinuity Using Spacings}
\JTshorttitle{Density Discontinuity Using Spacings}
\JTauthor{first=William T., last=Firmin, dept=Department of Economics,
          org=Northwestern University}
\JTabstract{In the regression discontinuity design, researchers assess the credibility of assumptions by testing the running variable's density for a discontinuity at the cutoff.  This paper proposes a novel such test.  The test uses $q$ spacings on each side of the cutoff, where spacings are the widths between ordered observations.  Spacings yield a continuous statistic that behaves well under weak assumptions.  I study its properties in two asymptotic frameworks: one where $q$ is fixed as the sample size increases and one where $q$ slowly diverges to infinity.  With $q$ fixed, the test achieves exact asymptotic size control without user randomization.  When $q$ increases slowly, the test becomes consistent without sacrificing exact asymptotic size.  Compared to the preferred version of Bugni and Canay's (2021) test, the only other test with comparably weak assumptions, this test reduces sensitivity to the tuning parameter and performs similarly when theirs is well-chosen.}
\JTkeywords{regression discontinuity, manipulation test, density discontinuity, spacings}
\JTthanks{I thank Federico Bugni and Ivan Canay for their feedback and advice.  I thank Nicolas Sourisseau for reading earlier versions of the proofs.  Anthropic's Claude, Opus 5.5, was used for checking proofs, debugging code, and other minor editing tasks.  All errors are my own.}
\JTfunding{}
\JTcompeting{}
\JTaidisclosure{}

\JTbibtex{refs}

\begin{document}
\JTmaketitle

\input{Sections/Body}

\JTappendix
\input{Sections/MainProofs}
\input{Sections/AuxilliaryProofs}
\input{Sections/Computation}
\input{Sections/MC}

\makeatletter
\ifdefined\old@section \let\section\old@section \fi
\makeatother
\clearpage
\nocite{*}
\JTprintreferences

\end{document}

%% file: macros.tex
\usepackage{enumitem}                 
\usepackage{caption}
\usepackage{subcaption}
\usepackage[flushleft]{threeparttable}
\usepackage{ragged2e}
\usepackage[export]{adjustbox}
\usepackage{bbm}

\usepackage{calrsfs}                  
\DeclareMathAlphabet{\pazocal}{OMS}{zplm}{m}{n}
\newcommand{\cala}[1]{\pazocal{#1}}

\newcommand{\R}{\mathbb{R}}

\newcommand{\N}{\mathbb{N}}

\newcommand{\E}{\mathbb{E}}

\newcommand{\ind}[1]{\mathbbm{1}\left\{#1\right\}}

\newcommand{\PP}{\text{I\kern-0.15em P}}

%% file: Sections/Body.tex
\section{Introduction}\label{sec:intro}

The regression discontinuity design (RDD) has only increased in popularity as a method for identifying causal treatment effects.  The sharp design is characterized by a deterministic treatment rule: individuals are treated if an observed score crosses a known threshold.  RDD nonparametrically identifies the average treatment effect for individuals at the cutoff under an arguably weak assumption: the potential outcomes must have continuous conditional expectations at the cutoff (Hahn et al., 2001).

The widespread adoption of RDD reflects the attractive credibility of its assumption, but there are situations where it fails.  For example, suppose individuals can freely choose their running variable based on knowledge of their potential outcomes.  The consequent sorting can create jumps in the conditional expectations.  Despite this, the continuity assumption is fundamentally untestable, so researchers tend to assess the credibility of RDD using a set of stronger assumptions proposed by Lee (2008).  Sacrificing a little generality buys interpretable conditions and two testable implications: (i) a continuous distribution of the covariates at the cutoff and (ii) a continuous density of the running variable (score) at the cutoff.  Bugni and Canay (2021) find tests of the implications in two thirds of empirical RDD papers across four major journals.  Rejecting the implications rejects the stronger assumptions.

This paper proposes a new test for the second implication, continuity of the running variable's density.  The test offers several advantages over competitors, including applicability, robustness, and simplicity.  First, the test is asymptotically valid under two asymptotic frameworks, both of which require similarly weak assumptions.  The test is asymptotically unbiased under one and becomes consistent under the other.  The second framework also accommodates random, data-dependent choices of the tuning parameter.  Second, the test is more robust to choices of the tuning parameter than the preferred version of Bugni and Canay's (2021) test, the only other test with comparably weak assumptions.  It achieves this through a continuous test statistic, which avoids the need for user randomization and simplifies methods for choosing the tuning parameter.  Finally, the test statistic is a ratio of two differences requiring only two order statistics and the cutoff point.  The test requires only one tuning parameter and does not require consistent estimators of the density.

The test builds on the intuition that spacings, the distances between ordered observations, reflect the density where they occur.  After all, a higher density implies that the observations should be more dense.  To formalize this, I show under weak assumptions that the spacings, defined using order statistics as $D_i:=X_{(i)}-X_{(i-1)}$, are approximately independent exponential random variables near the cutoff.  When scaled by $n$, the limiting rate parameter equals the limiting density on that side of the cutoff.  We exploit this fully with the optimal two-sample test of parameter equality for exponential distributions.  I show that this translates into asymptotic optimality among all tests that use a fixed number of observations on each side of the cutoff.

I study the spacings test in two asymptotic frameworks that depend on the size $q$ of the local samples.  For a chosen value of $q$, the test takes the $q$ closest observations to the cutoff on each side.  This gives an effective total sample size of $Q=2q$.  In terms of smoothness, we assume the density has one-sided limits at the cutoff; no derivatives are required.  For a fixed $q$ as $n$ increases to infinity, the test is asymptotically unbiased.  This reflects situations where the researcher is only comfortable using a relatively small number of observations near the cutoff.  The test's limiting power is uniformly higher than that of any other asymptotically unbiased test that uses the $q$ closest observations on each side of the cutoff.  The test is consistent (under an additional Lipschitz condition) as $q$ increases slowly to infinity.

To better understand the advantages of the spacings test, consider the existing alternatives.  McCrary (2008) requires three bounded and continuous derivatives of the density on each side of the cutoff.  Otsu et al. (2013) require two derivatives.  Cattaneo et al. (2020) require one continuous derivative.  Such smoothness assumptions effectively require the density to be approximately cubic, quadratic, or linear near the cutoff.  Importantly, higher assumed orders require the density to more tightly conform to that polynomial shape.  These methods perform poorly in the presence of spikes, kinks, and discontinuities near the cutoff point.  Assuming only one-sided continuity imposes the weakest requirements on the density.  Besides the present test, only Bugni and Canay (2021) make such weak assumptions.

Bugni and Canay (2021) use the $Q=2q$ closest observations to the cutoff and count how many are above.  Each local observation occurs above or below the cutoff with probability near 1/2 under a continuous density, so they approximate the count by a Binomial$(Q,1/2)$ distribution.  Statistics with discrete limiting distributions cannot provide exact asymptotic size control without randomization.  The problem is largest for smaller $Q$. The number of observations $n$ dwarfs the number used by the test (e.g., $Q$ proportional to $\sqrt{n}$), so $n$ must be very large for $Q$ to be large.  The preferred version of their test is nonrandomized and is asymptotically conservative with a fixed $Q$.  The spacings test overcomes this challenge through a continuous statistic, providing exact asymptotic size control without user randomization or restrictions on $Q$ that can have adverse effects in smaller samples.  When the density is twice as high on one side of the cutoff, this results in 5.6 more percentage points of limiting power, averaged across $2\le Q\le 100$.  For 97\% of these $Q$, the spacings test provides at least as much limiting power, uniformly across alternatives.  Bugni and Canay's (2021) test is competitive when requiring a heavily restricted choice of the tuning parameter.  The restrictions can harm performance and account for most differences between the tests in simulations.

The remainder of the paper is organized as follows.  Section \ref{sec:setup} introduces notation and describes the testing problem.  Next, Section \ref{sec:test} presents the test and a method for choosing the tuning parameter $q$.  Section \ref{sec:spacings} offers the intuition behind approximating spacings with exponential random variables.  Section \ref{sec:asymptotic} follows with formal asymptotic results.  Section \ref{sec:compare} compares the spacings test to alternatives using both asymptotic results and Monte Carlo simulations.  Finally, Section \ref{sec:conclusion} concludes.

\section{Setup and Notation}\label{sec:setup}
This section introduces notation and formalizes the testing problem.  Let $X$ be a random variable for which we observe an iid sample $\{X_i\}_{i=1}^n$.  Denote its CDF by $F:\mathbb{R}\to[0,1]$.  We are interested in the behavior of $F$ near a pre-specified point $c$, the cutoff in RDD.  Specifically, we would like to know whether the density $f=\frac{d}{dx}F$ is continuous at $c$.  To this end, we will first assume that $f$ actually exists near $c$.  We will also assume that $f$ has one-sided limits at $c$:
\begin{equation}
    f^+(c):=\lim_{x\to c^+}f(x)\hspace{1in}f^-(c):=\lim_{x\to c^-}f(x).
\end{equation}
This leads to a simple expression of the testing problem:
\begin{equation}\label{eqn:hypothesis}
    H_0:f^+(c)=f^-(c)\hspace{1in}H_1:f^+(c)\not =f^-(c).
\end{equation}

We will test the hypothesis using spacings, defined as follows.  Order the observations as $X_{(1)}\le X_{(2)}\le ... \le X_{(n)}$.  For $i=2,...,n$, the $i^{th}$ spacing is defined as
\begin{equation}\label{eqn:defn_spacings}
    D_i:=X_{(i)}-X_{(i-1)}
\end{equation}

The problem at hand only concerns the behavior of $F$ near $c$, so we focus on the spacings near $c$.  This paper adopts notation reversing the order statistics below $c$ to have $X_{(i)}^+$ and $X_{(i)}^-$ be the $i^{th}$ closest order statistics to $c$ from above and below.  Reversing the order below $c$ allows symmetric notation that improves clarity.  With this, write the $q$ closest order statistics on each side as
\begin{equation}\label{eqn:spacings_notation}
    X_{(q)}^-\le ...\le X^-_{(1)}<c\le X^+_{(1)}\le ...\le X^+_{(q)}
\end{equation}
and their spacings as $(D^-_q,...,D^-_1,D^+_1,...,D^+_q)$, where $D_1^+,D_1^-$ are the distances between $c$ and $X^+_{(1)},X^-_{(1)}$.  We also use $n^+:=\sum_{i=1}^n\ind{X_i\ge c}$ and $n^-:=\sum_{i=1}^n\ind{X_i< c}$ to denote the number of observations above and below $c$.

\section{Spacings Test}\label{sec:test}
The test uses spacings near the cutoff $c$.  Specifically, it only considers the $q$ closest spacings above and below $c$ for a total of $2q$ spacings.  I first describe how the test will use these spacings.  Next, I propose a data-based method for choosing $q$.

\subsection{The Test}

Before presenting the testing procedure, let us first preview the results of Section \ref{subsec:fixed-q}.  When considering a fixed number of local spacings, we may approximate them with independent exponential random variables.  In fact, the approximating rate parameters equal the density limits at $c$, exactly the objects of interest in (\ref{eqn:hypothesis}).

The approximation suggests a natural procedure: a two-sample test of parameter equality for exponential distributions.  The dual sample of exponential random variables is the limiting experiment; it directly reflects the ``fixed-q" asymptotic framework of Section \ref{subsec:fixed-q} where $q$ is fixed and sample size $n$ increases to infinity.  The test is designed for this limiting experiment, but Section \ref{subsec:increasing-q} shows that it also works with many (an increasing number of) local spacings.

Han et al. (2012) derive the optimal two-sample test for exponential random variables using the theory in Lehmann and Romano (2005, pp. 150-152).  The test statistic simplifies in our setting, as we will soon discuss.  For a significance level $\alpha\in(0,1)$, define
\begin{equation}\label{eqn:defn_test}
    T_n=\frac{c-X^-_{(q^-)}}{X_{(q^+)}^+-X_{(q^-)}^-}\hspace{1in}\phi(T)=\left\{\begin{array}{cl}
     1&:T\not\in[C_1,C_2]  \\
     0&:T\in[C_1,C_2]
\end{array}\right.
\end{equation}
where $C_1$ and $C_2$ are uniquely determined via the following two conditions: for $V\sim\text{Beta}(q^-,q^+)$,
\begin{equation}\label{eqn:defn_critical_values}
    \E[\phi(V)]=\alpha\hspace{0.5in}\text{and}\hspace{0.5in}\E[V\phi(V)]=\alpha \frac{q^-}{q^-+q^+}.
\end{equation}
The spacings test allows the usage of different numbers of observations on each side of the cutoff.  The formal results in Appendix \ref{appendix:proofs} maintain this generality, but here I will consider $q:=q^+=q^-$ for simplicity and clarity.  Although the test defines its critical values implicitly, a bisection search can quickly solve for $C_1$ and $C_2$ in practice.  I give computational details in Appendix \ref{appendix:computation}.

\begin{figure}[h!]
    \centering
    \includegraphics[width=0.45\linewidth]{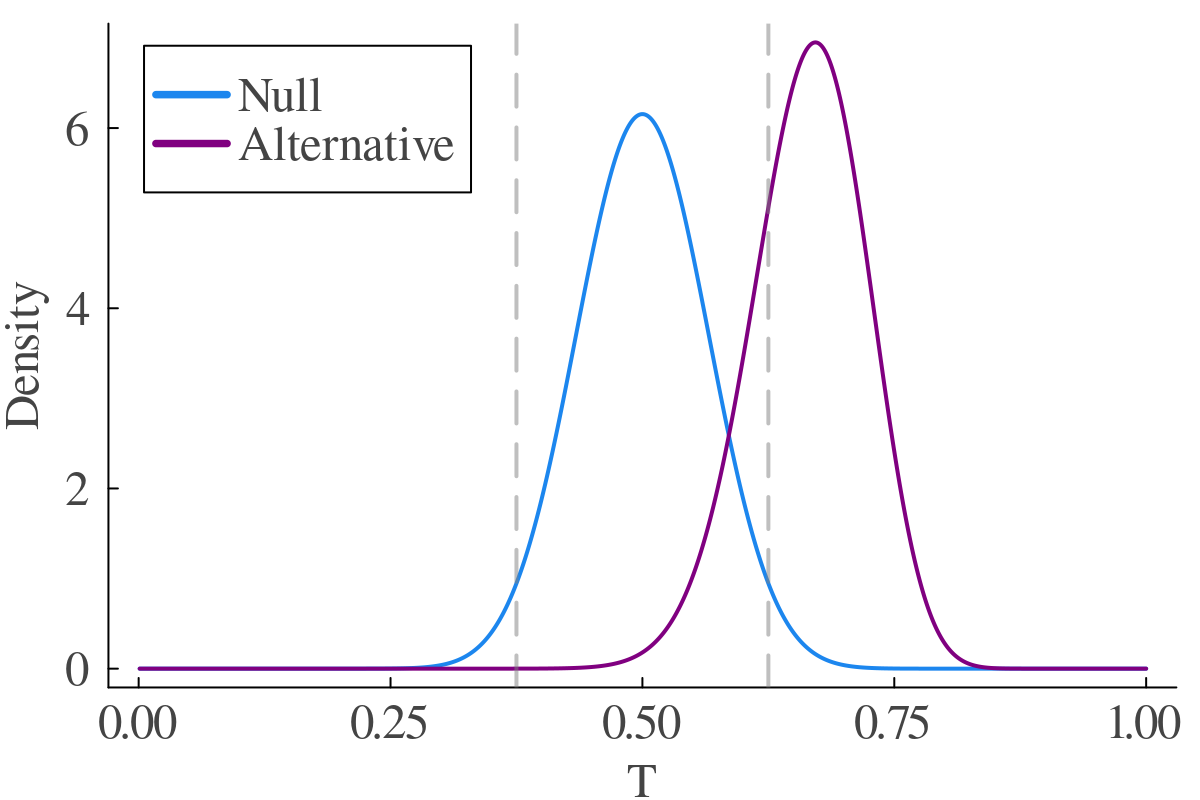}
    \caption{Approximate distribution of $T_n$ and $\alpha=5\%$ critical values with $q=30$.}
    \label{fig:statDensity}
\end{figure}

The test is simple but peculiar: Figure \ref{fig:statDensity} displays the statistic's approximate distribution for $q=30$ under the null hypothesis and when the density above $c$ is twice as high as that below.  First, the critical values are not symmetric around zero.  They both lie in $[0,1]$, as does the test statistic.  The test rejects the null when $T_n$ falls too far from $1/2$.  When $q^+=q^-$, the critical values are symmetric around $1/2$.  

The order statistics representation in (\ref{eqn:defn_test}) offers computational simplicity, but we can rewrite the test using spacings to help build intuition:
\begin{equation}
    T_n=\frac{\sum_{i=1}^{q}D_i^-}{\sum_{i=1}^{q}D_i^++\sum_{i=1}^{q}D_i^-}.
\end{equation}
In this form, the limiting experiment involves truly exponential scaled spacings $nD_i^+$ and $nD_i^-$ with parameters $f^+(c)$ and $f^-(c)$.  The sums of each exponential sample are sufficient statistics, allowing $T_n$ to simplify to (\ref{eqn:defn_test}) instead of explicitly using each spacing.

In reality, the test compares the reach of the $q$ observations just below $c$ to that of the $q$ observations just above.  Specifically, it takes the ratio of the reach below to the total span.  If the density takes similar values on both sides of the cutoff, then the nearest observations will be similarly dense, and the reach of the $q^{th}$ farthest observations should roughly match.  If the density takes higher values on the right of the cutoff, then the nearest observations on the right will tend to be more dense, and the reach of the $q^{th}$ farthest observation on the right should be smaller.  This would produce a larger test statistic.

For example, suppose the density function takes values twice as large above $c$.  The nearest observations above $c$ will be twice as dense, and the reach of the $q^{th}$ nearest observation above $c$ will be roughly half that of the $q^{th}$ nearest below $c$.  The statistic should take a value near $2/3$ as in Figure \ref{fig:statDensity}.  Statistics near 0 or 1 indicate that one side is much denser than the other, suggesting a violation of the null hypothesis.  Even densities on each side would yield a statistic near $1/2$.

Given $q$, the test may be implemented as:

\begin{enumerate}[label=\textbf{Step \arabic*.}]
    \item Find the $q^{th}$ closest observations above and below the cutoff, $X^+_{(q)}$ and $X^-_{(q)}$.
    \item Compute $T_n$ as in (\ref{eqn:defn_test}).
    \item Compute $C_1$ and $C_2$ as described in Appendix \ref{appendix:computation}.
    \item Reject the null hypothesis in (\ref{eqn:hypothesis}) if $T_n$ falls outside the interval $[C_1,C_2]$.
\end{enumerate}

\subsection{Choosing $q$}\label{subsec:choosing_q}

The test still requires a choice of $q$.  The results of Section \ref{subsec:increasing-q} allow a method for choosing $q$ based on the asymptotic bias of the standardized version of $T_n$ in (\ref{eqn:defn_test}).  Under the null, for $q=O_p(n^{2/3})$, the bias is asymptotically bounded by 
\begin{equation}t:=\frac{(2q)^{3/2}}{n}\frac\kappa {4f(c)^2}\end{equation}
where $\kappa$ is a local Lipschitz constant for the density in the increasing-$q$ asymptotic framework.  This bound may be interpreted as the worst-case bias to standard deviation ratio for the standardized statistic---in other words, a bound on the asymptotic bias from a N(0,1) approximation.  Following Bugni and Canay (2021), the researcher may specify an allowable amount of bias $t^*$ and solve for the corresponding value of $q$.  Rearranging,
\begin{equation}
    q^*=\frac12 n^{2/3}(t^*)^{2/3}\left(\frac{4f(c)^2}\kappa\right)^{2/3}.
\end{equation}
Undersmoothing, i.e. requiring $q=o(n^{2/3})$, corresponds to decreasing the allowable bias $t^*$ as $n$ increases.  This removes any bias asymptotically.

Despite a clean formula, we cannot control asymptotic bias exactly.  The parameters $f(c)$ and $\kappa$ are unknown, and we cannot estimate $\kappa$ precisely enough to control the worst-case bias (Low 1997, Armstrong and Koles\'ar 2018).  Instead, we can use a pilot distribution to solve for a reasonable value of $q$.  Consider the following rule-of-thumb for choosing $q$:
\begin{equation}\label{eqn:rot}
    \hat q_{rot}:=\frac12\sqrt{n}\left(\frac{4\hat\sigma\phi^2_{\hat\mu,\hat\sigma}(c)}{\phi_{\hat\mu,\hat\sigma}(\hat\mu+\hat\sigma)}\right)^{2/3}
\end{equation}
where $\phi_{\mu,\sigma}(\cdot)$ is the density function of a Normal$(\mu,\sigma^2)$ random variable and $\hat\mu,\hat\sigma$ are the sample mean and standard deviation of $X$.  As discussed in Bugni and Canay (2021), the rule-of-thumb provides an initial guess of $\kappa$ and $f(c)$ using a normal distribution approximation.  We could estimate the $f(c)$ with sufficient precision, but the above rule opts for simplicity instead of introducing a new density estimator.  If $X$ does take the normal distribution, the rule-of-thumb limits the worst-case bias to standard deviation ratio to $n^{-1/4}$.  Then for 500, 1000, and 5000 draws, the worst-case bias to standard deviation ratio is bounded by 0.22, 0.18, and 0.12.  For testing with level $\alpha=0.05$, these biases imply worst-case sizes of $P(|\xi +\text{bias}|>z_{\alpha/2})=5.56\%$, $5.37\%$, and $5.17\%$  under the null (let $\xi\sim N(0,1)$).

Realistically, $X$ does not follow a normal distribution.  If the normal approximation overestimates $\frac{f(c)^2}{\kappa}$ by a factor of $a>1$, then the worst-case bias becomes $t^*=an^{-1/4}$.  Overestimating by a factor of 2 with 1000 observations yields a bias of 0.36 and size of $6.50\%$.  Only slightly higher than the targeted rejection rate of 5\%, the rule-of-thumb provides a reasonable choice for $q$.  Monte Carlo simulations in Section \ref{subsec:monte-carlo} demonstrate good control of size in a variety of designs.  In a particularly difficult design, however, the rule-of-thumb rejects with probability 61.6\% under the null.  In the case that severe non-normality is suspected, we can choose a smaller, more cautious value of $q$ than suggested by the rule-of-thumb.  This would improve size control at the cost of power.

\section{Spacings Theory}\label{sec:spacings}

This section provides intuition for the test through a discussion on spacings.  As defined in (\ref{eqn:defn_spacings}), spacings are the distances between consecutive ordered observations.  I present a key result for uniformly distributed data and extensions to the general testing problem.  Pyke (1965) provides an extensive discussion of spacings in statistics.

Central to this paper, the spacings of iid uniform random variables possess the same distribution as iid exponential random variables divided by their sum.  Formally, let $\{U_i\}_{i=1}^n$ be iid Uniform(0,1).  For order statistics $0=:U_{(0)}\le U_{(1)}\le ...\le U_{(n)}\le U_{(n+1)}:=1$, the uniform spacings are
\begin{equation}\label{eqn:uniform-spacings}
    d_i:=U_{(i)}-U_{(i-1)}\overset d=\frac{Y_i}{\sum_{j=1}^{n+1}Y_j}
\end{equation}
for $\{Y_i\}_{i=1}^{n+1}$ iid Exp(1).  This provides a nice representation for spacings, but it is limited in two ways.  First, it requires uniformly distributed data. Although this is unrealistic in general RDD settings, we can extend it later using asymptotics.

Second, we must account for the cutoff.  The spacings test utilizes the space between nearby observations and the cutoff, but the representation in (\ref{eqn:uniform-spacings}) only considers spacings between observations.  Suppose the cutoff $\tilde c$ sits between the $i^{th}$ and $(i+1)^{th}$ ordered observations, so $U_{(i)}<\tilde c\le U_{(i+1)}$.  The representation in (\ref{eqn:uniform-spacings}) says nothing about the distribution of $\tilde c-U_{(i)}$ and $U_{(i+1)}-\tilde c$.  Even further, the index $i$ that is closest to $\tilde c$ is random.

Lemma \ref{representation} provides an analogous representation that accounts for the cutoff:  

\begin{lemma}\label{representation}
    Let $U_i\sim\text{Uniform}(0,1)$ iid for $i=1,...,n$ and $\tilde c\in(0,1)$.  Fix $n\in\N$ and consider the spacings between $\{U_i\}_{i=1}^n\cup\{0,\tilde c,1\}$.  There exist independent Exp(1) random variables $\{Y_i^+\}_{i=1}^{n+1}$ and $\{Y_i^-\}_{i=1}^{n+1}$ such that
\[\left(\begin{array}{c}
        (d_i^+)_{i=1}^{n^++1}\\
        (d_i^-)_{i=1}^{n^-+1}
    \end{array}\right)=\left(\begin{array}{c}
        \left((1-\tilde c)\frac{Y_i^+}{\sum_{j=1}^{n^++1}Y_{j}^+}\right)_{i=1}^{n^++1}\\
        \left(\tilde c\frac{Y_{i}^-}{\sum_{j=1}^{n^-+1}Y_{j}^-}\right)_{i=1}^{n^-+1}
    \end{array}\right)\;\text{a.s.}\]
    Furthermore, $\{Y_i^+\}_{i=1}^{n+1}$ and $\{Y_i^-\}_{i=1}^{n+1}$ are independent of $n^+$ and $n^-$.
\end{lemma}

Lemma \ref{representation} has several differences from the classic result in (\ref{eqn:uniform-spacings}).  Noticeably, the expression differs on opposite sides of the cutoff $\tilde c$.  While the form of a normalized exponential random variable remains, the sums only contain the variables from the respective side.  Each term is also multiplied by the total probability of its side.  Intuitively, a Uniform(0,1) distribution is a mixture of Uniform(0,$\tilde c$) and Uniform($\tilde c$,1) distributions.  The connection to (\ref{eqn:uniform-spacings}) becomes clear after conditioning on $n^+$ and $n^-$.  The proof of Lemma \ref{representation} constructs the exponential random variables and verifies through density transforms.

The spacings literature contains several similar results.  Nagaraja et al. (2015), for example, study the behavior of spacings near an order statistic.  Lemma \ref{representation} instead considers spacings near a pre-specified point.  In contrast to the work of Nagaraja et al. (2015), we must deal with the artificial spacings created by the cutoff.  This escapes the usual study of spacings.

Now that we've accounted for the cutoff, we can extend the representation to study non-uniform data.  The quantile function provides a method to do this.  The rest of this section provides a sketch of the proof for the exponential approximation.  

For our random sample $\{X_i\}_{i=1}^n$, spacings take the form
\begin{equation}\label{eqn:inv_cdf_transform}
    D_i:=X_{(i)}-X_{(i-1)}=F^{-1}(U_{(i)})-F^{-1}(U_{(i-1)})\;\text{ almost surely.}
\end{equation}
A similar representation accounts for the cutoff $c$ by replacing a $U$ with $\tilde c$, where $\tilde c=F(c)$.  Going forward, we will focus on the spacings between observations.

Here are a few technical remarks on the above.  Equation (\ref{eqn:inv_cdf_transform}) relies on the random variables $\{X_i\}_{i=1}^n$ existing in a probability space with independent uniform random variables such that $X_i=F^{-1}(U_i)$ almost surely.  The assumption has no bearing on the behavior of $\{X_i\}_{i=1}^n$, as a suitable probability space can always be constructed while maintaining all relevant properties of $\{X_i\}_{i=1}^n$.  Also, the right-side equality in (\ref{eqn:inv_cdf_transform}) only holds almost surely.  For example, $X$ could take values outside its support on a measure-zero subset of the sample space, but $F^{-1}$ will never map anything above the support of $X$.

The unknown quantile function $F^{-1}$ prevents implementing a simple finite-sample representation like in (\ref{eqn:uniform-spacings}).  Turning to asymptotics, we can derive a similar representation for more general spacings by linearizing (\ref{eqn:inv_cdf_transform}) with a mean value theorem.  For continuous distributions with density $f$,
\begin{equation}
    D_i=\frac{d_i}{\xi_i}\hspace{0.3in}\text{with}\hspace{0.3in}\xi_i=\frac{\int_{X_{(i-1)}}^{X_{(i)}}f(x)dx}{X_{(i)}-X_{(i-1)}}.
\end{equation}
Each spacing therefore equals the corresponding uniform spacing from Lemma \ref{representation}, divided by the average density over the interval the spacing spans.
If $f$ is continuous and $n$ is large, then
\begin{equation}\label{eqn:spacing_approx}
    nD_i\approx \frac{Y_i}{f(X_{(i)})}.
\end{equation}
In other words, the spacings are approximately iid exponential random variables with rate proportional to the density where they occur.  This motivates the proposed procedure as a two-sample exponential test of parameter equality.  Section \ref{sec:asymptotic} provides the formal results.

\section{Asymptotic Results}\label{sec:asymptotic}
This section presents formal results for the spacings test.  We study the spacings test under two asymptotic frameworks.  Under the first, $n$ increases to infinity while $q$ remains fixed.  This framework motivates the design of the test.  It represents finite-sample applications where the researcher is only comfortable using a relatively small number of observations near the cutoff.  The second framework allows $q$ to increase slowly with $n$.  It represents finite-sample applications where the researcher is comfortable using enough local observations to credibly use additional ``large $q$" approximations.  In this second framework, we also consider how $q$ is chosen by the researcher: in addition to increasing, $q$ may be a random variable.  The results both accommodate and inform data-dependent choices of $q$, such as the rule-of-thumb in Section \ref{subsec:choosing_q}.

I study the test's performance in two different asymptotic frameworks, but even if one framework seems more relevant, the implementation does not change.  Both frameworks study the same test with the same implementation.  The spacings test is based on the fixed-$q$ framework but has desirable properties in the increasing-$q$ framework as well.  Section \ref{subsec:fixed-q} will show that the test is asymptotically unbiased when given a fixed value of $q$.  Section \ref{subsec:increasing-q} will consider $q$ increasing to infinity and random, showing consistency and deriving expressions for local power and asymptotic bias.  Both frameworks require similar and relatively weak assumptions on the density of the running variable.  The assumptions are as follows:

\begin{assumption}\label{assumptions}
    $\{X_i\}_{i=1}^n$ is distributed according to a CDF $F$ with $0<P(X<c)<1$.  $F$ is absolutely continuous in a neighborhood $(c-\eta,c+\eta)$ of $c$.  On this set, the density function $f(x)$ satisfies the following:
    \begin{enumerate}[label=(\roman*)]
    \item $f(x)$ has one-sided limits $f^+(c)=\lim_{x\to c^+}f(x)$ and $f^-(c)=\lim_{x\to c^-}f(x)$.
    \item $\max\{f^+(c),f^-(c)\}>0$ and $\min\{f^+(c),f^-(c)\}<\infty$.
    \item When both limits $f^+(c),f^-(c)$ are finite and nonzero, there exists a constant $\kappa\in[0,\infty)$ such that $|f(x)-f^+(c)|\le \kappa|x-c|$ for all $x\in(c,c+\eta)$ and $|f(x)-f^-(c)|\le \kappa|x-c|$ for all $x\in(c-\eta,c)$.
\end{enumerate}
\end{assumption}

Assumption \ref{assumptions}(i) guarantees the existence of both one-sided limits, the equality of which is being tested.  Assumption \ref{assumptions}(ii) rules out distributions in which both limits are 0 or both limits are infinity.  As is, the procedure would have trouble if the limits matched in this way.  The relevant limiting distributions would depend on the rate of decrease or increase to 0 or infinity.  This paper aims for minimal assumptions on the smoothness of the density, so such cases are assumed away.  The first two assumptions suffice for the fixed-$q$ framework.

Assumption \ref{assumptions}(iii) imposes an additional smoothness requirement on the density.  Slightly weaker than requiring a first derivative, the Lipschitz assumption allows the density at the $q^{th}$ nearest observations to converge sufficiently quickly to its limits at $c$.  Such smoothness guarantees asymptotic normality when $q$ increases slowly to infinity.  In case the researcher possesses additional knowledge about the density, the more general results in Appendix \ref{appendix:proofs} allow the constant $\kappa$ to differ on each side of the cutoff.

Assumption \ref{assumptions} essentially matches that of Bugni and Canay (2021).  The only difference is that we allow one of $f^+(c),f^-(c)$ to be infinite.  Compared to other papers that test for density discontinuity, the main difference is Assumption \ref{assumptions}(iii): the Lipschitz assumption is weaker than assuming derivatives.

Without these assumptions, the test is still consistent against a mass point at the cutoff:
\begin{theorem}\label{mass}
    Suppose $X$ has a mass point at $c$ and $P(X<c)>0$.  Let $q=o_p(n)$.  Then 
    \[\lim_{n\to\infty}\E[\phi(T_n)]=1.\]
\end{theorem}

This result handles the case where agents manipulate their score just enough to equal the cutoff.  It creates a mass point at $c$, which the test can detect with hardly any assumptions. In this case, the test is consistent whether $q$ is fixed or increases (at a rate slower than $n$).  With a mass point at $c$, the $q$ closest observations above the cutoff will eventually equal $c$ with probability 1.  The spacings will equal 0 and the test statistic will equal 1, leading to rejection of the null.

Without a mass point, the properties of the test are not so trivial.  The rest of this section presents formal results divided into the two asymptotic frameworks.

\subsection{Results for Fixed $q$}\label{subsec:fixed-q}
This section presents asymptotic theory for a fixed value of $q$ as $n$ increases to infinity.  The fixed-$q$ framework represents a situation in which $q$ observations on each side are very close to the cutoff.  In this setting, $q$ may be too small to use a central limit theorem approximation for the sum of local spacings.  The increasing-$q$ framework, on the other hand, will consider using many observations.  The following results describe the limiting behavior of local spacings and show asymptotic unbiasedness of the test.  

First, a brief remark on notation.  The number of observations above or below the cutoff takes a binomial distribution.  As a consequence, there may be fewer than $q$ spacings above or below the cutoff.  The lack of usable spacings occurs with positive but decreasing probability.  To simplify notation, consider ill-defined spacings $D_i^s$ for $i>n^s$, $s\in\{+,-\}$ to equal 0.  This is asymptotically negligible and reduces clutter in stating the results.

Theorem \ref{fixed-q-dist} presents the limiting distribution of a fixed sample of local spacings.  It states its assumptions independently of Assumption \ref{assumptions} because of some technical differences.  Theorem \ref{unbiasedness} later reconciles these differences in showing the unbiasedness of the spacings test.

\begin{theorem}\label{fixed-q-dist}
    Let Assumption \ref{assumptions}(i) hold.  Further assume that $P(X\in(c-\varepsilon,c))>0$ and $P(X\in[c,c+\varepsilon))>0$ for all $\varepsilon>0$.  For a fixed integer $q>0$,
    \[n\cdot\left(\begin{array}{c}
        (D_i^+)_{i=1}^{q}\\
        (D_i^-)_{i=1}^{q}
    \end{array}\right)\overset{d}\to \left(\begin{array}{c}
        \left(\frac 1{f^+(c)}Y_i^+\right)_{i=1}^{q}\\
        \left(\frac 1{f^-(c)}Y_i^-\right)_{i=1}^{q}
    \end{array}\right)\]
where $\{Y_i^+\}_{i=1}^q,\{Y_i^-\}_{i=1}^q$ are independent Exponential(1) random variables.
\end{theorem}

\begin{remark}
    In the case that $f^+(c)$ or $f^-(c)$ are $0$, their reciprocal in the above result should be interpreted as $\infty$, i.e. the normalized spacings diverge to infinity.
\end{remark}

Theorem \ref{fixed-q-dist} formalizes the exponential approximation sketched in Section \ref{sec:spacings}.  It builds on the finite-sample representation from Lemma \ref{representation}, which extends a core result from the theory of spacings.  To the best of my knowledge, no previous results study the behavior of spacings split by a pre-specified non-random point $c$.  Previous results only apply to the spacings between ordered observations of $X_i$.  With those results, we could only study the spacings $D_1^+$ and $D_1^-$ through their sum, the spacing between $X_{(1)}^-$ and $X_{(1)}^+$.  We would have to disregard this spacing since it mixes information about both sides of $c$.  Theorem \ref{fixed-q-dist} accounts for how $c$ splits the spacing, allowing us to extract more information about the density near $c$.  

In the proof of Theorem \ref{fixed-q-dist}, linearizing $F^{-1}$ and using Lemma \ref{representation} reveal that
\begin{equation}
    n\left(\begin{array}{c}
        (D_i^+)_{i=1}^{q}\\
        (D_i^-)_{i=1}^{q}
    \end{array}\right)\overset d\approx\left(\begin{array}{c}
        \left(\frac 1{\xi_i^+}Y_i^+\right)_{i=1}^{q}\\
        \left(\frac 1{\xi_i^-}Y_i^-\right)_{i=1}^{q}
    \end{array}\right)
\end{equation}
where $\xi_i^s$, $s\in\{+,-\}$, equals the average density over the interval associated with $D_i^s$.  The full expression in the proof is more complicated, but this simplification conveys important intuition.  The accuracy of the first-order approximation depends primarily on how quickly $f(X_{(q)}^+)$ converges to $f^+(c)$, and likewise for $X_{(q)}^-$.  The spacings test will benefit from flatter and higher densities near $c$.  Higher densities quicken the convergence of the nearest observations, and flatter densities on each side quicken the convergence of the density at these points.

\begin{theorem}\label{unbiasedness}
    Let Assumptions \ref{assumptions}(i)-(ii) hold.  The following hold for fixed $q\ge 1$ and $\alpha\in (0,1)$:
    \begin{enumerate}[label=\Alph*)]
        \item Under $H_0$ in (\ref{eqn:hypothesis}), $\lim_{n\to\infty}\E[\phi(T_n)]=\alpha$.
        \item Under $H_1$ in (\ref{eqn:hypothesis}), $\lim_{n\to\infty}\E[\phi(T_n)]>\alpha$.
    \end{enumerate}
\end{theorem}

Theorem \ref{unbiasedness} shows asymptotic unbiasedness of the spacings test: exact asymptotic size control and asymptotic power greater than the nominal level.  The spacings test uses the approximation from Theorem \ref{fixed-q-dist}, treating the spacings as two samples of independent exponential random variables.  This limiting experiment admits a unique uniformly most powerful unbiased test, which was derived by Han et al. (2012).  By using this test, we have a stronger guarantee for power than in Theorem \ref{unbiasedness}(B):

\begin{theorem}\label{ump}
    Fix $q$ and consider any sequence of tests $\{\tilde\phi_n\}_{n=1}^\infty$ that are measurable functions of the $q$ closest observations on each side of $c$.  Let $\cala{M}$ be a set of data generating processes that satisfy Assumptions \ref{assumptions}(i)-(ii) and includes all pairs of limiting densities $(f^-(c),f^+(c))\in(0,\infty)^2$.  Suppose the sequence of tests is asymptotically level and unbiased under $\cala{M}$.  Then for any data generating process that satisfies Assumptions \ref{assumptions}(i)-(ii) and $H_1$ in (\ref{eqn:hypothesis}),
    \[\limsup_{n\to\infty}\E[\tilde\phi_n(X_{(q)}^-,...,X_{(q)}^+)]\le \lim_{n\to\infty}\E[\phi(T_n)].\]
\end{theorem}

Theorem \ref{ump} states that, in this asymptotic framework, the spacings test is efficient among tests that use $q$ observations on each side of the cutoff.  The result relates optimality (UMP unbiased) in the limiting experiment to optimality asymptotically.  Theorem \ref{ump} provides strong theoretical justification for the test, but comparisons to alternatives require care.  Optimality in the limit does not make the spacings test optimal over all.  First, this is a pointwise asymptotic result in a particular asymptotic framework.  Finite-sample behavior may differ, and a different test may perform better in the increasing-$q$ framework discussed below.  Second, optimality in the limiting experiment provides no comparison for tests that don't operate in the limiting experiment.  The result restricts information to the $q$ closest observations on each side.  This rules out nearest neighbor methods that don't distinguish sides, data-dependent choices of $q$, and methods that use more observations.  For example, a kernel-based test takes in all observations when using a Gaussian kernel.  Comparing the tests under the limiting experiment could strip the kernel-based test of any advantage (or disadvantage) from considering all observations.  The same reason hinders a direct comparison to the test of Bugni and Canay (2021), as we will discuss later.  In short, optimality in the limit is a nice property but doesn't paint the full picture.

The result is similar to that of M\"uller (2011), which provides a similar link between optimality in limiting experiments and optimality asymptotically.  M\"uller (2011) shows that optimal tests in the limiting experiment are asymptotically optimal among tests that are asymptotically level and unbiased for all data generating processes that result in convergence to the limiting experiment, as in Theorem \ref{fixed-q-dist}.  This is a larger class of data generating processes than considered by Assumptions \ref{assumptions}(i)-(ii), so the unbiasedness requirement can rule out tests that fail for these additional DGPs but are otherwise of interest.  To provide a more direct result, I show that convergence to the limiting experiment holds in total variation, which yields stronger guarantees.

The optimality result still holds if we make stronger smoothness assumptions and use tests that account for slope.  In this case, $\cala{M}$ would be the subset of allowable DGPs in which $f$ has one-sided derivatives at $c$.  Without further restrictions, $\cala{M}$ would still satisfy the conditions of Theorem \ref{ump}: no scaling or function of the $q$ closest points on each side can find an asymptotic advantage over the spacings test.  The strength of convergence to the limiting experiment means that any other information contained in the local observations is asymptotically negligible.  Taking advantage of smoothness requires another asymptotic framework altogether, such as increasing $q$ with $n$ or considering a sequence of DGPs with increasing slope.

\subsection{Results for Increasing q}\label{subsec:increasing-q}
This section presents asymptotic results when $q$ diverges to infinity.  Although the test was designed under the fixed-$q$ framework, the following results on asymptotic bias and local power are useful in evaluating its qualities.  Theorem \ref{increasing-q-dist} shows asymptotic normality and provides a bound for the asymptotic bias.  Whereas assuming a first derivative would produce an explicit formula for the bias, Assumption \ref{assumptions}(iii) only yields worst-case bounds.  Section \ref{subsec:choosing_q} uses these bounds to provide a data-dependent method of choosing $q$.

\begin{theorem}\label{increasing-q-dist}
    Suppose there exists some sequence $a_n\to\infty$, $a_n=o(n)$ such that $q/a_n\overset p\to 1$.  Denote $\beta:=f^-(c)/f^+(c)$.  Under Assumption \ref{assumptions} and the additional assumption that both limits $f^+(c),f^-(c)$ are finite and nonzero,
    \begin{equation}\label{eqn:thm_bias}
        \frac{(1+\beta)^2}{\beta\sqrt{2}}\sqrt{q}\left(T_n-\frac1{1+\beta}\right)=Z_n+B_n\left(\frac{[2q]^{3/2}}{n}\right)[1+o_p(1)]
    \end{equation}
    where $Z_n\overset d\to N(0,1)$ and $|B_n|\le \frac{\kappa}{8}\left[\frac{1}{f^+(c)^2}+\frac{1}{f^-(c)^2}\right]$.  
    
\end{theorem}

\begin{remark}
    The 2 on the right-hand side of (\ref{eqn:thm_bias}) scales $q$ to represent the effective sample size, allowing direct comparison of the worst-case bias to that in Bugni and Canay (2021).
\end{remark}

If $q=o_p(n^{2/3})$, then the left hand side of Theorem \ref{increasing-q-dist} converges in distribution to $N(0,1)$.  Such a requirement on $q$ would correspond to the common practice of under-smoothing in density inference.  The bounds on the bias are tight and achievable by a density with derivative equal to $\kappa$ at $c$.  This tightness implies that faster growth rates of $q$ cannot guarantee asymptotic normality without further assumptions.  A rate proportional to $n^{2/3}$ could still produce asymptotic normality if $B_n$ were more controlled within its bounds (e.g. existence of a first derivative of the density), but even then the limiting distribution would have an unknown center. 

The requirements for $q$ allow it to be data-dependent.  With the rule-of-thumb in Section \ref{subsec:choosing_q}, we can take $a_n$ to be the population analogue of (\ref{eqn:rot}).  As long as the running variable has a finite, nonzero variance, $\hat q_{rot}/a_n$ converges to 1 in probability.

Theorem \ref{increasing-q-dist} builds on the finite-sample representation from Lemma \ref{representation}, which facilitates arguments that use central limit theorems for $X^+_{(q)}-c$ and $c-X^-_{(q)}$ (as sums of spacings).  This is particularly helpful for data-dependent $q$, as central limit theorems for a random number of summands require only that $q/a_n\overset p \to 1$ for some deterministic $a_n\to\infty$.

Theorem \ref{consistency} establishes the consistency of the spacings test as $q$ increases to infinity.  It incorporates the under-smoothing requirement that $q=o_p(n^{2/3})$.

\begin{theorem}\label{consistency}
   Fix $\alpha\in (0,1)$.  Let Assumption \ref{assumptions} hold and suppose there exists some sequence $a_n\to\infty$, $a_n=o(n^{2/3})$ such that $q/a_n\overset p\to 1$.  Then:
    \begin{enumerate}[label=\Alph*)]
        \item Under $H_0$ in (\ref{eqn:hypothesis}), $\lim_{n\to\infty}\E[\phi(T_n)]=\alpha$.
        \item Under $H_1$ in (\ref{eqn:hypothesis}), $\lim_{n\to\infty}\E[\phi(T_n)]=1$.
    \end{enumerate}
\end{theorem}

Theorem \ref{local} considers sequences of local alternatives in which only the density limits change.  The relative magnitude of the limits determines the difficulty of the testing problem, so the sequence considers ratios that approach 1.

\begin{theorem}\label{local}
    Fix $\alpha\in(0,1)$.  Consider a sequence of CDFs $F_n$, each of which satisfies the assumptions of Theorem \ref{increasing-q-dist} with $c$, $\tilde c$, $\eta$, and $\kappa$ as absolute constants.  Let $q$ be as in Theorem \ref{consistency} with a fixed sequence $a_n$.  Suppose the CDFs have density limits at $c$ such that $f_n^+(c)\to f^+(c)\in(0,\infty)$, $f_n^-(c)\to f^-(c)\in(0,\infty)$, and $\sqrt{2a_n}\left(\frac{f_n^-(c)}{f_n^+(c)}-1\right)\to \Delta\not = 0$.  Then
    \[\E[\phi_n(T_n)]\to P\left(\left|Z+\frac\Delta2\right|>z_{1-\alpha/2}\right)>\alpha\]
    where $Z\sim N(0,1)$ and $z_{1-\alpha/2}$ is the $(1-\frac{\alpha}2)$-quantile of $Z$.
\end{theorem}

Theorem \ref{local} describes how the spacings test performs in difficult scenarios: when the density limits are close but not equal, it can be hard to distinguish from the null.  This description provides another way to compare different tests.  In the next section, I use this to compare the spacings test to that of Bugni and Canay (2021).

\section{Comparison}\label{sec:compare}
This section compares the spacings test to alternative methods.  I compare formal asymptotic results and use the tests in Monte Carlo simulations to study their finite-sample behavior.  Special attention is given to the spacings test's closest alternative, the g-order statistics test of Bugni and Canay (2021), henceforth referred to as BC.  I first describe their test.

The g-order statistic approach of BC uses a fixed number of the observations closest to $c$.  I will use $Q$ to denote the number of observations, which reflects the effective sample size.  For the spacings test, the effective sample size is $Q=2q$.  I will also use $\pi_f=f^-(c)/[f^-(c)+f^+(c)]$ to denote the relative size of the discontinuity.  Values near $\frac12$ reflect small discontinuities, whereas values near 0 or 1 reflect large ones.  I define $\pi_f$ slightly differently than BC do: their version uses $f^+(c)$ in the numerator.

The g-order statistic test counts how many of the $Q$ local observations fall above the cutoff.  Under the null hypothesis the densities take approximately equal values on both sides, so the local observations are evenly distributed around $c$.  The count approximately follows a Binomial$(Q,1-\pi_f)$ distribution.  The limiting experiment, reflected by the fixed-$q$ framework, uses one draw from the Binomial distribution to test whether $\pi_f=1/2$.  

The test is simple, but the discrete test statistic yields asymptotically conservative inference with a fixed $Q$ and without user randomization.  Since the test statistic only takes $Q$ values, rejecting based on the statistic alone offers only $Q$ asymptotic sizes $\alpha$.  If $Q=21$, for example, the closest limiting sizes to $\alpha=5\%$ for a equal-tailed test are 2.6\% and 7.8\%.  BC consider two approaches to deal with this.  The first involves user randomization: if the statistic is on the edge of rejection, we can use an auxiliary random variable (e.g. flip a biased coin) to complete the test.  Correct randomization closes the gap between 2.6\% and 7.8\%, e.g., to the desired 5\%.  The non-randomized version, which avoids questioning whether a researcher honestly randomized their test, accepts that it cannot exactly control size in the limit.  BC deal with this by altering their rule for choosing $Q$.  In addition to bias, the altered rule considers which values of $Q$ offer limiting sizes close to the target $\alpha$.  They call this the ``informed rule-of-thumb."  The non-randomized version provides the most realistic alternative to the spacings test, so I focus my comparisons on it.

\subsection{Tuning Parameters}\label{subsec:tuning}

In this subsection, I compare the spacings test to its closest alternative: that of BC.  Specifically, I discuss the proposed methods for choosing $Q$ and the tests' sensitivity to such choices.

The primary advantage of the spacings test over that of BC lies in choosing the number of local observations.  With the spacings test, choosing $q$ only trades off power and bias.  Using a smaller number of local observations will reduce bias at the expense of power.  Critically, the limiting rejection rate under the null does not depend on $q$.  This is not the case for the preferred test of BC.  Consequently, choosing $Q$ becomes a size problem, even in the absence of bias.

\begin{figure}[h!]
    \centering
    \includegraphics[width=0.45\linewidth]{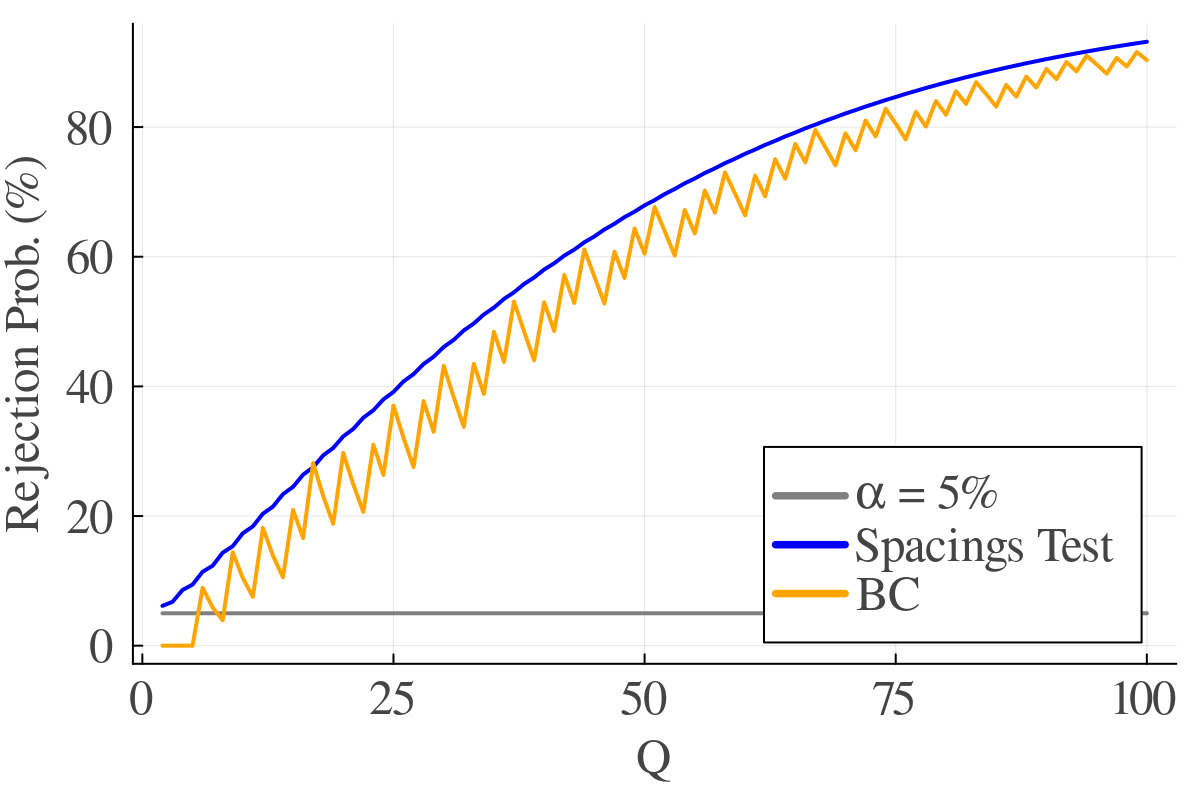}
    \caption{Limiting Power against $\pi_f=1/3$ with $\alpha=5\%$ and $2\le Q \le 100$}
    \label{fig:power033}
\end{figure}

Figure \ref{fig:power033} shows the limiting power of each test in the fixed-$q$ framework against a density that jumps twice as high above $c$.  In this design, the asymptotic power of the spacings test increases monotonically in $q$.  With exact asymptotic size control, a larger $q$ buys power without affecting size until bias becomes non-negligible.  In contrast, the asymptotic power of BC oscillates with the choices of $Q$.  Increasing power is not as simple as increasing $Q$: it is more sensitive to the precise choice.  Furthermore, the power is only comparable to that of the spacings test when $Q$ is well-chosen.  

BC address this triple tradeoff by picking an initial $Q$ that aims to control bias, like in Section \ref{subsec:choosing_q}, and then switching it to a ``nearby'' choice that provides limiting size close to nominal.  For $\alpha=5\%$, there are only 7 admissible choices of $Q$ between 2 and 100, with gaps of up to 26 between them.  As a result, the restricted choice of $Q$ can substantially change the tradeoff between bias and power in an undesirable way.  In fact, simulations show that most of the differences in performance between BC and the spacings test come from this restriction (Tables \ref{table:500} and \ref{table:1000}).

Additionally, the rule in BC targets choices of $Q$ that achieve the peaks observed in Figure \ref{fig:power033}.  Since it's based on the limiting probabilities, the rejection rates could look different in practice, and BC provide no guarantees about how well the rule navigates the oscillations in finite samples.  The rejection rates are sensitive to small changes in $Q$, so even a slight misstep in choosing $Q$ can have large effects.

The spacings test avoids all of this.  The test is always calibrated to an exact limiting size, regardless of $q$.  With the spacings test, researchers can focus on the usual tradeoff between bias and power, and the test's properties are less sensitive to these choices.  Rules for choosing $q$, perhaps based on believed underlying distributions or shape restrictions, remain unhindered from achieving their goal.  The additional steps in BC complicate the procedure just to achieve limiting rejection rates that are close to nominal.

\begin{figure}[h!]
        \centering
    \begin{subfigure}[t]{0.45\textwidth}
    \includegraphics[width=\textwidth]{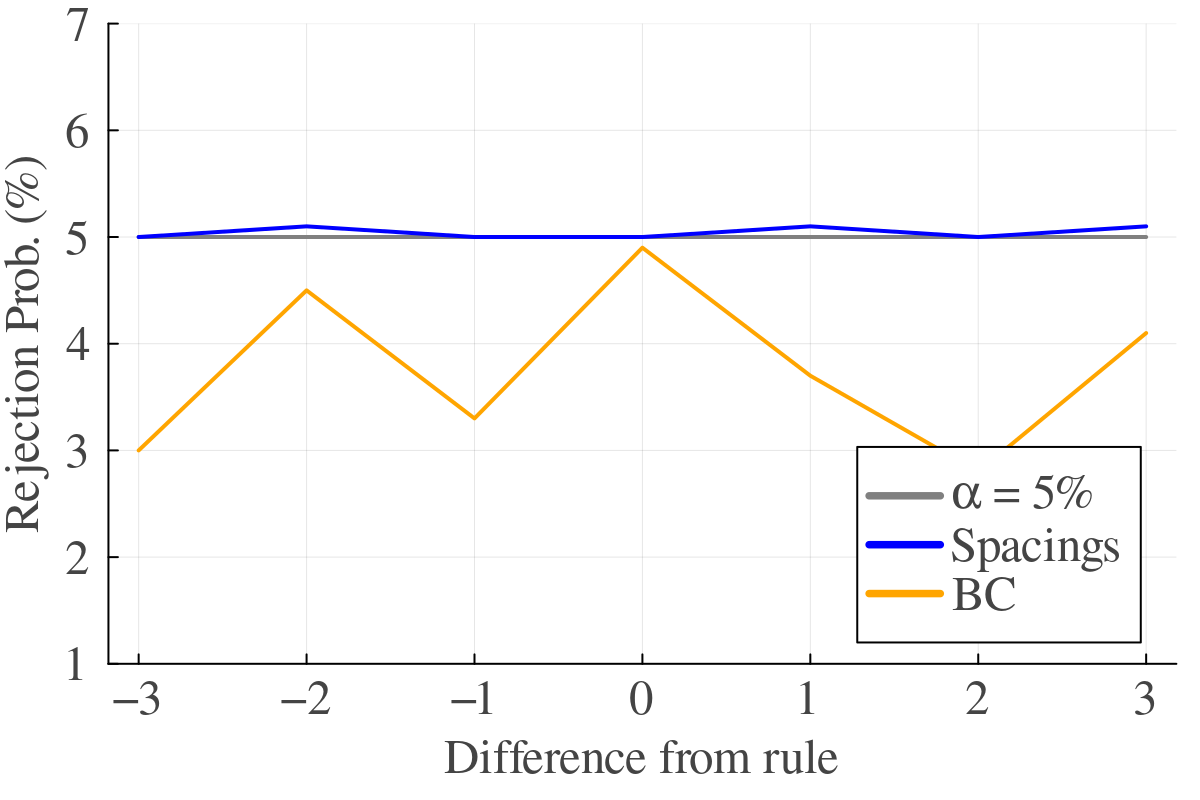}
    \caption{Design 1.a}
\end{subfigure}
\hspace{0.3in}
\begin{subfigure}[t]{0.45\textwidth}
    \includegraphics[width=\textwidth]{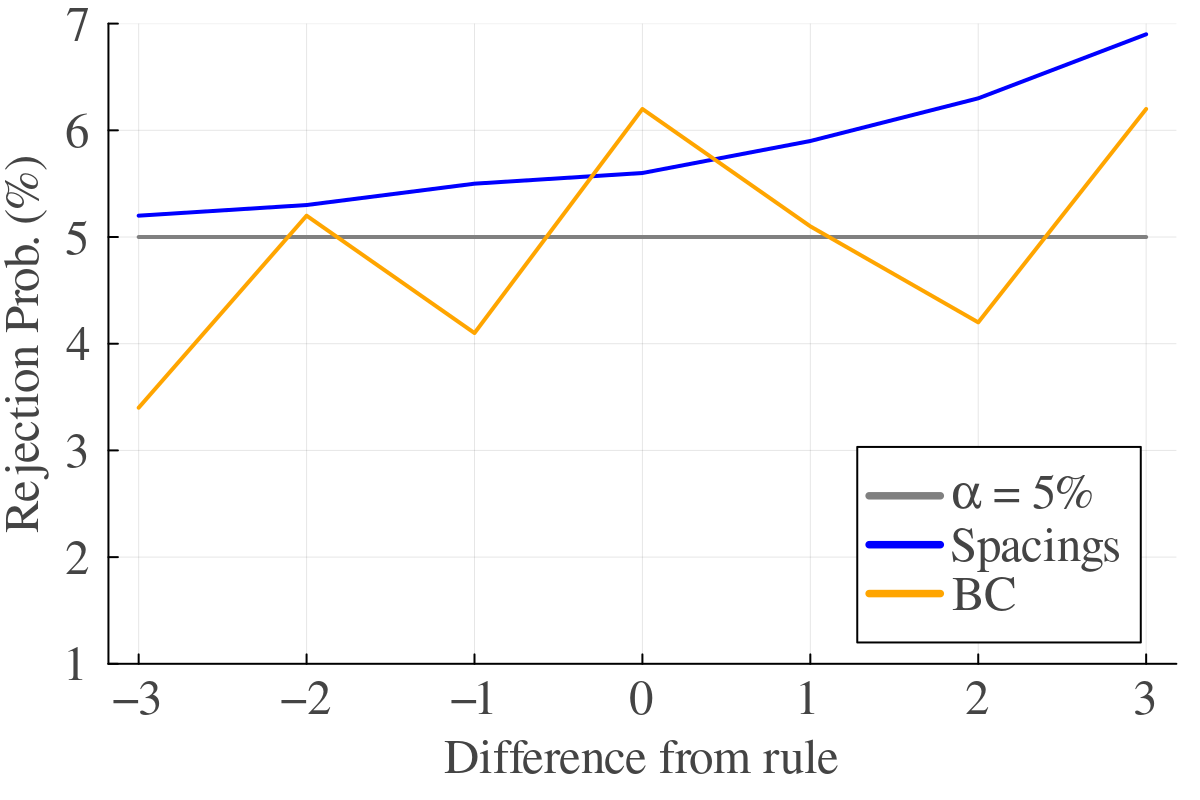}
    \caption{Design 5.b}
\end{subfigure}
    \caption{Null rejection rate relative to respective rule. 100,000 simulations, sample size 500.}
    \label{fig:sensitivity}
\end{figure}

To illustrate the sensitivity, we briefly preview some simulations.  Figure \ref{fig:sensitivity} displays finite-sample examples of the discussed behavior.  With draws from a normal distribution in panel (a), both tests' rules precisely characterize the asymptotic bias.  Furthermore, centering the normal distribution at the cutoff removes bias for BC.  The rule-of-thumb recommends $q=22$ for $\alpha=5\%$.  Even if we choose $q$ slightly higher or lower, the spacings test rejects at nearly 5\% under the null.  BC, on the other hand, must further adjust $Q$ to compensate for the discrete statistic.  In the end, BC rejects just slightly under 5\%, but slight deviations from the rule yield rates closer to 3 or 4\%.

To contrast, panel (b) uses a design with significant bias.  The rule-of-thumb suggests $q=19$ for the spacings test, which rejects 5.6\% of the time.  Instead of using the analogous $Q=2q=38$, the rule of BC chooses $Q=51$ to compensate for limiting conservativeness.  The sizable increase takes on more bias, leading to over-rejection at 6.2\%.  We see that other choices could have provided better size control than the spacings test, but panel (a) highlights that deviating is risky without precise knowledge of the bias.

Finally, the results in Section \ref{subsec:increasing-q} explicitly allow $q$ to depend on the data, treating it as a sequence of random variables rather than a deterministic sequence.  BC only consider deterministic sequences of $Q$.

\subsection{Power}\label{subsec:power}
We now turn to a more detailed analysis of power, where I discuss the relative performance of the spacings test and BC in both asymptotic frameworks.  In the fixed-$q$ framework, we find that the spacings test provides uniformly more asymptotic power in many cases, though the differences are mild when $Q$ is well-chosen for BC.

As discussed after Theorem \ref{ump}, the optimality property of the spacings test in the fixed-$q$ framework only holds among tests that consider $q$ observations on each side of $c$.  Since BC consider $Q$ local observations with a random number on each side, the optimality property offers no comparison.  We can instead directly compare the limiting properties of the tests in the fixed-$q$ framework.  

I compute the limiting power for all choices of $Q$ from 2 to 100 with $\alpha=5\%$ and find that the spacings test provides uniformly more limiting power for 97\% of them.  Since the spacings test provides exact asymptotic size control, higher limiting power is beneficial and not caused by over-rejection under the null.  Only three values find more power against some alternative, and they are some (but not all) of the values considered by the informed rule-of-thumb in BC.  Similar results emerge for different choices of $\alpha$.  Even for some well-chosen values of $Q$, BC provide less asymptotic power against all alternatives.

For each $Q$,  I compute the limiting rejection probabilities over a grid of $\pi_f$ with width $10^{-6}$ between 0 and 1.  When $Q$ is odd, the spacings test uses one more observation below the cutoff than above.  The following figures show the limiting power in a variety of ways.

\begin{figure}[h!]
    \centering
    \begin{subfigure}[t]{0.45\textwidth}
    \includegraphics[width=\textwidth]{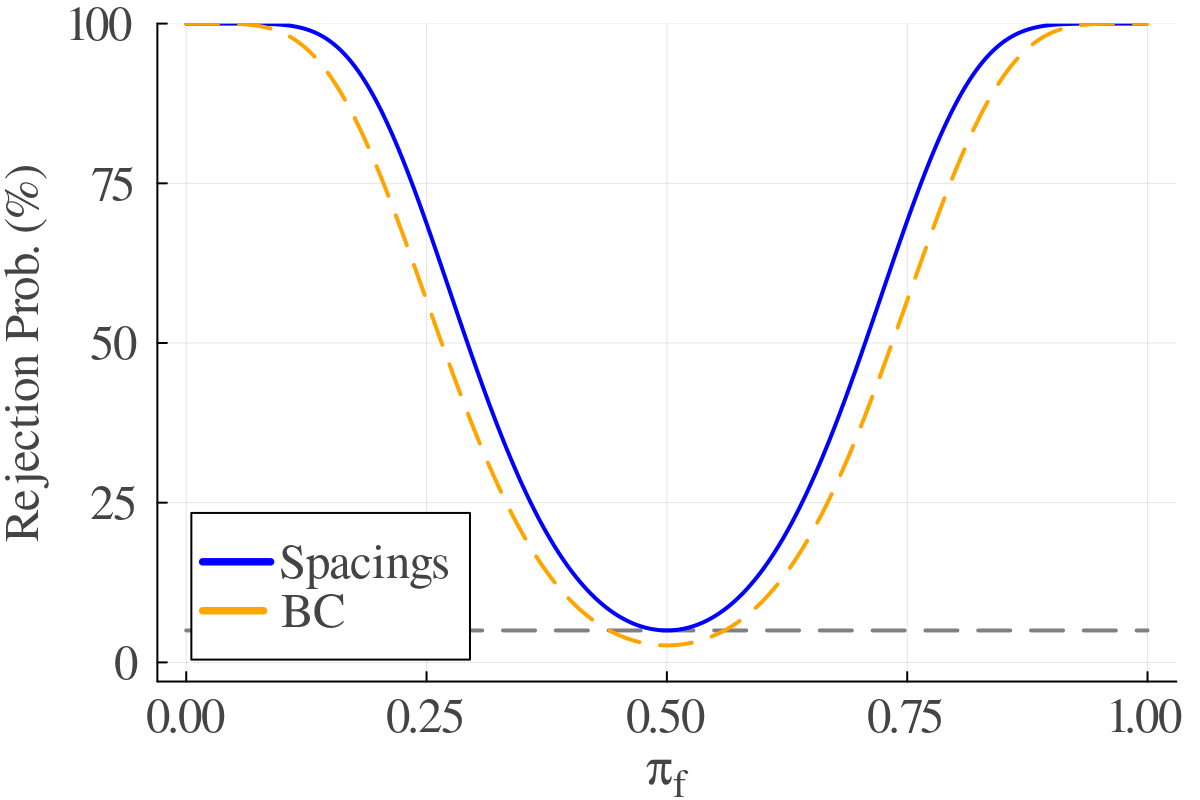}
    \caption{$Q=21$}
\end{subfigure}
\hspace{0.3in}
\begin{subfigure}[t]{0.45\textwidth}
    \includegraphics[width=\textwidth]{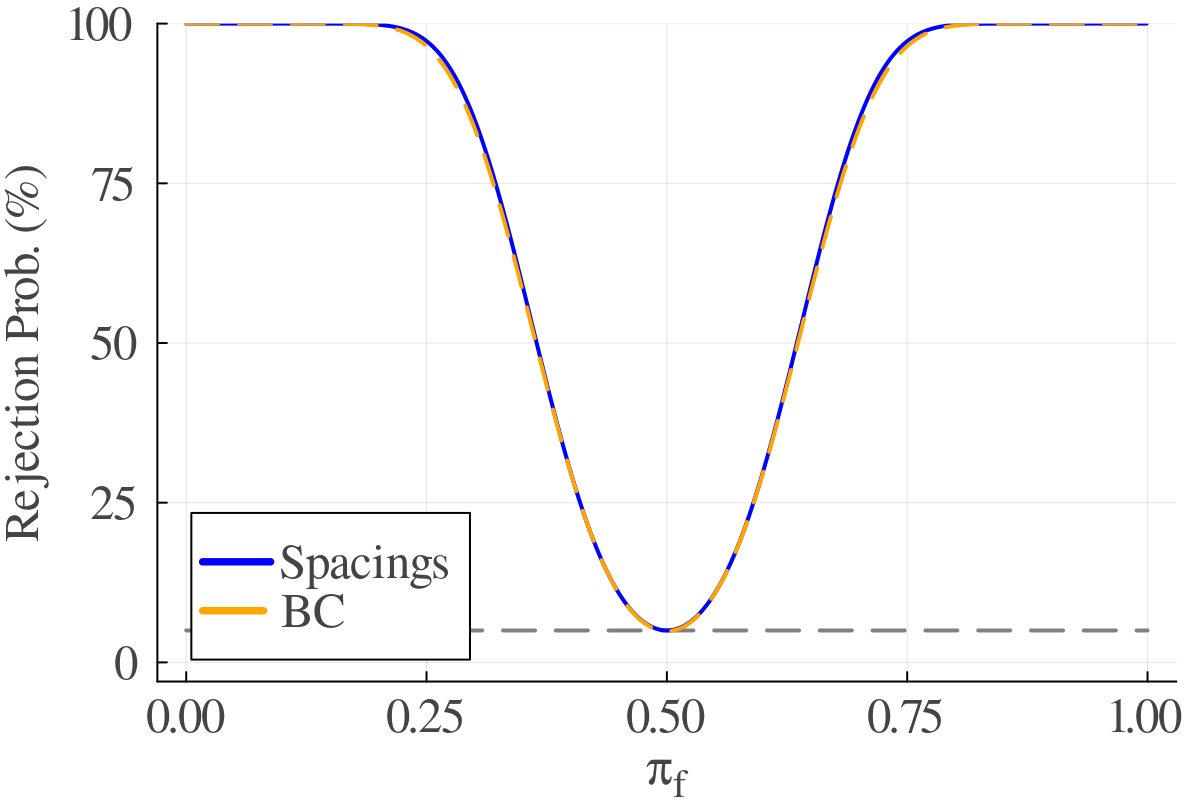}
    \caption{$Q=51$}
\end{subfigure}
\vspace{0.1in}
\caption{Limiting Power with $\alpha=5\%$}
\label{fig:limitingPower}
\end{figure}

Figure \ref{fig:limitingPower} shows the limiting rejection probabilities for selected values of $Q$.  For completeness, Figure \ref{fig:limitingPowerExtra} in Appendix \ref{section:MC} presents all considered choices of $Q$, separately for each test.  In both panels, the power curves for the spacings test hit the nominal rejection rate $\alpha=5\%$ under the null hypothesis ($\pi_f=1/2$).  The power curves rise towards 100\% as $Q$ increases.  The curves for the g-order statistics test don't achieve the nominal rate exactly.  The rejection probabilities vary substantially under the null, reflecting poor control with discrete statistics.  The curve for $Q=21$ represents the typical unrestricted choice of $Q$.  It lies below the curve for the spacings test, losing substantial power to poor size control.  Panel (b) displays a well-chosen value of $Q$ for BC.  Choosing $Q=51$ almost achieves the nominal rejection rate under the null.  The curves mostly lie on top of each other, but the spacings test finds slightly more power against moderate-sized discontinuities.

Figure \ref{fig:power033} shows the limiting power against a particular alternative: $\pi_f=1/3$.  This corresponds to a density that is twice as high above $c$.  Averaged across the 99 choices of $Q$, the spacings test provides 5.6 more percentage points of power.  Its power curve lies strictly above that of BC except at $Q=17$, where BC provide 0.65 more percentage points of power.  Appendix \ref{section:MC} displays the differences in power across all alternatives and precisely where BC can find more power.

The difference in power is smaller for large $Q$.  In fact, the increasing-$q$ results prove that any differences disappear as $Q$ increases.  Intuitively, the spacings test benefits from the continuity of its test statistic, whereas the order statistics test suffers from its discreteness.  Increasing $Q$ to infinity smooths out the discontinuities, and the Binomial($Q,1-\pi_f$) approximation becomes approximately normal.  In this framework, the disadvantages of a discrete test statistic are effectively ignored.

Consider the asymptotic bias of the normal approximation in Theorem \ref{increasing-q-dist}.  BC bound their asymptotic bias under the null by $\kappa/[4f^2(c)]$, which matches that of Theorem \ref{increasing-q-dist}.  This paper and BC both choose $q$ to control asymptotic bias using a normal approximation.  Since both tests produce the same asymptotic bias, the choice of $q$ is nearly the same.  BC must alter this choice to accommodate the discreteness of their statistic.

Now consider the limiting power under local alternative hypotheses.  BC state their local power result in terms of the scaled and centered limit of $f_n^+(c)/[f_n^+(c)+f_n^-(c)]$ as opposed to $f^-_n(c)/f^+_n(c)$.  Adjusting their notation, their centering term matches Theorem \ref{local} as $\Delta/2$.  Both tests yield the same asymptotic bias and local power in the increasing-$q$ framework.  For large numbers $q$ of local observations, the two tests should behave similarly.

\subsection{Monte Carlo Simulations}\label{subsec:monte-carlo}

This subsection compares the spacings test with those of BC and Cattaneo et al. (2020), henceforth referred to as CJM.  CJM test for a discontinuity by smoothing the empirical CDF from both sides using local polynomials.

The following figures display Monte Carlo simulations for the spacings test, order statistics tests of BC, and local polynomial test of CJM.  Appendix \ref{section:MC} contains tables with these results and simulations for additional choices of $Q$.  Each design runs 100,000 simulations for the spacings and order statistics tests.  Each design runs 10,000 simulations for CJM since it distinguishes itself with fewer runs.  The spacings and order statistics tests share many characteristics and thus require greater precision.  Each test relies on some tuning parameter.  The order statistics tests use $Q_{BC}$, the method of choice proposed by BC.  The spacings test uses the rule-of-thumb $\hat q_{rot}$ from Section \ref{subsec:choosing_q}.  CJM is implemented via their Python package ``rddensity" with the default parameters.

\begin{figure}[h!]
\centering
\begin{subfigure}[t]{0.25\textwidth}
    \includegraphics[width=\textwidth]{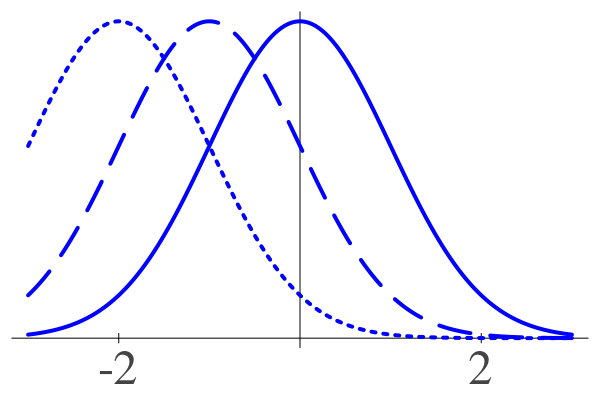}
    \caption{Design 1: $\mu=(0,-1,-2)$ in solid, dotted, and dashed}
\end{subfigure}
\hspace{0.3in}
\begin{subfigure}[t]{0.25\textwidth}
    \includegraphics[width=\textwidth]{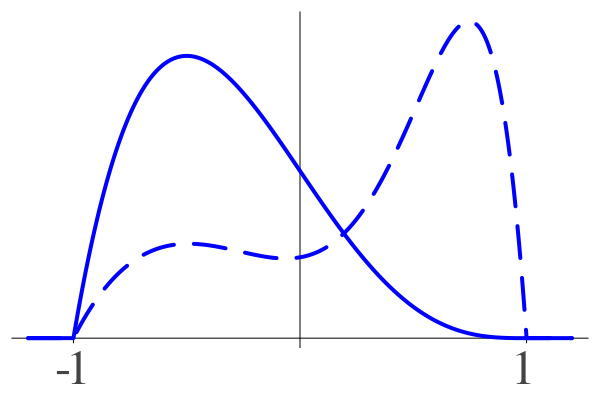}
    \caption{Design 2: $\lambda=(\frac13, 1)$ in solid and dashed}
\end{subfigure}
\hspace{0.3in}
\begin{subfigure}[t]{0.25\textwidth}
    \includegraphics[width=\textwidth]{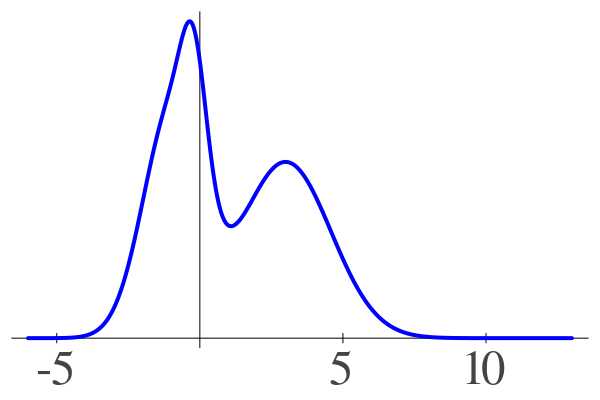}
    \caption{Design 3}
\end{subfigure}
\\ \vspace{0.4in}
\begin{subfigure}[t]{0.25\textwidth}
    \includegraphics[width=\textwidth]{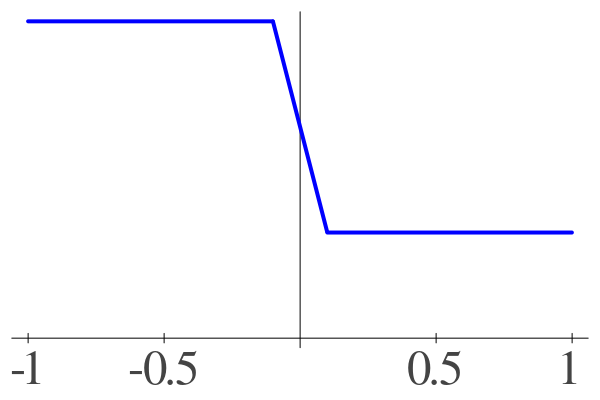}
    \caption{Design 4: $\lambda=0.1$}
\end{subfigure}
\hspace{0.3in}
\begin{subfigure}[t]{0.25\textwidth}
    \includegraphics[width=\textwidth]{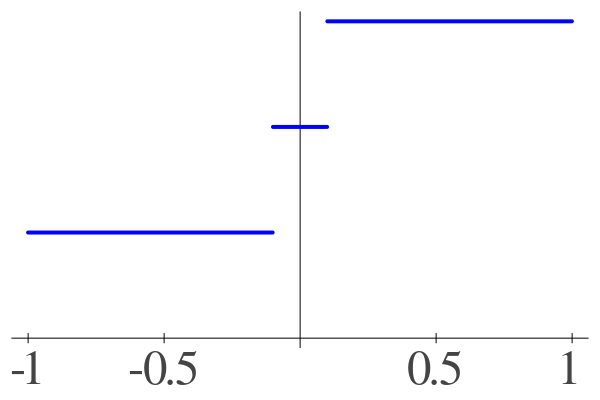}
    \caption{Design 5: $\lambda=0.1$}
\end{subfigure}
\vspace{0.1in}
\caption{Density functions for Monte Carlo simulation designs}
\label{fig:designs}
\end{figure}

The simulations use designs from BC.  They chose these designs to display both the strengths and weaknesses of their order statistics test.  Since the spacings test shares many characteristics with theirs, the designs provide a fair basis for comparison.  The design densities are plotted in Figure \ref{fig:designs}.  The spacings test treats the density as approximately flat, so spikes, discontinuities, and steep curves near the cutoff will cause problems.  Spikes, discontinuities, and kinks will also cause problems for methods with stronger smoothness assumptions.  Consequently, the tests will perform best on designs 1 and 2.  The following discussion will give informal descriptions of each design, and readers may go to BC for density expressions.

Design 1, the simplest, considers normally distributed data.  Three different values of $\mu$ show performance for cutoffs in the center or tails of the distribution.  Design 2 originates from Canay and Kamat (2018).  Design 3 comes from Saez (2010) and represents the distribution of taxable income.  Designs 4 and 5 feature flat densities except within a central interval of length $2\lambda$, pitting the methods against highly nonsmooth densities.  Figure \ref{fig:designs} plots these densities for $\lambda=0.1$, but the simulations also use values of $0.05$ and $0.25$.  Each design satisfies the null hypothesis.  To simulate alternatives, observations in the interval $x\in[0,0.1]$ are reflected across 0 with probability $0.2-2x$.

\begin{figure}[h!]
    \centering
    \begin{subfigure}[t]{0.6\textwidth}
    \includegraphics[width=\textwidth]{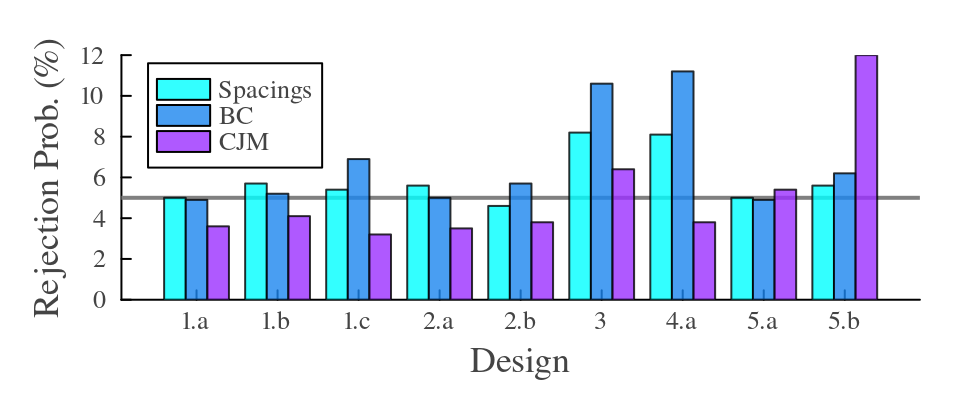}
\end{subfigure}
\hspace{0.3in}
\begin{subfigure}[t]{0.3\textwidth}
    \includegraphics[width=\textwidth]{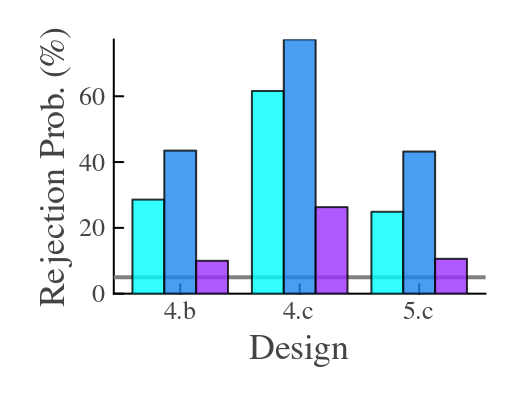}
\end{subfigure}
\vspace{0.1in}
\caption{Rejection Rates under $H_0$ with $n=500$ and $\alpha=5\%$}
\label{fig:MC500Level}
\end{figure}

Figure \ref{fig:MC500Level} displays the results when $n=500$ and we target a nominal level of $\alpha=5\%$.  
The figures label designs by number and letter, where the letter indicates the parameter.  The letters follow the order of parameters in decreasing order.  For example, \textit{1c} refers to design 1 with $\mu=-2$.  

The spacings test controls size better than BC in most designs.  The rejection rate for the spacings test tends to sit right on the nominal level 5\% marked by the gray line.  The test of CJM displays even larger differences in rejection rates.  It significantly under-rejects in the smoother designs where the spacings test controls size well.  In the designs where the spacings test over-rejects, CJM tends to over-reject to a lesser extent.

\begin{figure}[h!]
    \centering
    \begin{subfigure}[t]{0.6\textwidth}
    \includegraphics[width=\textwidth]{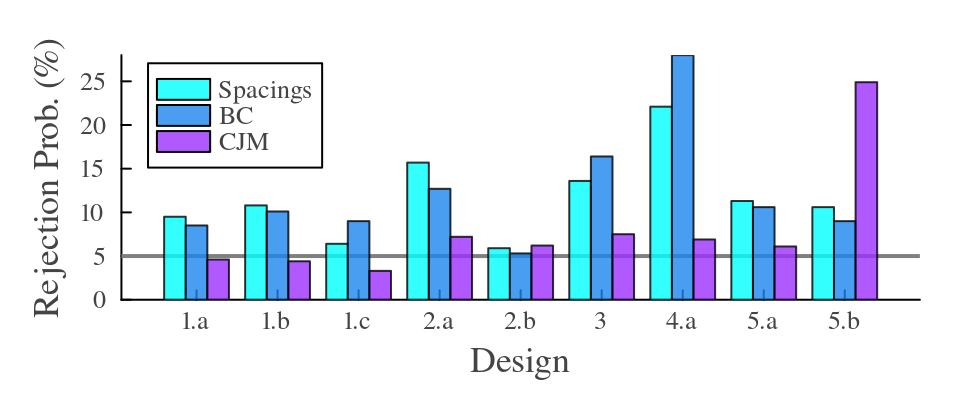}
\end{subfigure}
\hspace{0.3in}
\begin{subfigure}[t]{0.3\textwidth}
    \includegraphics[width=\textwidth]{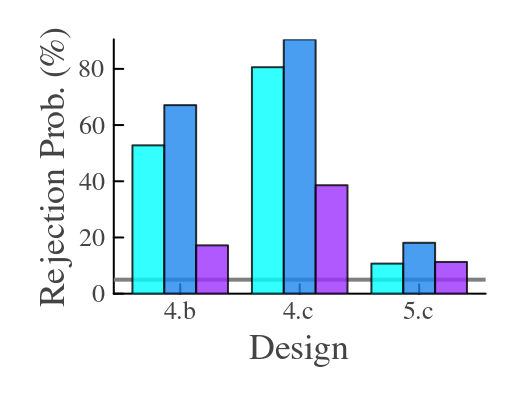}
\end{subfigure}
\vspace{0.1in}
\caption{Rejection Rates under $H_1$ with $n=500$ and $\alpha=5\%$}
\label{fig:MC500Power}
\end{figure}

Figure \ref{fig:MC500Power} shows the analogous results for an alternative hypothesis.  With a small change to the hypothesis and a low sample size, the changes in rejection rates are small.  Many of the differences in rejection rates reflect the differences under the null.  In designs 2.b and 5.b, the spacings test finds more power than BC despite higher over-rejection from BC under the null.  The test of CJM rejects below the nominal level in some designs due to its under-rejection under the null.  

Surprisingly, the spacings test and BC reject less frequently under the alternative of 5c than they did under the null.  The design subjects the tests to a piecewise constant function with steps near the cutoff that move upward as the running variable increases.  Since the alternative hypothesis moves mass from above the cutoff to below, it actually makes the density seem flatter near the cutoff.  

Failure to control size in designs 3-5 can arise from two sources: 1) nonsmoothness that slows convergence and 2) severe nonnormality that impairs the rule for choosing $q$.  For example, design 5 features a flat density near the cutoff.  The spacings test and BC should perform very well for small choices of $Q$, but the global shape of the distribution may cause the rule-of-thumb to pick an excessively large value.  If severe non-normality is suspected, we can use a cautiously small choice of $Q$.  This improves the size control of the spacings test at the expense of power.  Picking a smaller $Q$ is easy with the spacings test but requires more care with BC: some choices control size better than others, as targeted by their informed rule-of-thumb.  The tables in Appendix \ref{section:MC} show simulation results for small fixed choices of $Q$, and the spacings test controls size precisely.  These smaller values of $Q$ are not those targeted by BC, which is reflected in its performance.

\begin{figure}[h!]
    \centering
    \begin{subfigure}[t]{0.6\textwidth}
    \includegraphics[width=\textwidth]{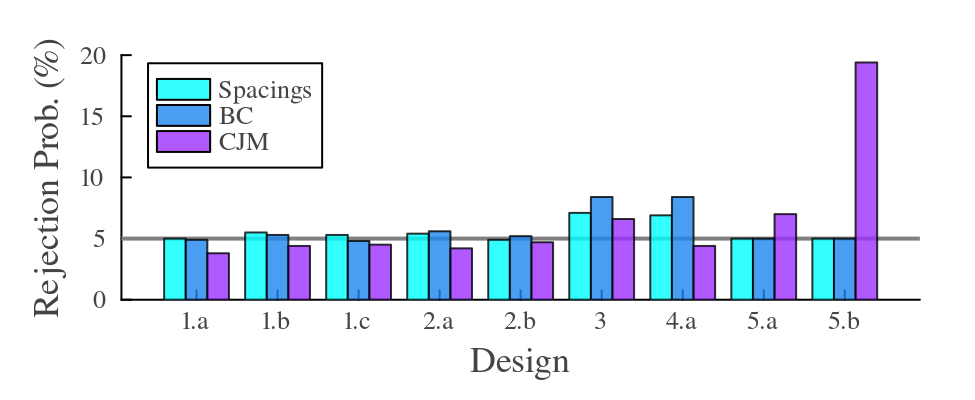}
\end{subfigure}
\hspace{0.3in}
\begin{subfigure}[t]{0.3\textwidth}
    \includegraphics[width=\textwidth]{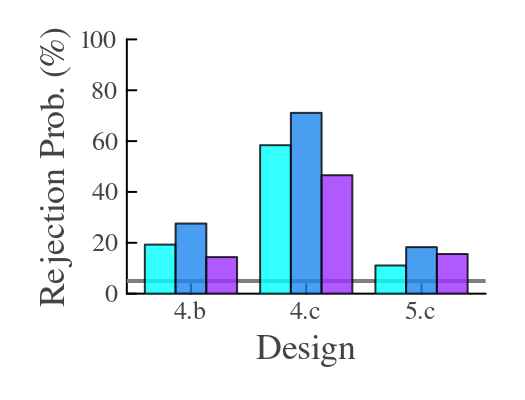}
\end{subfigure}
\vspace{0.1in}
\caption{Rejection Rates under $H_0$ with $n=1000$ and $\alpha=5\%$}
\label{fig:MC1000Level}
\end{figure}

\begin{figure}[h!]
    \centering
    \begin{subfigure}[t]{0.6\textwidth}
    \includegraphics[width=\textwidth]{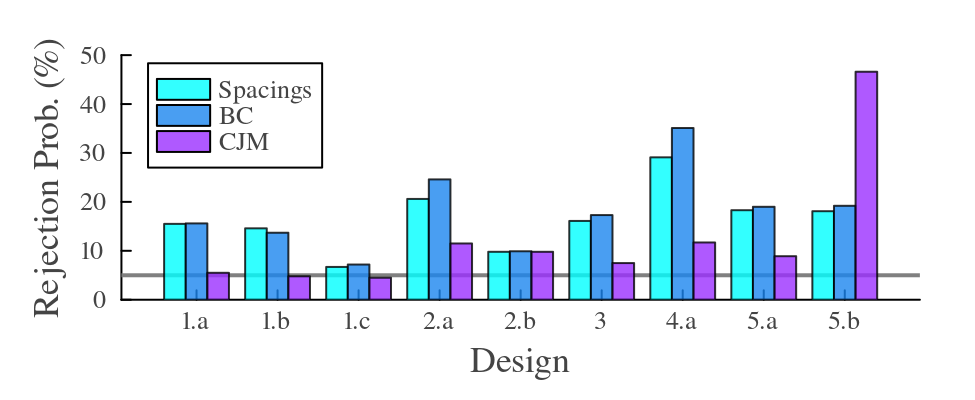}
\end{subfigure}
\hspace{0.3in}
\begin{subfigure}[t]{0.3\textwidth}
    \includegraphics[width=\textwidth]{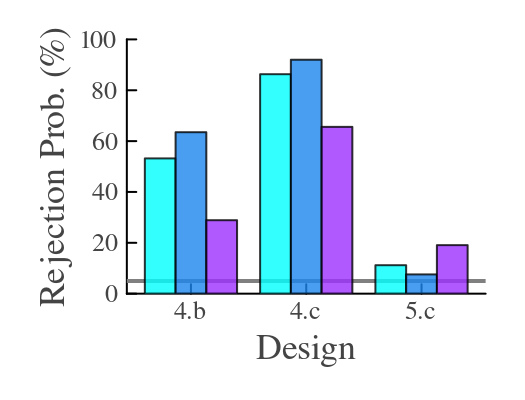}
\end{subfigure}
\vspace{0.1in}
\caption{Rejection Rates under $H_1$ with $n=1000$ and $\alpha=5\%$}
\label{fig:MC1000Power}
\end{figure}

I now increase the sample size to $n=1000$ and display the results in Figures \ref{fig:MC1000Level} and \ref{fig:MC1000Power}.  As expected, size control improves in most designs.  The tests still struggle with particularly nonsmooth designs, such as 4.b, 4.c, and 5.c.  The spacings test still controls size better than BC in most designs.  The rejection rates under the alternative show similar patterns as before.

\section{Conclusion}\label{sec:conclusion}

This paper proposes a new method to test for density discontinuity in RDD.  The test operates under relatively weak assumptions: it allows asymptotically valid inference even when the density has no derivatives.  It uses a number $q$ of local observations on each side of the cutoff, which we consider to be fixed or slowly increasing to infinity.  When fixed, the spacings test is asymptotically unbiased.  The test is designed to be optimal in the limiting experiment, which translates into asymptotic efficiency among asymptotically unbiased tests that use $q$ local observations on each side of the cutoff.  When $q$ increases slowly, the spacings test is consistent.

Relative to the g-order statistics test of Bugni and Canay (2021), the spacings test uses the same assumptions and eliminates the need for user randomization or restrictive choices of the tuning parameter.  This simplifies the choice of tuning parameter and prevents adverse effects from such restrictions.  The spacings test also provides more limiting power in many cases under the fixed-$q$ asymptotic framework.  When considering $q$ to slowly increase to infinity, the spacings test and g-order statistics test possess similar limiting properties.  In Monte Carlo simulations with smaller sample sizes, both tests perform comparably.

%% file: Sections/MainProofs.tex
\section{Proof of the main results}\label{appendix:proofs}

\subsection*{Lemma \ref{representation}}
\begin{proof}
For each $k\in\{0,1,...,n\}$, let
\begin{equation}\label{eqn:def:delta_k}
    \Delta_k=\left(\begin{array}{c}
    (d_1^+,d_2^+,...,d_{n-k+1}^+) \\
    (d_1^-,d_2^-,...,d_{k+1}^-)
\end{array}\right)
\end{equation}

Formally, the following arguments will construct iid Exponential(1) random variables $\{Y_i^+\},\{Y_i^-\}$ independent of $n^-$ such that $\Delta_{n^-}=W_{n^-}$ almost surely, where
\begin{equation}\label{eqn:def:w_k}
    W_k:=\left(\begin{array}{c}
    \frac{1-\tilde c}{\sum_{i=1}^{n-k+1}Y_i^+}(Y_i^+)_{i=1}^{n-k+1} \\
    \frac{\tilde c}{\sum_{i=1}^{k+1}Y_i^-}(Y_i^-)_{i=1}^{k+1}
\end{array}\right).
\end{equation}

Let $\Gamma_k^-\sim\text{Gamma}(k+1,1)$ and $\Gamma_k^+\sim\text{Gamma}(n-k+1,1)$ be independent of each other, $n^-$, and the spacings.  Defining $Y_i^-:=d_i^-\frac{\Gamma^-_{n^-}}{\tilde c}$ and $Y_i^+:=d_i^+\frac{\Gamma_{n^-}^+}{1-\tilde c}$ produces the appropriate random variables for $1\le i\le n^s+1$, $s\in\{+,-\}$.  For $i>n^s+1$, $Y_i^s$ can take arbitrary values as long as they're distributed as iid Exponential(1) random variables independent of $n^-$.  These values do not affect $W_{n^-}$.  It is easy to see that $\Delta_{n^-}=W_{n^-}$ almost surely by plugging in this construction.  All that remains is to prove that these variables have joint iid Exponential(1) distribution.

The remainder of the proof follows in two parts.  First, the arguments characterize the conditional distribution of $\Delta_k$.  Second, the arguments show that the constructed random variables have the correct exponential distribution.

Conditional on $n^-=k$ for fixed $k$, the distribution of $\Delta_k$ is singular, but all of its information can be recovered from an $n$-dimensional object with non-singular distribution.  Specifically, $d_{k+1}^-=\tilde c-\sum_{i=1}^kd^-_i$ and $d_{n-k+1}^+=1-\tilde c-\sum_{i=1}^{n-k}d_i^+$, and any further spacings are known to be $0$.  Define a truncated version as
\begin{equation}\label{eqn:def:tilde-delta-k}
    \tilde\Delta_k:=\left(\begin{array}{c}
    (d_1^+,...,d_{n-k}^+) \\
    (d_1^-,...,d_k^-)
\end{array}\right)
\end{equation}
Further, $\tilde \Delta_k$ can be rewritten using the order statistics of $\{U_i\}$.  Let
\begin{equation}\label{eqn:map:u->tilde_delta_k}
    \tilde \Delta_k(u):=\left(\begin{array}{c}
    (u_{k+1}-\tilde c,u_{k+2}-u_{k+1},...,u_{n}-u_{n-1}) \\
    (\tilde c-u_k,u_k-u_{k-1},...,u_2-u_1)
\end{array}\right)
\end{equation}
for $0\le u_1\le...\le u_n\le 1$, $u_k<\tilde c\le u_{k+1}$.
The next arguments will use the density of uniform order statistics to recover the density of $\tilde\Delta_k$. Inverting this mapping yields
\begin{equation}\label{eqn:map:tilde_delta_k->u}
    \mathbb U_k(\delta)=\left(\tilde c-\sum_{i=1}^k \delta_i^-,\tilde c-\sum_{i=1}^{k-1}\delta_i^-,...,\tilde c-\delta_1^-,\tilde c+\delta_1^+,\tilde c+\delta_1^++\delta_1^-,...,\tilde c+\sum_{i=1}^{n-k}\delta_i^+\right)
\end{equation}
for arguments $\delta_i^s\ge 0$ representing $d_i^s$ and satisfying $\sum_{i=1}^{k}\delta_i^-\le\tilde c$ and $\sum_{i=1}^{n-k}\delta_i^+\le1-\tilde c$.  Using the density transform method, the density of $\tilde\Delta_k$ conditional on $n^-=k$ is
\begin{equation}\label{eqn:density_transform:tilde_delta_k}
    f_{\tilde\Delta_k|n^-=k}(\delta)=f_{\mathbb U|n^-=k}(\mathbb U_k(\delta))|\det D_\delta \mathbb U_k(\delta)|.
\end{equation}

For convenience of computing the determinant above, order the arguments of $\delta$ as $(\delta_k^-,...,\delta_1^-,\allowbreak\delta_1^+,...,\delta_{n-k}^+)$.  This can be done freely, as swapping columns only changes the sign of the determinant.  The absolute value nullifies this effect.  With this ordering,
\begin{equation}\label{eqn:det:tilde_delta_k->u}
    D_\delta \mathbb{U}_k(\delta)=\left[\begin{array}{cc}
    \mathbf{A} & \mathbf{0} \\
    \mathbf{0} & \mathbf{B}
\end{array}\right]
\end{equation}
where $\mathbf{A}$ is upper triangular with all entries $-1$ and $\mathbf{B}$ is lower triangular with all entries $1$.  Their determinants are thus the product of their diagonals, which will be $\pm1$.  The determinant of the whole matrix is $\det\mathbf{A}\det\mathbf{B}=\pm1$ with absolute value $1$.

As for $f_{\mathbb{U}|n^-=k}(\mathbb{U}_k(\delta))$, it is well-known (e.g., Pyke (1965)) that the density of uniform order statistics is the constant value $n!$ for arguments $0\le u_1\le ...\le u_n\le 1$ (and 0 elsewhere).  This comes from there being $n!$ permutations of order statistics to get equiprobable draws from a uniform distribution.  Since $n^-$ is distributed as Binomial$(n,\tilde c)$ and can be recovered from the order statistics $\mathbb{U}$,
\begin{equation}\label{eqn:density:u,n-}
    f_{\mathbb{U},n^-}(u,k)=f_{\mathbb U}(u)\ind{u_k<\tilde c \le u_{k+1}}\hspace{1in}\eqbreak p_{n^-}(k)=\binom{n}{k}(\tilde c)^k(1-\tilde c)^{n-k}
\end{equation}

Therefore
\begin{equation}\label{eqn:density:tilde_delta_k}
    \begin{aligned}
    f_{\tilde\Delta_k|n^-=k}(\delta)&=f_{\mathbb U|n^-=k}(\mathbb U_k(\delta))\\
    &=\frac{n!}{\binom{n}{k}(\tilde c)^k(1-\tilde c)^{n-k}}\ind{\tilde c-\delta_1^-<\tilde c\le \tilde c+\delta_1^+}\eqbreakwith{\times}\ind{0\le \tilde c-\sum_{i=1}^k\delta_i^-\le ...\le \tilde c-\delta_1^-\le \tilde c+\delta_1^+\le ...\le \tilde c+\sum_{i=1}^{n-k}\delta_i^+\le 1}\\
    &=\frac{k!(n-k)!}{(\tilde c)^k(1-\tilde c)^{n-k}}\ind{\delta_i^s\ge 0\;\;\forall i,s;\;\sum_{i=1}^k\delta_i^-\le\tilde c;\;\sum_{i=1}^{n-k}\delta_i^+\le 1-\tilde c}\eqbreakwith{\times}\ind{\delta_1^->0}
\end{aligned}
\end{equation}
Note that the last indicator term $\ind{\delta_1^->0}$ can be dropped since it equals 1 almost everywhere with respect to the probability distribution.  This density captures the conditional distribution of $\tilde\Delta_k$ and thus the singular distribution of $\Delta_{n^-}$ conditional on $n^-=k$.

Now, the following arguments will show that the constructed random variables have the appropriate distribution.

Let
\begin{equation}\label{eqn:def:Y}
    \mathbb{Y}:=\left(\begin{array}{c}
    (Y_1^+,...,Y_{n+1}^+) \\
    (Y_1^-,...,Y_{n+1}^-)
\end{array}\right).
\end{equation}

Conditional on $n^-=k$, consider
\begin{equation}\label{eqn:def:Y_k}
    \mathbb{Y}_k:=\left(\begin{array}{c}
    (Y_1^+,...,Y_{n-k+1}^+) \\
    (Y_1^-,...,Y_{k+1}^-)
\end{array}\right).
\end{equation}

The following one-to-one mapping recovers the spacings and values of $\Gamma^-_k,\Gamma^+_k$ from $\mathbb Y_k$:
\begin{equation}\label{eqn:map:Y_k->tilde_delta_k,gamma}
    \tilde\Delta_k(\mathbb{Y}_k):=\left(\begin{array}{c}
    \left(\frac{(1-\tilde c) Y_i^+}{\sum_{i=1}^{n-k+1}Y_i^+}\right)_{i=1}^{n-k}\\
    \left(\frac{\tilde cY_i^-}{\sum_{i=1}^{k+1}Y_i^-}\right)_{i=1}^k
\end{array}\right)\hspace{0.4in}\eqbreak\Gamma_k^-(\mathbb Y_k):=\sum_{i=1}^{k+1}Y_i^-\hspace{0.4in}\Gamma_k^+(\mathbb Y_k):=\sum_{i=1}^{n-k+1}Y_i^+
\end{equation}
This allows use of the density transform method.  To simplify the calculations, first condition on $\Gamma_k^-=\gamma^-$ and $\Gamma_k^+=\gamma^+$.  $\mathbb Y_k$ has a singular distribution conditional on $\Gamma_k^-,\Gamma_k^+$, so consider
\begin{equation}\label{eqn:def:tilde_Y_k}
    \tilde{\mathbb{Y}}_k:=\left(\begin{array}{c}
    (Y_1^+,...,Y_{n-k}^+) \\
    (Y_1^-,...,Y_{k}^-)
\end{array}\right).
\end{equation}
With $\Gamma_k^-,\Gamma_k^+$ known, the dropped elements can be recovered, so all information is preserved.  Then $\tilde \Delta_k$ can be expressed as a one-to-one function of $\tilde{\mathbb Y}_k$ and $\Gamma_k^-,\Gamma_k^+$:
\begin{equation}\label{eqn:map:tilde_y_k,gamma->tilde_delta_k}
    \tilde\Delta_k(\tilde y,\gamma):=\left(\begin{array}{c}
    \left(\frac{(1-\tilde c)\tilde y_i^+}{\gamma^+}\right)_{i=1}^{n-k}\\
    \left(\frac{\tilde c \tilde y_i^-}{\gamma^-}\right)_{i=1}^k
\end{array}\right).
\end{equation}

Transforming the density of $\tilde \Delta_k$,
\begin{equation}\label{eqn:density_transform:tilde_Y_k}
    f_{\tilde{\mathbb Y}_k|n^-=k,\gamma^-,\gamma^+}(\tilde y)=f_{\tilde\Delta_k|n^-=k,\gamma^-,\gamma^+}(\tilde\Delta_k(\tilde y,\gamma))|\det D_{\tilde y}\tilde \Delta_k(\tilde y,\gamma)|.
\end{equation}
Since $\Gamma^-_k$ and $\Gamma^+_k$ are independent of the spacings,
\begin{equation}\label{eqn:density:tilde_delta_k-given-gamma}
    \begin{aligned}
    f_{\tilde\Delta_k|n^-=k,\gamma^-,\gamma^+}(\tilde\Delta_k(\tilde y,\gamma))&=f_{\tilde\Delta_k|n^-=k}(\tilde\Delta_k(\tilde y,\gamma))\\
    &=\frac{k!(n-k)!}{(\tilde c)^k(1-\tilde c)^{n-k}}\eqbreakwith{\times}\mathbbm{1}\Biggl\{\frac{\tilde c \tilde y_i^-}{\gamma^-}\ge 0\;\forall i;\;\frac{(1-\tilde c)\tilde y_i^+}{\gamma^+}\ge 0\;\forall i;
    \;\\
    &\hspace{1in}\sum_{i=1}^{n-k}\frac{(1-\tilde c) \tilde y_i^+}{\gamma^+}\le 1-\tilde c;\;\sum_{i=1}^k\frac{\tilde c\tilde y_i^-}{\gamma^-}\le\tilde c\Biggr\}\\
    &=\frac{k!(n-k)!}{(\tilde c)^k(1-\tilde c)^{n-k}}\eqbreakwith{\times}\ind{\tilde y_i^s\ge 0\;\forall i,s;\;\sum_{i=1}^{n-k}\frac{\tilde y_i^+}{\gamma^+}\le 1;\;\sum_{i=1}^k\frac{\tilde y_i^-}{\gamma^-}\le1}
\end{aligned}
\end{equation}
As for the determinant of the Jacobian, order the arguments as $\tilde y=\allowbreak(\tilde y_1^-,...,\tilde y_k^-,\allowbreak\tilde y_1^+,...,\tilde y_{n-k}^+)$.  This diagonalizes the Jacobian, making the absolute value of its determinant easily computable as $\left(\frac{\tilde c}{\gamma^-}\right)^{k}\left(\frac{1-\tilde c}{\gamma^+}\right)^{n-k}$.  Therefore
\begin{equation}\label{eqn:density:y_tilde-given-gamma}
    f_{\tilde{\mathbb Y}_k|n^-=k,\gamma^-,\gamma^+}(y)=\frac{k!(n-k)!}{(\gamma^-)^k(\gamma^+)^{n-k}}\ind{y_i^s\ge 0,\;\sum_{i=1}^{n-k}\frac{y_i^+}{\gamma^+}\le 1,\;\sum_{i=1}^k\frac{y_i^-}{\gamma^-}\le1}.
\end{equation}

The Gamma random variables have density
\begin{equation}\label{eqn:density:gamma}
    f_{\Gamma_k|n^-=k}(\gamma)=\frac{(\gamma^-)^k(\gamma^+)^{n-k}}{k!(n-k)!}\exp(-\gamma^+-\gamma^-)\ind{\gamma^s\ge 0},
\end{equation}

so combining with (\ref{eqn:density:y_tilde-given-gamma}) gives joint density
\begin{equation}\label{eqn:density:y_tilde,gamma}
    f_{\tilde{\mathbb Y}_k,\Gamma_k|n^-=k}(y,\gamma)=\exp(-\gamma^+-\gamma^-)\ind{y_i^s\ge 0,\;\sum_{i=1}^{n-k}\frac{y_i^+}{\gamma^+}\le 1,\;\sum_{i=1}^k\frac{y_i^-}{\gamma^-}\le1}\ind{\gamma^s\ge 0}.
\end{equation}
Now, there is a convenient invertible mapping between $(\tilde{\mathbb Y}_k,\Gamma_k)$ and $\mathbb Y_k$: let
\begin{equation}\label{eqn:map:G}
    G(\tilde y,\gamma):=\left(\begin{array}{c}
    (\tilde y^+_1,...,\tilde y_{n-k}^+,\gamma^+-\sum_{i=1}^{n-k}\tilde y_i^+)\\
    (\tilde y^-_1,...,\tilde y_{k}^-,\gamma^--\sum_{i=1}^k\tilde y_i^-)
\end{array}\right).
\end{equation}
The inverse of this mapping is trivial for $\tilde{\mathbb Y}_k$, and for $\Gamma_k^+,\Gamma_k^-$ it gives $\sum_{i=1}^{n-k+1}y_i^+$ and $\sum_{i=1}^{k+1}y_i^-$ respectively.

With a second density transformation,
\begin{equation}\label{eqn:density_transform:Y_k}
    f_{\mathbb Y_k|n^-=k}(y)=f_{\tilde{\mathbb Y}_k,\Gamma_k|n^-=k}(G^{-1}(y))|\det D_yG^{-1}(y)|.
\end{equation}

Ordering the inputs and outputs of $G^{-1}$ as $(y_1^+,...,y_{n-k+1}^+,y_1^-,...,y_{k+1}^-)$ and $(\tilde y_1^+,...,\tilde y_{n-k}^+,\allowbreak \gamma^+,\allowbreak \tilde y_1^-,...,\tilde y_k^-,\gamma^-)$, the Jacobian is
\begin{equation}\label{eqn:det:G}
    D_yG^{-1}(y)=\left[\begin{array}{cc}
    \mathbf{A} & \mathbf{0} \\
    \mathbf{0} & \mathbf{B}
\end{array}\right]
\end{equation}
where $\mathbf{A}$ and $\mathbf{B}$ have all entries as zero except for ones on their diagonal and last row.  The absolute value of the Jacobian's determinant is thus 1.  Finally, (\ref{eqn:density:y_tilde,gamma}) gives
\begin{equation}\label{eqn:density:Y_k}
    f_{\mathbb Y_k|n^-=k}(y)=\exp\left(-\sum_{i=1}^{n-k+1}y_i^+-\sum_{i=1}^{k+1}y_i^-\right)\ind{y_i^s\ge 0\;\forall i,s}.
\end{equation}
The other events in the indicator functions of (\ref{eqn:density:y_tilde,gamma}) are satisfied trivially.  The result is the joint distribution for $n+2$ iid Exponential(1) random variables.  As mentioned before, the values taken by $Y_i^s$ are not relevant when $i>n^s+1$.  Define them on the subset of the sample space with $n^-=k$ to be iid Exponential(1) and independent of the already-defined exponential random variables on this space.  Then 
\begin{equation}\label{eqn:density:Y}
    f_{\mathbb Y|n^-=k}(y)=\exp\left(-\sum_{i=1}^{n+1}y_i^+-\sum_{i=1}^{n+1}y_i^-\right),
\end{equation}
which implies the random variables in $\mathbb{Y}$ are iid Exponential(1) independent of $n^-$.
\end{proof}

\subsection*{Theorem \ref{mass}}
\begin{proof}

    If $F$ has a mass point at $c$, the closest $q$ observations weakly greater than $c$ equal $c$ with probability approaching 1.  This happens as long as $q^+=o_p(n)$: the mass point satisfies the part of Lemma \ref{X->c}'s hypothesis that guarantees convergence of the quantiles corresponding to observations above $c$.  The mass point captures all quantiles just above $P(X<c)$, so convergence in probability of the quantiles implies that the closest observations equal $c$ with probability tending to 1.  
    
    Then the spacings above $c$ equal 0 with probability tending to 1.  The assumption $P(X<c)>0$ guarantees that enough observations below $c$ will eventually exist.  These observations must be strictly less than $c$, so the sum of the $q^-$ nearest spacings must be positive when they exist.  Therefore the sum of the nearest spacings above and below will eventually be 0 and some positive number.  Plugging in, $T_n$ converges to 1 in probability.  This leads to rejection with probability approaching 1 as described at the end of the proofs of Theorems \ref{unbiasedness} and \ref{consistency}.
\end{proof}

\subsection*{Theorem \ref{fixed-q-dist}}
\begin{proof}
\begin{equation}
    \begin{aligned}
    nD_i^+&=n(X^+_{(i)}-X^+_{(i-1)})\ind{n^+\ge i}\\
    &=\frac n{\xi_{n,i}^+}[U_{(i)}^+-U^+_{(i-1)}]\ind{n^+\ge i}+o_{p}(1)\\
    &=\frac n{\xi_{n,i}^+}d_i^++o_{p}(1)
\end{aligned}
\end{equation}
for some $\xi_{n,i}^+\in[X_{(i-1)}^+,X_{(i)}^+]$.

The first equality uses the definition of $D_i^+$.  The second applies Lemmas \ref{X->c} and \ref{MVT}, with the little-o term capturing when the results don't apply.  The third uses the definition of $d_i^+$.  Recall that excessive spacings $D_i^s$ and $d_i^s$ with $i>n^s$ are defined to be 0.  The same arguments are used for $n(X^-_{(i-1)}-X_{(i)}^-)=nD^-_i$.  

Jointly, combine this with Lemma \ref{representation}:
\begin{equation}
    \begin{aligned}
    n\left(\begin{array}{c}
        (D_i^+)_{i=1}^{q^+}\\
        (D_i^-)_{i=1}^{q^-}
    \end{array}\right)&=\left(\begin{array}{c}
         \left(\frac n{\xi_{n,i}^+}d_i^+\right)_{i=1}^{q^+} \\
         \left(\frac n{\xi_{n,i}^-}d_i^-\right)_{i=1}^{q^-}
    \end{array}\right)+o_{p}(1)\\
    &\overset {a.s.}=\left(\begin{array}{c}
        \left(\frac n{\xi_{n,i}^+}(1-\tilde c)\frac{Y_{n,i}^+\ind{n^+\ge i}}{\sum_{j=1}^{n^++1}Y_{n,j}^+}\right)_{i=1}^{q^+}\\
        \left(\frac n{\xi_{n,i}^-}\tilde c\frac{Y_{n,i}^-\ind{n^-\ge i}}{\sum_{j=1}^{n^-+1}Y_{n,j}^-}\right)_{i=1}^{q^-}
    \end{array}\right)+o_{p}(1)\\
    &=\left(\begin{array}{c}
        \left(\frac 1{\xi_{n,i}^+}\frac{n}{n^++1}(1-\tilde c)\frac{Y_{n,i}^+\ind{n^+\ge i}}{\frac1{n^++1}\sum_{j=1}^{n^++1}Y_{n,j}^+}\right)_{i=1}^{q^+}\\
        \left(\frac 1{\xi_{n,i}^-}\frac{n}{n^-+1}\tilde c\frac{Y_{n,i}^-\ind{n^-\ge i}}{\frac1{n^-+1}\sum_{j=1}^{n^-+1}Y_{n,i}^-}\right)_{i=1}^{q^-}
    \end{array}\right)+o_{p}(1)\\
    &\overset{d}\to \left(\begin{array}{c}
        \left(\frac 1{f^+(c)}Y_i^+\right)_{i=1}^{q^+}\\
        \left(\frac 1{f^-(c)}Y_i^-\right)_{i=1}^{q^-}
    \end{array}\right)
\end{aligned}
\end{equation}

for $\{Y_i^+\},\{Y_i^-\}$ iid Exponential(1) random variables.  The last line results from the continuous mapping theorem and convergence of each term as follows: $\xi_{n,i}^+\overset {a.s.}\to f^+(c)$ by the bounds from Lemma \ref{MVT} that $\inf_{[X_{(i-1)}^+,X_{(i)}^+]}f(x)\le\xi_{n,i}^+\le \sup_{[X_{(i-1)}^+,X_{(i)}^+]}f(x)$, almost sure convergence of the local $X_{(i)}^+$'s to $c$ by Lemma \ref{X->c}, and one-sided continuity of $f$ at $c$.  The fraction $\frac{n}{n^++1}\to\frac1{1-\tilde c}$ a.s. by LLN and CMT.  This then cancels with the $(1-\tilde c)$ term next to it.  The denominator converges to 1 as established in the proof of Lemma \ref{X->c}.  The indicator terms equal 1 with probability tending to 1.  The same applies for the (-) versions below $c$.  Finally, $Y_{n,i}^+$ are all distributed as independent Exponential(1) random variables, so they trivially converge in distribution to independent Exponential(1) random variables.  The sequence $\{Y_{n,i}^+\}_{i=1}^{q^+}$ is also independent of $\{Y_{n,i}^-\}_{i=1}^{q^-}$, so the overall joint limiting distribution is independent between these two groups.
\end{proof}

\subsection*{Theorem \ref{unbiasedness}}
\begin{proof}

\textbf{Case A:} $P(X\in(c-\varepsilon,c))>0$ and $P(X\in[c,c+\varepsilon))>0$ for all $\varepsilon>0$.  

In conjunction with Assumption \ref{assumptions}(i), Theorem \ref{fixed-q-dist} gives that the local samples converge in distribution to two samples of independent exponential random variables, unless the limiting distribution is 0 or $\infty$.  Ignoring these cases for now, $T_n$ converges in distribution by the continuous mapping theorem to the respective statistic in the two-sample exponential problem of Han et al. (2012, Lemma 1).

Since $\phi$ is continuous almost everywhere and $T$ has a continuous distribution, $\phi(T_n)$ converges in distribution to $\phi(T)$ by the continuous mapping theorem.  The magnitude of the test $|\phi(T_n)|$ is dominated by 1, so $\phi(T_n)$ is uniformly integrable and thus $\E[\phi(T_n)]\to\E[\phi(T)]$ (Serfling (1980), Lemma 1.4A and Theorem 1.4A).  Lemma 1 of Han et al. (2012) gives the result in this case.

Now suppose that one of the limiting distributions is $0$ or $\infty$.  By Assumption \ref{assumptions}(ii), it cannot be that both limiting distributions are $0$ or that both are $\infty$.  If the upper spacings (above $c$) limit to 0 or the lower spacings (below $c$) limit to $\infty$, then $T_n\overset{d}\to 0$ by the continuous mapping theorem.  Similarly, if the upper spacings limit to $\infty$ or the lower spacings limit to 0, then $T_n\overset{d}\to 1$.  By Lemma \ref{critical_values} $0<C_1<C_2<1$, so $\phi$ is continuous almost everywhere with respect to the limiting distribution.  Therefore $\E[\phi(T_n)]\to1$ by the continuous mapping theorem and uniform integrability, as before.

\textbf{Case B:} Now consider the case in which $P(X\in(c-\varepsilon,c))=0$ or $P(X\in[c,c+\varepsilon))=0$ for some $\varepsilon>0$.  Suppose $P(X\in[c,c+\varepsilon))=0$.  By Assumption \ref{assumptions}(ii), this implies that $f^-(c)>0$, so the normalized local spacings below $c$ converge to an exponential sample or 0.  As a result, $n(c-X^-_{(q)})$ converges in distribution to a gamma distribution or 0.  With probability 1, there will eventually be an observation above (greater than or equal to) $c$.  Also with probability 1, none of the observations above $c$ will occur within a distance $\varepsilon$ of $c$.  

In these events, $n(X^+_{(q)}-c)\ge n\varepsilon\to\infty$.  By the continuous mapping theorem, $T_n$ converges in distribution to $0$.  If instead $P(X\in(c-\varepsilon,c))=0$, the same arguments reveal that $T_n$ converges in distribution to $1$.  Since $0<C_1$ and $C_2<1$, the continuous mapping theorem and uniform integrability guarantee that $\E[\phi(T_n)]\to 1$.

\end{proof}

\subsection*{Theorem \ref{ump}}

Fix $q^+,q^-$ and consider any sequence of tests $\{\tilde\phi_n\}_{n=1}^\infty$ that are measurable functions of the $q^+,q^-$ closest observations on the respective side of $c$.  Let $\cala{M}$ be a set of data generating processes that satisfy Assumptions \ref{assumptions}(i)-(ii) and includes all pairs of limiting densities $(f^-(c),f^+(c))\in(0,\infty)^2$.  Suppose the sequence of tests is asymptotically level and unbiased under $\cala{M}$.  Then for any data generating process that satisfies Assumptions \ref{assumptions}(i)-(ii) and $H_1$ in (\ref{eqn:hypothesis}),
    \[\limsup_{n\to\infty}\E[\tilde\phi_n(X_{(q)}^-,...,X_{(q)}^+)]\le \lim_{n\to\infty}\E[\phi(T_n)].\]

\begin{proof}
First suppose one of $f^+(c),f^-(c)$ are zero or infinite.  The proof of Theorem \ref{unbiasedness} shows the limiting power to be 1, so the conclusion holds.

Otherwise, $f^+(c)$ and $f^-(c)$ are finite and nonzero.  In this case, we can show that convergence of the spacings to the exponential distribution holds in total variation, and the result follows with some subsequence arguments.

Let $S_n=(X^-_{(q^-)},...,X^+_{(q^+)})$.  When all nearest observations are within the $\eta$ neighborhood of $c$ where $F$ is absolutely continuous, $S_n$ has density
\[\begin{aligned}
    h_n(S)=\binom{n}{q^++q^-}\binom{q^++q^-}{q^+}q^+!q^-!\left[\prod_{i=1}^{q^+}f(X_{(i)}^+)\right]\left[\prod_{i=1}^{q^-}f(X_{(i)}^-)\right]\\
    \times\left[F(X_{(q^-)}^-)+1-F(X^+_{(q^+)})\right]^{n-q^+-q^-},
\end{aligned}\]
which follows from a simple counting argument.

Transforming, the density of normalized spacings $(\Lambda_i^s) :=(nD_i^s)$ is
\[\begin{aligned}
    \tilde h_n(\Lambda)&=\frac{n!}{(n-q^+-q^-)!}\left[\prod_{i=1}^{q^+}f(c+n^{-1}\sum_{j=1}^i\Lambda_j^+)\right]\left[\prod_{i=1}^{q^-}f(c-n^{-1}\sum_{j=1}^i\Lambda_j^-)\right]\\
    &\quad\cdot \left[F(c-n^{-1}\sum_{i=1}^{q^-}\Lambda_i^-)+1-F(c+n^{-1}\sum_{i=1}^{q^+}\Lambda_i^+)\right]^{n-q^+-q^-}\cdot n^{-q^+-q^-}.
\end{aligned}\]
Expanding $F$ around $c$ gives
\[F(c+n^{-1}\sum_{i=1}^{q^+}\Lambda _i^+)=\tilde c +f^+(c)n^{-1}\sum_{i=1}^{q^+}\Lambda_i^++o\left(n^{-1}\right).\]
This and the same expansion for the other $F$ term give the final density
\[\begin{aligned}
    \tilde h_n(\Lambda)&=\frac{n!}{(n-q^+-q^-)!}\left[\prod_{i=1}^{q^+}f(c+n^{-1}\sum_{j=1}^i\Lambda_j^+)\right]\left[\prod_{i=1}^{q^-}f(c-n^{-1}\sum_{j=1}^i\Lambda_j^-)\right]\\
    &\quad\cdot \left[1-\frac{f^+(c)\sum_{i=1}^{q^+}\Lambda_i^++f^-(c)\sum_{i=1}^{q^-}\Lambda_i^-+o(1)}{n}\right]^{n-q^+-q^-}\cdot n^{-q^+-q^-}\\
    \to h_\theta(\Lambda):=& f^+(c)^{q^+}f^-(c)^{q^-}\exp\left(-f^+(c)\sum_{i=1}^{q^+}\Lambda_i^+-f^-(c)\sum_{i=1}^{q^-}\Lambda_i^-\right)
\end{aligned}\]
which is the PDF of the limiting distribution in Theorem \ref{fixed-q-dist}.  Denote this distribution by $H_\theta$ for $\theta=(f^+(c),f^-(c))$.

Since the nearest observations may not always fall within the $\eta$ neighborhood, let $A_n$ indicate the event that there are at least $q^+$,$q^-$ observations on the respective side and the nearest fall in the neighborhood.  Then we can write
\[\begin{aligned}
    P_n(\Lambda_n\in B)&=P_n(\Lambda_n\in B,A_n)+P_n(\Lambda_n\in B,A_n^c).
\end{aligned}\]
The second term on the RHS is bounded by $P_n(A_n^c)$, which converges to 0 by Lemma \ref{X->c}.  The first term on the RHS describes a distribution with density given by $\tilde h_n$ above multiplied by $\ind{\sum_i\Lambda_i^+<n\eta,\sum_i\Lambda_i^-<n\eta}\to 1$.  Then the product converges pointwise to $h_\theta$.  This implies convergence in total variation by Scheff\'e's Lemma.  Combined with the uniform convergence of the second term, we have
\[\sup_B|P_n(\Lambda_n\in B)-H_\theta(B)|\to 0.\]

WLOG, let $\tilde\phi_n$ be measurable functions of the normalized spacings $\Lambda_n$.
\[\begin{aligned}
    \E[\tilde\phi_n(\Lambda _n)]&=\int \tilde \phi_n(\Lambda_n)dH_\theta(\Lambda_n)+\left[\int \tilde \phi_n(\Lambda_n)dP_n(\Lambda_n)-\int \tilde \phi_n(\Lambda_n)dH_\theta(\Lambda_n)\right]
\end{aligned}\]
The bracketed term is bounded by the total variation distance between $P(\Lambda_n)$ and $H_\theta$, which converges to 0 as shown above.  Then
\[\begin{aligned}
    \E[\tilde\phi_n(\Lambda _n)]&=\int \tilde \phi_n(\Lambda_n)dH_\theta(\Lambda_n)+o(1)
\end{aligned}\]
Now let $n_k$ be a subsequence such that $\lim_{k\to\infty}\E[\tilde\phi_{n_k}(\Lambda_{n_k})]=\limsup_{n\to\infty}\E[\tilde\phi_n(\Lambda_n)]$.  

The map $g\mapsto\int\tilde\phi_{n_k}(\Lambda)g(\Lambda)d\lambda(\Lambda)$ for functions $g\in L^1(\R^{q^++q^-},\lambda)$ and Lebesgue measure $\lambda$ is linear and bounded and therefore in $(L^1)^*$.  Since the set of such maps defines a closed ball (as $0\le\tilde \phi_{n_k}\le 1$), this set is compact in the weak$^*$ topology (Folland, 1999 THM 5.18).  This set is also metrizable (Folland, 1999 Ch5 ex.50), so every sequence has a convergent subsequence in the sense that
\[\int \tilde\phi_{n_{k_j}}(\Lambda)g(\Lambda)d\lambda(\Lambda)\to\int \psi(\Lambda)g(\Lambda)d\lambda(\Lambda)\]
for every $g\in L^1(\R^{q^++q^-},\lambda)$, for some $\psi:\R^{q^++q^-}\to[0,1]$. 

Letting $g=h_{\theta'}$, the above convergence describes the limiting power of that sequence of tests under the parameters $\theta'$.  For $\theta'=\theta$, the first subsequence $n_k$ ensured that the limit is $\limsup_{n\to\infty}\E[\tilde\phi_n(\Lambda_n)]$.  The convergence above says that this limit equals the power of $\psi$ against $\theta'$ in the limiting experiment.  For $\theta'\not = \theta$ we know that $\tilde\phi_n$ is asymptotically level and unbiased, so $\psi$ must be level and unbiased in the limiting experiment.  Since the spacings test $\phi$ is the UMP unbiased test in the limiting experiment (Han et al., 2012), $\psi$ cannot have higher power than $\phi$ against $\theta$ in the limiting experiment.  Therefore,
\[\begin{aligned}
    \limsup_{n\to\infty}\E[\tilde\phi_n(\Lambda_n)]&=\lim_{j\to\infty}\int \tilde\phi_{n_{k_j}}(\Lambda)h_\theta(\Lambda)d\lambda(\Lambda)\\
    &=\int \psi(\Lambda)h_\theta(\Lambda)d\lambda(\Lambda)\\
    &\le \int\phi(\Lambda)h_\theta(\Lambda)d\lambda(\Lambda)\\
    &=\lim_{n\to\infty}\E[\phi(T_n)].
\end{aligned}\]

\end{proof}

\subsection*{Theorem \ref{increasing-q-dist}}
Suppose there exists some sequence $a_n\to\infty$, $a_n=o(n)$ such that $q_n^+/a_n\overset p\to 1$ and $q_n^+/q_n^--\gamma=o_p(1/\sqrt{q_n^-})$ for some $\gamma\in(0,\infty)$.  Denote $\beta:=\frac{f^-(c)}{f^+(c)}$.  Under Assumptions \ref{assumptions}(i-iii) and the additional assumption that both limits $f^+(c),f^-(c)$ are finite and nonzero,
    \[\frac{(1+\gamma\beta)^2}{\beta\sqrt{\gamma(1+\gamma)}}\sqrt{q_n^-}\left(T_n-\frac1{1+\beta\gamma}\right)=Z_n+B_n\left(\frac{[(\gamma+1)q_n^-]^{3/2}}{n}\right)[1+o_p(1)]\]
    where $Z_n\overset d\to N(0,1)$ and $|B_n|\le \frac{\sqrt\gamma}{2(\gamma+1)^2}\left[\frac{\gamma\kappa^+}{f^+(c)^2}+\frac{\kappa^-}{f^-(c)^2}\right]$.

\begin{proof}

The following arguments will use the asymptotic normal notation of Serfling (1980): a sequence of random variables $X_n$ is $AN(\mu_n,\sigma_n^2)$ if $\frac{X_n-\mu_n}{\sigma_n}\overset d\to N(0,1)$.

Notice that the positiveness of both limits $f^+(c),f^-(c)$ implies $P(X\in(c-\varepsilon,c))>0$ and $P(X\in[c,c+\varepsilon))>0$ for all $\varepsilon>0$.  With Assumption \ref{assumptions}(i), lemmas \ref{representation}, \ref{X->c}, and \ref{MVT} give that
\begin{equation}\label{eqn:avg_incr_q}
\begin{aligned}
    \frac{n}{q_n^-}(X^+_{(q_n^+)}-c)&=\frac{n}{q_n^-}\frac{U^+_{(q_n^+)}-\tilde c}{\xi_n^+}+o_p\left(1/\sqrt{q_n^-}\right)\\
    &=\frac{1-\tilde c}{\xi_n^+}\frac{n}{n^++1}\frac{q_n^+}{q_n^-}\frac{\frac1{q_n^+}\sum_{i=1}^{q_n^+}Y_{n,i}^+\ind{i\le n^+}}{\frac1{n^++1}\sum_{i=1}^{n^++1}Y_{n,i}^+}+o_{p}\left(1/\sqrt{q_n^-}\right).
\end{aligned} 
\end{equation}

Scaling by $\frac{n}{q_n^-}$ (and later normalizing by $\sqrt{q_n^-}$) as opposed to the more natural $\frac{n}{q_n^+}$ allows cancellation of these scalings in the overall statistic $T$.  Imposing that $q_n^+$ and $q_n^-$ are roughly proportional allows this to be dealt with.  The little-o term captures when the invoked results don't apply: Lemma \ref{MVT} requires $X_{(q_n^+)}^+$ to lie in the neighborhood where $f$ exists, which happens with probability tending to 1 by Lemma \ref{X->c}.  While any rate is valid, the equation displays $1/\sqrt{q_n^-}$ for later clarity.

By standard methods, we can show that the denominator of the right-most term will converge to 1 in probability and $\frac{n}{n^++1}$ will converge to $\frac1{1-\tilde c}$ almost surely.  The ratio $\frac{q_n^+}{q_n^-}$ converges to $\gamma$ in probability by assumption.  The remainder term $\frac1{\xi_n^+}$, with $\xi_n^+$ bounded by the density over $(c,X^+_{(q_n^+)})$, will converge to $f^+(c)$ in probability by Lemma \ref{X->c} and CMT.  Notice that without normalization the numerator will converge in probability to 1.  Then $\frac{n}{q_n^+}(X^+_{(q_n^+)}-c)$ converges in probability to $\frac{\gamma}{f^+(c)}$.  This fact will be useful later.

When normalized, the numerator is approximately normal.  Specifically, it is $AN(1,1/q_n^+)$:
\begin{equation}
    \begin{aligned}
        \sqrt{q_n^+}\left(\frac1{q_n^+}\sum_{i=1}^{q_n^+}Y_{n,i}^+\ind{i\le n^+}-1\right)&=\frac1{\sqrt{q_n^+}}\sum_{i=1}^{q_n^+}(Y_{n,i}^+-1)-\ind{q_n^+>n^+}\frac1{\sqrt{q_n^+}}\sum_{i=n^++1}^{q_n^+}Y_{n,i}^+\\
        &=\frac1{\sqrt{q_n^+}}\sum_{i=1}^{q_n^+}(Y_{n,i}^+-1)+o_{p}(1)\\
        &\overset{d}\to N(0,1)
    \end{aligned}
\end{equation}
Convergence in distribution follows from a CLT for a random number of summands.  Chung (1974, THM 7.3.2) provides the CLT for a single array of random variables, and the same proof methods based on Kolmogorov's inequality apply for triangular arrays such as the one above.  The method shows that $\frac1{\sqrt{q_n^+}}\sum_{i=1}^{q_n^+}(Y_{n,i}^+-1)=\frac1{\sqrt{a_n}}\sum_{i=1}^{a_n}(Y_{n,i}^+-1)+o_p(1)$, allowing us to apply the usual methods in turn. 

Similar arguments apply for the analogous expansion of $\frac{n}{q_n^-}(c-X^-_{(q_n^-)})$.  Since $\{Y_{n,i}^+\}$ and $\{Y_{n,i}^-\}$ are independent, the numerators will jointly limit to independent normal distributions.

Decompose 
\[\frac{n}{q_n^-}(X^+_{(q_n^+)}-c)=\frac1{\xi_n^+}W_n^++o_{p}\left(1/\sqrt{q_n^-}\right)\]
where
\[W_n^+:=(1-\tilde c)\frac{n}{n^++1}\frac{q_n^+}{q_n^-}\frac{\frac1{q_n^+}\sum_{i=1}^{q_n^+}Y_{n,i}^+\ind{i\le n^+}}{\frac1{n^++1}\sum_{i=1}^{n^++1}Y_{n,i}^+}\]

By Lemma \ref{AN}, asymptotic normality transfers to $W_n^+$ as a whole if 
\[(1-\tilde c)\frac{n}{n^++1}\frac{q_n^+}{q_n^-}\frac1{\frac1{n^++1}\sum_{i=1}^{n^++1}Y_{n,i}^+}=\gamma+o_p\left(\frac1{\sqrt{q_n^-}}\right).\]
In this case, $W_n^+$ is $AN(\gamma,\frac{\gamma^2}{q_n^+})$, which is equivalent to $AN(\gamma,\frac{\gamma}{q_n^-})$ by Lemma 1.5.5A of Serfling (1980).  

In fact,
\[\begin{aligned}
    \left[(1-\tilde c)\frac{n}{n^++1}\right]\left[\frac{q_n^+}{q_n^-}\right]\left[\frac1{\frac1{n^++1}\sum_{i=1}^{n^++1}Y_{n,i}^+}\right]&=\left[1+O_p(1/\sqrt{n})\right]\left[\gamma+o_p\left(1/\sqrt{q_n^-}\right)\right]\left[1+O_p(1/\sqrt{n})\right]\\
    &=\gamma+o_p\left(1/\sqrt{q_n^-}\right).
\end{aligned}\]

The middle term's rate of convergence comes from assumptions.  The outer terms are reciprocals of averages minus their nonzero expected value.  Since the number of summands grows at a rate of $n$, the delta method gives their nondegenerate normalization as $\sqrt{n}$.  Since $q_n^-=o_p(n)$, the limiting rate is $1/\sqrt{q_n^-}$.

Having established the asymptotic normality of $W_n^+$, now turn to $\frac1{\xi_n^+}$.  First, consider $\xi_n^+$.

\[\begin{aligned}
    |\xi_n^+-f^+(c)|&=\left |\frac{\int_c^{X^+_{(q_n^+)}}f(x)dx}{(X_{(q_n^+)}^+-c)}-f^+(c)+o_{p}\left(1/\sqrt{q_n^-}\right)\right|\\
    &=\left|(X_{(q_n^+)}^+-c)^{-1}\int_c^{X^+_{(q_n^+)}}[f(x)-f^+(c)]dx+o_{p}\left(1/\sqrt{q_n^-}\right)\right|\\
    &\le \left|(X_{(q_n^+)}^+-c)^{-1}\int_c^{X^+_{(q_n^+)}}\kappa ^+(x-c)dx\right|+o_{p}\left(1/\sqrt{q_n^-}\right)\\
    &=\kappa^+(X_{(q_n^+)}^+-c)^{-1}\frac12(X_{(q_n^+)}^+-c)^{2}+o_{p}\left(1/\sqrt{q_n^-}\right)\\
    &=\frac{\kappa^+}2(X_{(q_n^+)}^+-c)+o_{p}\left(1/\sqrt{q_n^-}\right)
\end{aligned}\]
The first line uses the expression for $\xi_n^+$ from Lemma \ref{MVT}.  The little-o term captures when the result doesn't apply and equals 0 with probability tending to 1.  Again, any rate is valid, but $1/\sqrt{q_n^-}$ is displayed for clarity with the upcoming steps.  The rest of the lines use the Lipschitz assumption and triangle inequality.

The previous steps relate to the reciprocal through a Taylor expansion.  For $\tilde \xi_n^+$ between $\xi_n^+$ and $f^+(c)$, $\tilde\xi_n^+\overset{p}\to f^+(c)$ and

\[\begin{aligned}
    \sqrt{q_n^-}\left|\frac1{\xi_n^+}-\frac1{f^+(c)}\right|&=\left|\frac{-\sqrt{q_n^-}}{(\tilde \xi_n^+)^2}(\xi_n^+-f^+(c))\right|\\
    &\le \frac{\sqrt{q_n^-}}{(\tilde \xi_n^+)^2}\left[\frac{\kappa^+}2(X_{(q_n^+)}^+-c)+o_{p}\left(1/\sqrt{q_n^-}\right)\right]\\
    &=\frac{\sqrt{q_n^-}}{(\tilde \xi_n^+)^2}\left[\frac{\kappa^+}2\frac{q_n^-}n\left (\frac\gamma{f^+(c)}+o_{p}(1)\right)+o_{p}\left(1/\sqrt{q_n^-}\right)\right]\\
    &=\left[\frac{\kappa^+\gamma }{2f^+(c)^3}+o_{p}(1)\right]\left(\frac{(q_n^-)^{3/2}}n\right)+o_{p}(1)
\end{aligned}\]

The third line uses that $\frac{n}{q_n^-}(X^+_{(q_n^+)}-c)\overset{p}\to \frac\gamma{f^+(c)}$, which comes from applying standard LLN methods to (\ref{eqn:avg_incr_q}).

If $(q_n^-)^{3/2}/n\overset p\to h\in(0,1)$, then the above bound limits to $\frac{\kappa^+\gamma h}{2f^+(c)^3}$.  This limiting bound may be achieved if $f(x)=f^+(c)+\kappa^+(x-c)$ near $c$, so the bound is tight.

Now,
\[\begin{aligned}
    \frac{n}{q_n^-}(X^+_{(q_n^+)}-c)&=\frac{W_n^+}{\xi_n^+}+o_{p}\left(1/\sqrt{q_n^-}\right)\\
    &=W_n^+\left[\frac1{f^+(c)}+\left(\frac1{\xi_n^+}-\frac1{f^+(c)}\right)\right]+o_{p}\left(1/\sqrt{q_n^-}\right)\\
    &= W_n^+\left[\frac1{f^+(c)}+R_n^+/\sqrt{q_n^-}\right]+o_{p}\left(1/\sqrt{q_n^-}\right)
\end{aligned}\]
where 
\[\begin{aligned}
    R_n^+:&=\sqrt{q_n^-}\left(\frac1{\xi_n^+}-\frac1{f^+(c)}\right)\\
    |R_n^+|&\le \left[\frac{\kappa^+\gamma}{2f^+(c)^3}+o_{p}(1)\right]\left(\frac{(q_n^-)^{3/2}}{n}\right)+o_{p}\left(1\right)\\
\end{aligned}\]

Nearly identical steps show that
\[\begin{aligned}
    \frac{n}{q_n^-}(c-X_{(q_n^-)}^-)&=W_n^-\left[\frac1{f^-(c)}+R_n^-/\sqrt{q_n^-}\right]+o_{p}\left(1/\sqrt{q_n^-}\right)\\
    R_n^-&:=\sqrt{q_n^-}\left(\frac1{\xi_n^-}-\frac1{f^-(c)}\right)\\
    |R_n^-|&\le \left[\frac{\kappa^-}{2f^-(c)^3}+o_{p}(1)\right]\left(\frac{(q_n^-)^{3/2}}n\right)+o_{p}(1)
\end{aligned}\]
where $W_n^-$ is similarly defined and $AN(1,\frac1{q_n^-})$.  The two variables $W_n^+$ and $W_n^-$ jointly limit to independent normal distributions: the numerator sums in their definitions are asymptotically independent as previously mentioned, and the other factors converge to constants at a faster rate, maintaining asymptotic independence via Slutsky's Theorem.

Finally, $T_n=g(\frac{n}{q_n^-}[c-X^-_{(q_n^-)}], \frac{n}{q_n^-}[X^+_{(q_n^+)}-c])$ for $g(x,y):=\frac{x}{x+y}$.  Taylor expanding $g$ around $(x_0,y_0)$ gives
\[g(x,y)=g(x_0,y_0)+g_y(x_0,\tilde y)(y-y_0)+g_x(\tilde x,y)(x-x_0)\]
\[g_x(x,y):=\frac{y}{(x+y)^2}\quad\quad g_y(x,y):=\frac{-x}{(x+y)^2}\]
for values $\tilde x$ and $\tilde y$ between $x$ and $x_0$ and $y$ and $y_0$.  Setting $x_0=1/f^-(c)$ and $y_0=\gamma/f^+(c)$, $x$ and $\tilde x$ converge in probability to $x_0$, and $y$ and $\tilde y$ converge in probability to $y_0$.  Letting $\beta:=f^-(c)/f^+(c)$, $g(x_0,y_0)=\frac{1}{1+\beta\gamma}$.  Then
\begin{equation}\label{eqn:T_inc_q}\begin{aligned}
    T_n=\frac{1}{1+\beta\gamma}&-\left[\frac{f^-(c)}{(1+\beta\gamma)^2}+o_{p}(1)\right]\left[\frac{W_n^+-\gamma}{f^+(c)}+\frac{W_n^+R_n^+}{\sqrt{q_n^-}}+o_{p}\left(1/\sqrt{q_n^-}\right)\right]\\
    &+ \left[\frac{\gamma\beta f^-(c)}{(1+\beta\gamma)^2}+o_{p}(1)\right]\left[\frac{W_n^--1}{f^-(c)}+\frac{W_n^-R_n^-}{\sqrt{q_n^-}}+o_{p}\left(1/\sqrt{q_n^-}\right)\right]
\end{aligned}\end{equation}
Denote $Z_n^+:=\sqrt{q^-_n/\gamma}(W_n^+-\gamma)\overset d\to N(0,1)$ and $Z_n^-:=\sqrt{q_n^-}(W_n^--1)\overset d\to N(0,1)$.  The limiting distributions are independent as described earlier.  Centering and scaling,
\begin{equation}\label{eqn:center_scale_T}\begin{aligned}
    \left(1+\beta\gamma\right)^2\sqrt{q_n^-}\left(T_n-\frac1{1+\beta\gamma}\right)&=-[f^-(c)+o_{p}(1)]\left[\frac{\sqrt{\gamma}}{f^+(c)}\sqrt{\frac{q_n^-}\gamma}(W_n^+-\gamma)+W_n^+R_n^+\right]\\
    &\quad\;+\left[\gamma\beta f^-(c)+o_{p}(1)\right]\left[\frac{1}{f^-(c)}\sqrt{q_n^-}(W_n^--1)+W_n^-R_n^-\right]+o_{p}(1)\\
    &=-\sqrt{\gamma}\beta Z_n^++\gamma\beta Z_n^--[f^-(c)+o_{p}(1)]W_n^+R_n^+\\
    &\hspace{1.4in}+[\gamma\beta f^-(c)+o_{p}(1)]W_n^-R_n^-\\
    \frac{(1+\beta\gamma)^2}{\beta\sqrt{\gamma(\gamma+1)}}\sqrt{q_n^-}\left(T_n-\frac1{1+\beta\gamma}\right)&=\frac{\gamma  Z_n^--\sqrt{\gamma} Z_n^+}{\sqrt{\gamma(\gamma+1)}}-\left[\frac{f^-(c)}{\beta\sqrt{\gamma(\gamma+1)}}+o_{p}(1)\right]W_n^+R_n^+\\
    &\hspace{1.25in}+\left[\frac{\beta\gamma f^-(c)}{\beta\sqrt{\gamma(\gamma+1)}}+o_{p}(1)\right]W_n^-R_n^-\\
    &=: Z_n+B_n\left(\frac{[(\gamma+1)q_n^-]^{3/2}}{n}\right)(1+o_{p}(1))
\end{aligned}\end{equation}
for $Z_n\overset d\to N(0,1)$ and
\[\begin{aligned}
    |B_n|&\le \frac1{(\gamma+1)^{3/2}}\frac{1}{\beta\sqrt{\gamma(\gamma+1)}}\left[\frac{\beta \gamma^2\kappa^+}{2f^+(c)^2}+\frac{\beta\gamma \kappa^-}{2f^-(c)^2}\right]\\
    &=\frac{\sqrt\gamma}{2(\gamma+1)^2}\left[\frac{\gamma\kappa^+}{f^+(c)^2}+\frac{\kappa^-}{f^-(c)^2}\right]
\end{aligned}\]

In the simple case where $\gamma=1$ and $\kappa^+=\kappa^-=:\kappa$,
\[|B_n|\le \frac{\kappa(\beta^2+1)}{8f^-(c)^2}.\]

\end{proof}

\subsection*{Theorem \ref{consistency}}

Fix $\alpha\in (0,1)$.  Suppose there exists some sequence $a_n\to\infty$, $a_n=o(n^{2/3})$ such that $q_n^+/a_n\overset p\to 1$ and $q_n^+/q_n^--\gamma=o_p(1/\sqrt{q_n^-})$ for some $\gamma\in(0,\infty)$.  Let Assumptions \ref{assumptions}(i-iii) hold.  Then:
    \begin{enumerate}[label=\Alph*)]
        \item Under $H_0$ in (\ref{eqn:hypothesis}), $\lim_{n\to\infty}\E[\phi(T_n)]=\alpha$.
        \item Under $H_1$ in (\ref{eqn:hypothesis}), $\lim_{n\to\infty}\E[\phi(T_n)]=1$.
    \end{enumerate}
\begin{proof}
Suppose first that both limits $f^+(c),f^-(c)$ are finite and nonzero so that Theorem \ref{increasing-q-dist} applies.  Then consistency can be shown using standard Slutsky arguments as follows.

\begin{equation}
    \E[\phi(T_n)]=P(T_n>C_{n,2})+P(T_n<C_{n,1})
\end{equation}
We will work directly with the first term.  The arguments for the second term are the same.
\begin{equation}
\begin{aligned}
    P(T_n>C_{n,2})
&=P\left(\frac{\sqrt{q_n^-(1+\gamma)^3}}{\sqrt{\gamma}}(T_n-\frac1{1+\gamma})>\frac{\sqrt{q_n^-(1+\gamma)^3}}{\sqrt{\gamma}}(C_{n,2}-\frac1{1+\gamma})\right)\\
    &=P\Biggl (\frac{\beta(1+\gamma)^2}{(1+\beta\gamma)^2}Z+\sqrt{a_n}\left[\frac{(1+\gamma)^{3/2}}{\gamma}+o_p(1)\right]\left[\frac1{1+\beta\gamma}-\frac1{1+\gamma}\right]+o_p(1)\\
    &\hspace{1in}>z_{1-\alpha/2}+o_p(1)\Biggr)
\end{aligned}
\end{equation}

where $Z\sim N(0,1)$.  The second line uses Theorem \ref{increasing-q-dist} and Lemma \ref{critical_values}.  When $\beta=1$, the above converges to $P(Z>z_{1-\alpha/2})=\alpha/2$ by Slutsky's Theorem.  Combined with same arguments for $P(T_n<C_{n,1})$, the test rejects with probability tending to $\alpha$ under the null.

When $\beta\not=1$, we can divide both sides of the inequality by $\sqrt{a_n}$ to get
\begin{equation}
    P(T_n>C_{n,2})=P\left(\left[\frac{(1+\gamma)^{3/2}}\gamma+o_p(1)\right]\left[\frac1{1+\beta\gamma}-\frac1{1+\gamma}\right]+o_p(1)>o_p(1)\right).
\end{equation}
By Slutsky's Theorem, the above probability converges to one if $\beta<1$ and zero if $\beta>1$.  The same arguments show that the other term, $P(T_n<C_{n,1})$, converges one if $\beta>1$ and zero if $\beta<1$.  Together, their sum converges to one in both cases, proving consistency against these alternatives.

Now suppose that $f^+(c)$ is zero or infinity.  In equation (\ref{eqn:avg_incr_q}) and subsequent discussion in the proof of Theorem \ref{increasing-q-dist}, $\frac{n}{q_n^-}(X^+_{(q_n^+)}-c)\overset{p}\to\frac{\gamma}{f^+(c)}$.  In this case, the limit is infinity or zero, respectively.  Since $f^-(c)$ cannot be the same value, $T_n$ will converge to zero or one in probability.  The critical values converge to $\frac1{1+\gamma}\in(0,1)$ in probability, so the test rejects $T_n$ with probability approaching one by Slutsky's Theorem.  The arguments are nearly identical if $f^-(c)$ is zero or infinity.

\end{proof}

\subsection*{Theorem \ref{local}}
    Fix $\alpha\in(0,1)$.  Consider a sequence of CDFs $F_n$, each of which satisfy the assumptions of Theorem \ref{increasing-q-dist} with $c$, $\tilde c$, $\eta$, and $\kappa$ as absolute constants.  Let $q_n^-,q_n^+$ be as in Theorem \ref{consistency} with a fixed sequence $a_n$.  Suppose the CDFs have density limits at $c$ such that $f_n^+(c)\to f^+(c)\in(0,\infty)$, $f_n^-(c)\to f^-(c)\in(0,\infty)$, and $\sqrt{(1/\gamma+1)a_n}\left(\frac{f_n^-(c)}{f_n^+(c)}-1\right)\to \Delta\not = 0$.  Then
    \[\E[\phi_n(T_n)]\to P\left(\left|Z+\frac{\sqrt{\gamma}\Delta}{\gamma+1}\right|>z_{1-\alpha/2}\right)>\alpha\]
    where $Z\sim N(0,1)$ and $z_{1-\alpha/2}$ is the $(1-\frac{\alpha}2)$-quantile of $Z$.

\begin{proof}
Since the density limits $f^+(c),f^-(c)$ are positive, and each density satisfies the Lipschitz assumption with a universal constant $\kappa$, the regularity conditions for Lemma \ref{X->c} apply.  Thus the proof of Theorem \ref{increasing-q-dist} goes through for the most part: the constant $\tilde c$ means that $n^+$ has a fixed distribution, the remainder terms $\xi_n^+,\xi_n^-$ converge with the fixed Lipschitz constant, and $R_n^+,R_n^-$ converge since the densities are bounded away from zero by the nonzero limit and fixed Lipschitz constant.

The sequence of density limits at $c$ only enter through $\xi_n^+$ and $\xi_n^-$.  Resuming from (\ref{eqn:T_inc_q}) and (\ref{eqn:center_scale_T}), 

\[\begin{aligned}
    T_n&=\frac{1}{1+\beta_n\gamma}&-\left[\frac{f_n^-(c)}{(1+\beta_n\gamma)^2}+o_{p}(1)\right]\left[\frac{W_n^+-\gamma}{f^+_n(c)}+o_{p}\left(1/\sqrt{q_n^-}\right)\right]\\
    &&+\left[\frac{\gamma\beta_n f^-_n(c)}{(1+\beta_n\gamma)^2}+o_{p}(1)\right]\left[\frac{W_n^--1}{f^-_n(c)}+o_{p}\left(1/\sqrt{q_n^-}\right)\right]
\end{aligned}\]
The bias terms $R_n^+$ and $R_n^-$ are absorbed by the little-o terms since this theorem assumes $q_n^-$ increases sufficiently slowly. 

Centering and scaling according to the null hypothesis,
\[\begin{aligned}
    \frac{(1+\gamma)^2}{\sqrt{\gamma(\gamma+1)}}\sqrt{q_n^-}\left(T_n-\frac1{1+\gamma}\right)&=\frac{(1+\gamma)^2}{\sqrt{\gamma(\gamma+1)}}\sqrt{q_n^-}\left(\frac1{1+\beta_n\gamma}-\frac1{1+\gamma}\right)\\
    &\quad\quad  -\frac{(1+\gamma)^2}{\sqrt{\gamma+1}}\frac{\beta_n}{(1+\beta_n\gamma)^2}\sqrt{\frac{q_n^-}\gamma}(W_n^+-\gamma)\\
    &\quad\quad  +\frac{(1+\gamma)^2}{\sqrt{\gamma(\gamma+1)}}\frac{\gamma\beta_n}{(1+\beta_n\gamma)^2}\sqrt{q_n^-}(W_n^--1)+o_{p}(1)\\
    &=\frac{(1+\gamma)^2}{\sqrt{\gamma(\gamma+1)}}\sqrt{q_n^-}\left(\frac{-\gamma}{(1+\tilde \beta_n\gamma)^2}(\beta_n-1)\right)\\
    &\quad\quad-\frac{Z_n^+}{\sqrt{\gamma+1}}+\frac{\gamma Z_n^-}{\sqrt{\gamma(\gamma+1)}}\\
    &= \frac{-\sqrt\gamma}{(\gamma+1)}\Delta+Z_n
\end{aligned}\]
The intermediary variable $\tilde \beta_n$ takes values between $\beta_n$ and 1 by Taylor's theorem.  $Z_n\overset d\to N(0,1)$, and $Z_n^+,Z_n^-$ behave as in the proof of Theorem \ref{increasing-q-dist}.

As in the proof of Theorem \ref{consistency}, 
\[\E[\phi_n(T_n)]\to P\left(\left|Z+\frac{\sqrt{\gamma}\Delta}{\gamma+1}\right|>z_{1-\alpha/2}\right)\]
which is greater than $\alpha$ by well-known properties of the normal distribution.

\end{proof}

%% file: Sections/AuxilliaryProofs.tex
\section{Auxiliary Results}
\subsection{Convergence to $c$}
This paper's test relies on the spacings $D_i^+=X^+_{(i)}-X^+_{(i-1)}$ and $D_i^-=X^-_{(i-1)}-X^-_{(i)}$ occurring near $c$ for $1\le i\le q$.  The following lemma provides conditions for convergence to c and allows $q=q_n$ to grow with $n$.  The statement of the lemma uses $q_n$ spacings on each side for simplicity, but the result also applies for $q_n^-$ spacings below $c$ and $q_n^+$ above by taking $q_n=\max\{q_n^+,q_n^-\}$ and using the monotonic relationship of the order statistics.

\begin{lemma}\label{X->c}
    Let $\{X_{in}\}_{i=1}^n$ be a triangular array of independent real-valued random variables with CDF $F_n$ and $c\in\R$ such that $\lim_{n\to\infty}P(X_{in}<c)\in(0,1)$, and let $q_n=o_p(n)$.  If there exists a function $\nu$ such that $P(X_{in}\in(c-\varepsilon,c))>\nu(\varepsilon)>0$ and $P(X_{in}\in[c,c+\varepsilon))>\nu(\varepsilon)>0$ for all $\varepsilon>0$ and $n$, then $X^+_{(q_n)}$ and $X^-_{(q_n)}$ converge to $c$ in probability.
\end{lemma}
\begin{proof}
Let $X_{in}\overset{a.s.}=F_n^{-1}(U_{in})$ for $U_{in}$ iid Uniform(0,1).  Let $\tilde c_n:=P(X_{in}<c)$, and let $\tilde c:=\lim_{n\to\infty}\tilde c_n$.  Since there may not always be $q_n$ observations on each side, let any ill-defined order-statistics (e.g. $U^+_{n,(n^++2)}$ and $U^-_{n,(n^-+2)}$) be $U^+_{n,(n^+)}$ and $U^-_{n,(n^-)}$, and let any ill-defined spacings be 0.  This is asymptotically negligible:
\begin{equation}\label{eqn:conv:n+/n}
    \frac{n^+}n=\frac{\sum_{i=1}^n\ind{U_{n,i}\ge \tilde c_n}}{n}\overset{p}\to 1-\tilde c.
\end{equation} 

Then
\begin{equation}\label{eqn:conv:q/n+}
    \frac{q_n}{n^+}=\frac{q_n}n\frac{n}{n^+}=o_p(1)O_{p}(1)=o_{p}(1).
\end{equation}

Symmetric arguments apply for $n^-$. 

\begin{equation}\label{eqn:conv:U_q_n}
    \begin{aligned}
    U^+_{n,(q_n)}&=\tilde c_n+\sum_{i=1}^{\min\{q_n,n^+\}}d^+_{n,i}\\
    &\overset{a.s.}=\tilde c_n+(1-\tilde c_n)\frac{\sum_{i=1}^{\min\{q_n,n^+\}}Y^+_{n,i}}{\sum_{i=1}^{n^++1}Y^+_{n,i}}\\
    &=\tilde c_n+(1-\tilde c_n)\frac{\frac1{n}\sum_{i=1}^{\min\{q_n,n^+\}}Y^+_{n,i}}{\frac1{n^++1}\sum_{i=1}^{n^++1}Y^+_{n,i}}\frac{n^+}{n^++1}\frac{n}{n^+}\\
    &=\tilde c_n+(1-\tilde c_n)\frac{\frac1{n}\sum_{i=1}^{\min\{q_n,n^+\}}Y^+_{n,i}}{\frac1{n^++1}\sum_{i=1}^{n^++1}Y^+_{n,i}}O(1)O_{p}(1)\\
\end{aligned}
\end{equation}
The first equality decomposes the order statistics into a sum of spacings.  The second uses Lemma \ref{representation}.  The third multiplies and divides by several terms.  The fourth uses (\ref{eqn:conv:n+/n}).  The numerator and denominator converge in probability to 0 and 1, which follows from standard methods.  Therefore $U_{n,(q_n)}^+$ converges to $\tilde c$ in probability.  Similar arguments apply for $U_{n,(q_n)}^-$.  

Now, $X^+_{n,(q_n)}\overset{a.s.}=F_n^{-1}(U^+_{n,(q_n)})$ converges to $c$ under some uniformity conditions.  By assumption, $F_n(c+\varepsilon)-\tilde c_n=P_n(X_{in}\in[c,c+\varepsilon))>\nu(\varepsilon)$.  Then $F_n(c+\varepsilon)>\tilde c_n+\nu(\varepsilon)$ implies that for $u\in(\tilde c_n,\tilde c_n+\nu(\varepsilon))$ we have $F_n^{-1}(u)\in(c,c+\varepsilon)$.  Then convergence of $U^+_{n,(q_n)}$ to $\tilde c$ gives convergence of $X^+_{n,(q_n)}$ to $c$.  The same arguments go through for $X_{n,(q_n)}^-$.
\end{proof}

\subsection{MVT for Inverse CDF}
\begin{lemma}\label{MVT}
    Let $\{X_i\}_{i=1}^n$ be a random sample from a CDF $F$ that is absolutely continuous on $(a,b)$ with density $f$.  Suppose that there exist independent Uniform(0,1) random variables $\{U_i\}_{i=1}^n$ such that $X_i=F^{-1}(U_i)$ almost surely.  Then the following holds on a set with probability 1:

    For all $a<X_i\le X_j<b$, $i\not=j$ define $\xi_{ij}:=\frac{\int_{X_i}^{X_j}f(t)dt}{X_j-X_i}$.  Then
    \begin{enumerate}[label=\alph*)]
        \item $\xi_{ij}>0$
        \item $\inf_{x\in[X_i,X_j]}f(x)\le \xi_{ij}\le \sup_{x\in[X_i,X_j]}f(x)$, and
        \item $X_j-X_i=\frac{U_j-U_i}{\xi_{ij}}$.
    \end{enumerate}
    Furthermore, the result also holds for all $a<X_i<b$ with $X_j$ replaced by a constant $c\in(a,b)$ and $U_j$ replaced by $F(c)$.
    \end{lemma}

\begin{proof}
    On a set $A$ with probability 1, $X_i=F^{-1}(U_i)$ for all $i$ by assumption.  By continuity of the distributions of $U$ and $X$, $X_i\not=X_j$ and $U_i\not = U_j$ for all $i\not=j$ on a set $B$ with probability 1.  The intersection of $A$ and $B$ has probability 1, and the following arguments will apply to this set.  Since $F$ is absolutely continuous on $(a,b)$,
    \begin{equation}
        F(X_j)-F(X_i)=\int_{X_i}^{X_j}f(t)dt.
    \end{equation}
    The integral is trivially bounded by
    \begin{equation}
        (X_j-X_i)\inf_{x\in[X_i,X_j]}f(x)\le\int_{X_i}^{X_j}f(t)dt\le (X_j-X_i)\sup_{x\in[X_i,X_j]}f(x)
    \end{equation}
    Define $\xi_{ij}$ as
    \begin{equation}
        \xi_{ij}=\frac{\int_{X_i}^{X_j}f(t)dt}{X_j-X_i}.
    \end{equation}
    From the above bounds,
    \begin{equation}
        \inf_{x\in[X_i,X_j]}f(x)\le\xi_{ij}\le \sup_{x\in[X_i,X_j]}f(x).
    \end{equation}
    Furthermore, 
    \begin{equation}\label{eqn:mvt_integral}
        0\not=U_j-U_i=F(X_j)-F(X_i)=\int_{X_i}^{X_j}f(t)dt
    \end{equation}
    by continuity of $F$, so $\xi_{ij}>0$.  Finally, dividing (\ref{eqn:mvt_integral}) by $(X_j-X_i)$ yields
    \begin{equation}
        \frac{U_j-U_i}{\xi_{ij}}=X_j-X_i.
    \end{equation}

    The proof for a constant $c$ simply replaces $X_i$ or $X_j$ with $c$.
    
\end{proof}

\subsection{Asymptotic Normality}
\begin{lemma}\label{AN}
    Suppose $X_n$ is AN($\mu,\sigma^2_n$) in that $\frac{X_n-\mu}{\sigma_n}\overset{d}\to N(0,1)$.  Let $Y_n\overset{p}\to y\not =0$.  If $\frac{Y_n-y}{\sigma_n}\overset{p}\to h\in\R$, then $X_nY_n$ is AN($\mu (y+\sigma_nh),\sigma^2_ny^2$).
    
\end{lemma}
\begin{proof}
    By Slutsky's theorem,
    \begin{equation}\label{eqn:slutsky}
        \frac{Y_n}y\frac{X_n-\mu}{\sigma_n}\overset{d}\to N(0,1).
    \end{equation}
    \begin{equation}
        \frac{X_nY_n-\mu (y+\sigma_nh)}{y\sigma_n}=\frac{Y_n}y\frac{X_n-\mu}{\sigma_n}+\frac\mu y\left(\frac{Y_n-y}{\sigma_n}-h\right).
    \end{equation}
    The first term converges in distribution to N(0,1) by (\ref{eqn:slutsky}), and the second converges in probability to 0.  By Slutsky's theorem, the result holds.
\end{proof}

\subsection{Convergence to Normal}

\begin{lemma}\label{ideal->normal}
    Let $\{Y_i^+\},\{Y_i^-\}$ be independent samples from exponential random variables with parameters $\lambda_+$ and $\lambda_-$, respectively.  Define
    \[T_n^*=\frac{\sum_{i=1}^{q_n^-}Y_i^-}{\sum_{i=1}^{q_n^-}Y_i^-+\sum_{i=1}^{q_n^+}Y_i^+}\]
    for $q_n^+,q_n^-\to\infty$ and $\sqrt{q_n^-}\left(\frac{q_n^+}{q_n^-}-\gamma\right)\to0$ for $\gamma\in(0,\infty)$.  Then for $\beta:=\frac{\lambda_-}{\lambda_+}$,
    \[T_n^*\text{ is }AN\left(\frac{1}{1+\beta\gamma},\frac1{q_n^-}\frac{\gamma(1+\gamma)\beta^2}{(1+\beta\gamma)^4}\right).\]
\end{lemma}

\begin{proof}

By the central limit theorem, $\frac{1}{q_n^-}\sum_{i=1}^{q_n^-}Y_i^-$ is $AN(\frac1{\lambda_-},\frac1{q_n^-\lambda_-^2})$.  Similarly, $\frac{1}{q_n^-}\sum_{i=1}^{q_n^+}Y_i^+=\frac{q_n^+}{q_n^-}\frac{1}{q_n^+}\sum_{i=1}^{q_n^+}Y_i^+$ is $AN(\frac{\gamma}{\lambda_+},\frac{\gamma}{q_n^-\lambda_+^2})$ by Lemma \ref{AN} and Lemma 1.5.5A of Serfling (1980).

Notice that $T_n^*=g\left(\frac{1}{q_n^-}\sum_{i=1}^{q_n^-}Y_i^-, \frac{1}{q_n^-}\sum_{i=1}^{q_n^+}Y_i^+\right)$ for $g(x,y)=\frac{x}{x+y}$.  The derivative of $g$ is 
\begin{equation}
    \nabla g(x,y)=\frac1{(x+y)^2}\left(y,-x\right),
\end{equation}
which evaluates at the means to
\begin{equation}
    \nabla g\left(\frac1{\lambda_-},\frac\gamma{\lambda_+}\right)=\frac1{\left(\frac1{\lambda_-}+\frac{\gamma}{\lambda_+}\right)^2}\left(\frac{\gamma}{\lambda_+},\frac{-1}{\lambda_-}\right).
\end{equation}
The independence of the random variables allows joint asymptotic normality with means $\frac1{\lambda_-},\frac{\gamma}{\lambda_+}$ and covariance matrix
\begin{equation}
    \frac{1}{q_n^-}\left[\begin{array}{cc}
        \lambda_-^{-2} & 0 \\
        0 & \gamma\lambda_+^{-2}
    \end{array}\right].
\end{equation}
By the delta method (Serfling 1980, Theorem 3.3A),
\begin{equation}
    T_n^*\text{ is }AN\left(\frac{\frac1{\lambda_-}}{\frac1{\lambda_-}+\frac{\gamma}{\lambda_+}},\frac1{q_n^-}\frac{\gamma(1+\gamma)}{\lambda_+^2\lambda_-^2(\frac1{\lambda_-}+\frac{\gamma}{\lambda_+})^4}\right).
\end{equation}
Rearranging gives the form stated in the lemma.
\end{proof}

\subsection{Critical Values}
\begin{lemma}\label{critical_values}
    Let $q_n^+,q_n^-$ be positive integers diverging to infinity such that $\frac{q_n^+}{q_n^-}=\gamma+o(1/\sqrt{q_n^-})$ for some $\gamma\in(0,\infty)$, and let $\mu=\frac1{1+\gamma}$ and $\sigma^2=\frac{\gamma}{(1+\gamma)^3}$.  For a fixed $\alpha\in(0,1)$, the critical values $C_{n,1},C_{n,2}$ defined in (\ref{eqn:defn_critical_values}) satisfy
    \begin{enumerate}[label=\alph*)]
        \item $0<C_{n,1}<C_{n,2}<1$
        \item $\left(\frac{\sqrt{q_n^-}}{\sigma}(C_{n,1}-\mu),\frac{\sqrt{q_n^-}}{\sigma}(C_{n,2}-\mu)\right)\to (z_{\alpha/2},z_{1-\alpha/2})$
    \end{enumerate}
    where $z_{\alpha/2},z_{1-\alpha/2}$ are critical values of the standard normal distribution.

    Furthermore, suppose instead that $q_n^-,q_n^+$ are random in that for some sequence $a_n\to\infty$, $q_n^+/a_n\overset p\to 1$ and $q_n^+/q_n^--\gamma=o_p(1/\sqrt{q_n^-})$.  Then the same results hold with convergence replaced by convergence in probability.
\end{lemma}

\begin{proof}
Theorem 1 of Han et al. (2012) connects the present test to likelihood ratio testing.  As a result (see their proof of Theorem 1),
\begin{equation}\label{eqn:likelihood}
    C_{n,1}^{q_n^-}(1-C_{n,1})^{q_n^+}=C_{n,2}^{q_n^-}(1-C_{n,2})^{q_n^+}.
\end{equation}
Therefore $C_{n,1}=0$ if and only if $C_{n,2}\in\{0,1\}$.  However, this would violate the level constraint, as the test would reject with probability 0 or 1 under the null.  The same is true if $C_{n,1}=C_{n,2}$ in general.  This proves (a).

The proof of part (b) builds in several steps: first showing $C_{n,1},C_{n,2}\to\mu$, next showing that this convergence has rate $1/\sqrt{q_n^-}$, and finally using this fact to establish the result.  Normalizing $C_{n,i}$ as $\tilde C_{n,i}:=\frac{\sqrt{q_n^-}}{\sigma}(C_{n,i}-\mu)$, the goal is to show that $\tilde C_{n,i}$ converge to symmetric critical values of the standard normal distribution.

First, note that $V_n\sim\text{Beta}(q_n^-,q_n^+)$ is a special case of the distribution in Lemma \ref{ideal->normal} with $\beta=1$.  Therefore $\frac{\sqrt{q_n^-}}{\sigma}\left(V_n-\mu\right)\overset{d}\to N(0,1)$, and by Polya's theorem,
\begin{equation}\label{eqn:v_polya}
    \sup_{t\in\R}\left|P\left(V_n\le \frac{\sigma}{\sqrt{q_n^-}}t+\mu\right)-\Phi(t)\right|\to0.
\end{equation}
Plugging in $\tilde C_{n,i}$ for $t$ yields
\begin{equation}\label{eqn:c_polya}
    \left|P\left(V_n\le C_{n,i}\right)-\Phi(\tilde C_{n,i})\right|\to0.
\end{equation}

Now suppose that $C_{n,2}$ does not converge to $\mu$.  Then  infinitely many satisfy $|C_{n,2}-\mu|\ge \varepsilon$ for some $\varepsilon>0$.  We can strengthen this to a one-sided inequality: if infinitely many satisfy $C_{n,2}\le \mu-\varepsilon$, then the normalized critical value $\tilde C_{n_k,2}$ diverges to $-\infty$, which contradicts the Polya result and level requirement since $C_{n,1}\le C_{n,2}$.  Therefore, if $C_{n,2}$ does not converge to $\mu$, then there exists some subsequence such that
\begin{equation}\label{eqn:c_suppose}
    C_{n_k,2}\ge \mu+\varepsilon.
\end{equation}
Thus the normalized critical value $\tilde C_{n_k,2}$ diverges to $\infty$.  To satisfy the level constraint, (\ref{eqn:c_polya}) shows that $\tilde C_{n_k,1}\to \Phi^{-1}(\alpha)$, which implies $C_{n,1}\to\mu$.

Taking logs of (\ref{eqn:likelihood}) and rearranging,
\begin{equation}\label{eqn:log_c_likelihood}
    \ln(C_{n_k,2})+\frac{q_{n_k}^+}{q_{n_k}^-}\ln(1-C_{n_k,2})=\ln(C_{n_k,1})+\frac{q_{n_k}^+}{q_{n_k}^-}\ln(1-C_{n_k,1}).
\end{equation}
Since $C_{n_k,1}\to\mu$, it must be that
\begin{equation}\label{eqn:c2_conv}
    \ln(C_{n_k,2})+\frac{q_{n_k}^+}{q_{n_k}^-}\ln(1-C_{n_k,2})\to\ln(\mu)+\gamma\ln(1-\mu).
\end{equation}
$C_{n_k,2}$ takes values in the compact interval $[\mu+\varepsilon,1]$, so there exists a further subsequence $n_{k_l}$ such that $C_{n_{k_l},2}\to C_2$ for some $C_2\in[\mu+\varepsilon,1]$.  Clearly, $C_2\not=1$, as this would result in the left-hand side of (\ref{eqn:c2_conv}) blowing up, whereas the right-hand side is finite.  Instead, the left-hand side must converge (on this subsequence) to 
\begin{equation}\label{eqn:c_limit_equality}
    \ln(C_2)+\gamma\ln(1-C_2)=\ln(\mu)+\gamma\ln(1-\mu).
\end{equation}
The function $g(x):=\ln(x)+\gamma\ln(1-x)$ has derivative $g'(x)=\frac1x-\frac{\gamma}{1-x}$, which is strictly less than 0 for $x\in(\mu,1)$.  The function $g$ is strictly decreasing here, so $C_2>\mu$ implies the equality can't hold.  Thus (\ref{eqn:c_limit_equality}) contradicts (\ref{eqn:c_suppose}), and it must be that $C_{n,2}\to\mu$.  

Symmetric arguments show $C_{n,1}\to\mu$.

The following will now show that this convergence has rate $1/\sqrt{q_n^-}$.  Taylor expanding log as 
\begin{equation}
    \ln(x)=\ln(a)+\frac{x-a}{a}-\frac{(x-a)^2}{2a^2}+O(|x-a|^3),
\end{equation}
each side of (\ref{eqn:log_c_likelihood}) becomes
\begin{equation}\label{eqn:c_taylor_oneside}
\begin{aligned}
    \ln(C_{n,i})+\frac{q_n^+}{q_n^-}\ln(1-C_{n,i})&=\ln(\mu)\eqbreak+\frac{C_{n,i}-\mu}{\mu}-\frac{(C_{n,i}-\mu)^2}{2\mu^2}\eqbreak+\frac{q_n^+}{q_n^-}\left[\ln(1-\mu)+\frac{\mu-C_{n,i}}{1-\mu}-\frac{(\mu-C_{n,i})^2}{2(1-\mu)^2}\right]\eqbreak+O(|C_{n,i}-\mu|^3).
\end{aligned}
\end{equation}

Now, suppose for contradiction that $\tilde C_{n,2}$ is unbounded, i.e. $|\mu-C_{n,2}|$ is not $O(1/\sqrt{q_n^-})$.  Then there exists a subsequence $n_k$ such that $\tilde C_{n_k,2}\to\infty$.  The level constraint and (\ref{eqn:c_polya}) imply that $\tilde C_{n_k,1}$ must converge to $\Phi^{-1}(\alpha)$.  Consequently, $C_{n_k,1}-\mu=O(1/\sqrt{q_{n_k}^-}).$

Equating the two expansions and rearranging,
\begin{equation}\label{eqn:c_taylor_equality}
\begin{aligned}
    &(C_{n_k,2}-\mu)\left(\frac1{\mu}-\frac{q_{n_k}^+}{q_{n_k}^-}\frac1{1-\mu}\right)-(C_{n_k,2}-\mu)^2\frac12\left(\frac1{\mu^2}+\frac{q_{n_k}^+}{q_{n_k}^-}\frac{1}{(1-\mu)^2}\right)\\
    &=O(|C_{n_k,1}-\mu|^2)+O(|C_{n_k,2}-\mu|^3).
\end{aligned}
\end{equation}
The big-O terms absorb the quadratic term for $C_{n,1}$ above.  Note that $\left(\frac1{\mu}-\frac{q_n^+}{q_n^-}\frac1{1-\mu}\right)=o(1/\sqrt{q_n^-})$ and $\left(\frac1{\mu^2}+\frac{q_n^+}{q_n^-}\frac1{(1-\mu)^2}\right)= \frac{(1+\gamma)^3}{\gamma}+o(1/\sqrt{q_n^-})$.
Then the above equation becomes
\begin{equation}
\begin{aligned}
    &(C_{n_k,2}-\mu)o\left(1/\sqrt{q_{n_k}^-}\right)-(C_{n_k,2}-\mu)^2\left(\frac{(1+\gamma)^3}{2\gamma}+o\left(1/\sqrt{q_{n_k}^-}\right)\right)\eqbreak=O(1/q_{n_k}^-)+O(|C_{n_k,2}-\mu|^3)\\
    &o\left(1/\tilde C_{n_k,2}\right)-\left(\frac{(1+\gamma)^3}{2\gamma}+o\left(1/\sqrt{q_{n_k}^-}\right)\right)\eqbreak=O\left(1/\tilde C_{n_k,2}^2\right)+O(|C_{n_k,2}-\mu|)
\end{aligned}
\end{equation}
where the second equality results from dividing each side by $(C_{n_k,2}-\mu)^2$.  Since $C_{n,2}\to\mu$ and $\tilde C_{n_k,2}\to\infty$, the right-hand side converges to 0 whereas the left-hand side converges to $-\frac{(1+\gamma)^3}{2\gamma}$.  This is a contradiction, so it must be that $\tilde C_{n,2}$ is bounded.

Symmetric arguments show $\tilde C_{n,1}$ is bounded.

Finally, equating both sides of (\ref{eqn:log_c_likelihood}) via (\ref{eqn:c_taylor_oneside}),
\begin{equation}
\begin{aligned}
    &\left[(C_{n,1}-\mu)^2-(C_{n,2}-\mu)^2\right]\left(\frac{1}{2\mu^2}+\frac{q_n^+}{q_n^-}\frac{1}{2(1-\mu)^2}\right)\eqbreak=[(C_{n,1}-\mu)-(C_{n,2}-\mu)]\left(\frac1\mu-\frac{q_n^+}{q_n^-}\frac1{1-\mu}\right)\eqbreak +O(|C_{n,1}-\mu|^3)+O(|C_{n,2}-\mu|^3)\\
    &\left[(C_{n,1}-\mu)^2-(C_{n,2}-\mu)^2\right]\left(\frac{(1+\gamma)^3}{2\gamma}+o\left(1/\sqrt{q_n^-}\right)\right)\eqbreak=O\left(1/\sqrt{q_n^-}\right)o\left(1/\sqrt{q_n^-}\right)+O\left((q_n^-)^{-3/2}\right)\\
    &\sigma^2\left[\tilde C_{n,1}^2-\tilde C_{n,2}^2\right]\left(\frac{(1+\gamma)^3}{2\gamma}+o\left(1/\sqrt{q_n^-}\right)\right)=o(1)\\
    &\tilde C_{n,1}^2-\tilde C_{n,2}^2=o(1)
\end{aligned}
\end{equation}
The second equality applies the established convergence results, and the third multiplies by $q_n^-$.  The fourth simplifies to show that the squares of the normalized critical values converge.

The level constraint and Polya result imply that $\tilde C_{n,1}-\tilde C_{n,2}$ cannot converge to 0 for any subsequence, or else the rejection probability would limit to 1 under the null.  With the level constraint and boundedness of $\tilde C_{n,i}$, it must be that $\tilde C_{n,i}$ converge to the appropriate symmetric critical values of the standard normal distribution.

Now we consider when the sequences $q_n^-,q_n^+$ are random.  The deterministic version proved above implies that for any $\varepsilon>0$, there exists $Q,\delta>0$ such that for all $q_n^-,q_n^+>Q$ and $\sqrt{q_n^-}|q_n^+/q_n^--\gamma|<\delta$,
\begin{equation}
    \left|\frac{\sqrt{q_n^-}}\sigma(C_{n,1}-\mu)-z_{\alpha/2}\right|<\varepsilon\quad\text{and}\quad \left|\frac{\sqrt{q_n^-}}\sigma(C_{n,2}-\mu)-z_{1-\alpha/2}\right|<\varepsilon.
\end{equation}
If no such $Q,\delta$ existed, then for each $k$ there would be a pair $q_k^-,q_k^+>k$ with $\sqrt{q_k^-}|q_k^+/q_k^--\gamma|<1/k$ for which one of the inequalities above fails.  This sequence satisfies the assumptions of the deterministic version, so the result implies that the inequalities must hold eventually.  By contradiction, such $Q,\delta$ must exist.

Convergence in probability holds if the above inequalities hold with probability approaching 1 for any $\varepsilon>0$.  Using the deterministic results,
\begin{equation}
    \begin{aligned}
        &P\left(\left|\frac{\sqrt{q_n^-}}\sigma(C_{n,1}-\mu)-z_{\alpha/2}\right|<\varepsilon\quad\text{and}\quad \left|\frac{\sqrt{q_n^-}}\sigma(C_{n,2}-\mu)-z_{1-\alpha/2}\right|<\varepsilon\right)\\
        \ge&P(q_n^-,q_n^+> Q\text{ and }\sqrt{q_n^-}|q_n^+/q_n^--\gamma|<\delta)\\
        \ge&1-P\left(\frac{q_n^-}{a_n}\le \frac{Q}{a_n}\right)-P\left(\frac{q_n^+}{a_n}\le \frac{Q}{a_n}\right)-P\left(\sqrt{q_n^-}|q_n^+/q_n^--\gamma|\ge\delta\right)\\
        \to& 1
    \end{aligned}
\end{equation}
since $\frac{q_n^-}{a_n}\overset p\to \frac1\gamma$, $\frac{q_n^+}{a_n}\overset p\to 1$, $\frac{Q}{a_n}\to 0$, and $q_n^+/q_n^--\gamma=o_p(1/\sqrt{q_n^-})$.  This proves result \textit{(b)}.

Result \textit{(a)} holds for any $q_n^-,q_n^+$ in the deterministic version, so it holds for every realization of the random variables as well.

\end{proof}

%% file: Sections/Computation.tex
\section{Computational Details}\label{appendix:computation}
This section provides Julia code for calculating the critical values.  In essence, we are just performing a bisection search.

The following function \verb+criticalValues()+ returns the critical values for the spacings test.  We may specify the number of observations on each side: \verb+q0+ below the cutoff and \verb+q1+ above.  We also specify the level $\alpha\in(0,1)$.  As in the paper, we find the critical values by solving $\E[V\phi(V)]=\alpha\frac{q^-}{q^-+q^+}$ for $V\sim\text{Beta}(q^-,q^+)$.  

The last few lines require a search method; I implement a basic bisection search for zeros.  The first argument is the function to search, the second and third specify the interval to search, and the fourth describes whether the function is increasing or decreasing.

\begin{verbatim}
function criticalValues(q0,q1, alpha)
    # Returns the critical values for the spacings test.

    # Define V so we may use its quantiles, cdf, etc.
    V = Distributions.Beta(q0,q1)
    
    # we can solve for one critical value 
    #    given the other to control size:
    c2 = c -> Distributions.quantile(V, 
                1-alpha + Distributions.cdf(V,c))
    
    # helper function for below:
    betac = (c,j) -> c^(q0+j)*(1-c)^(q1-j)
    
    # The expected value of V*phi(V):
    VphiV = c -> q0*factorial(big(q0+q1-1))
            *sum([(betac(c,j) - betac(c2(c),j))
                /(factorial(big(q1-j))*factorial(big(q0+j))) 
                for j in 1:q1]) 
            + q0/(q0+q1)
                
    # Solve for the critical values:
    c1 = bisection_search(c -> VphiV(c) -  alpha*q0/(q0+q1),
                0,
                Distributions.quantile(V,alpha);
                increasing=false)
                
    return (c1, c2(c1))
end
\end{verbatim}

The rest of the spacings test simply requires finding two order statistics and checking whether the statistic in (\ref{eqn:defn_test}) falls outside the returned critical values.

%% file: Sections/MC.tex
\section{Further Comparisons}\label{section:MC}

\begin{figure}[h!]
    \centering
    \begin{subfigure}[t]{0.45\textwidth}
    \includegraphics[width=\textwidth]{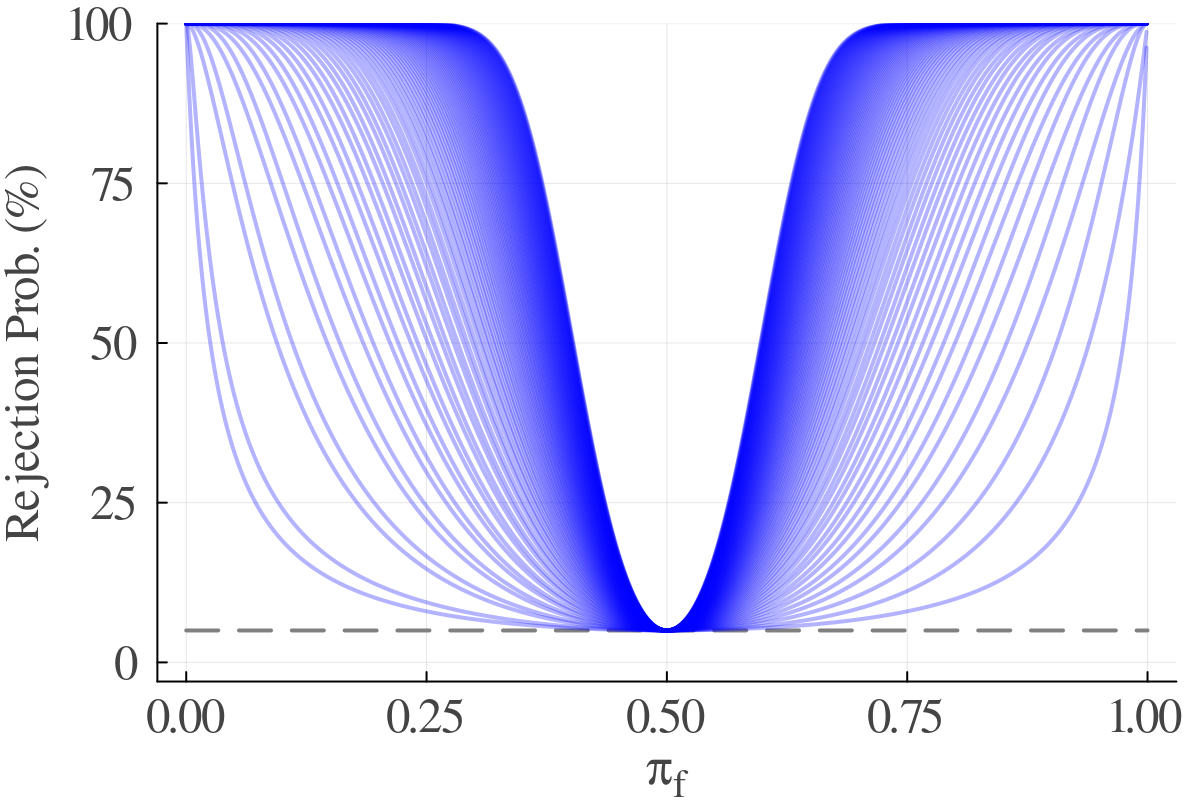}
    \caption{Spacings Test}
\end{subfigure}
\hspace{0.3in}
\begin{subfigure}[t]{0.45\textwidth}
    \includegraphics[width=\textwidth]{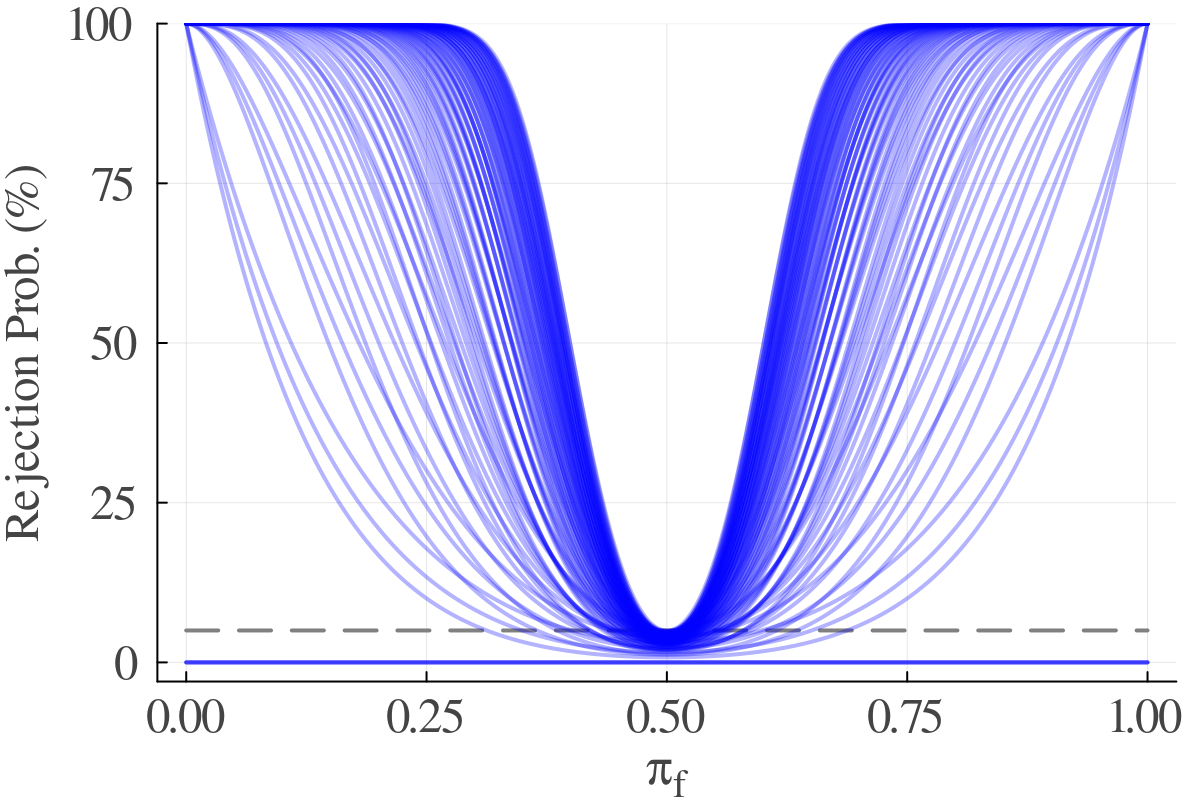}
    \caption{BC}
\end{subfigure}
\vspace{0.1in}
\caption{Limiting Power with $\alpha=5\%$ and $2\le Q \le 100$}
\label{fig:limitingPowerExtra}
\end{figure}

\begin{figure}[h!]
    \centering
    \begin{subfigure}[t]{0.45\textwidth}
    \includegraphics[width=\textwidth]{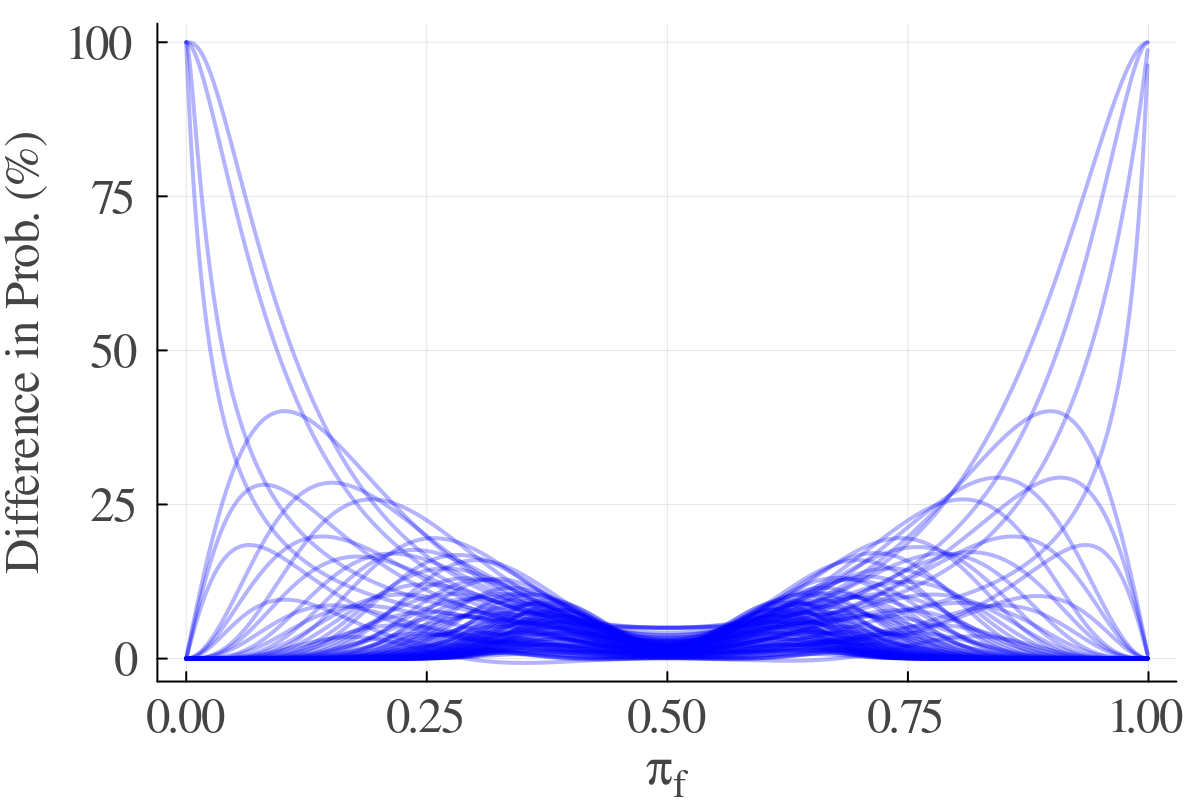}
    \caption{$2\le Q\le 100$}
\end{subfigure}
\hspace{0.3in}
\begin{subfigure}[t]{0.45\textwidth}
    \includegraphics[width=\textwidth]{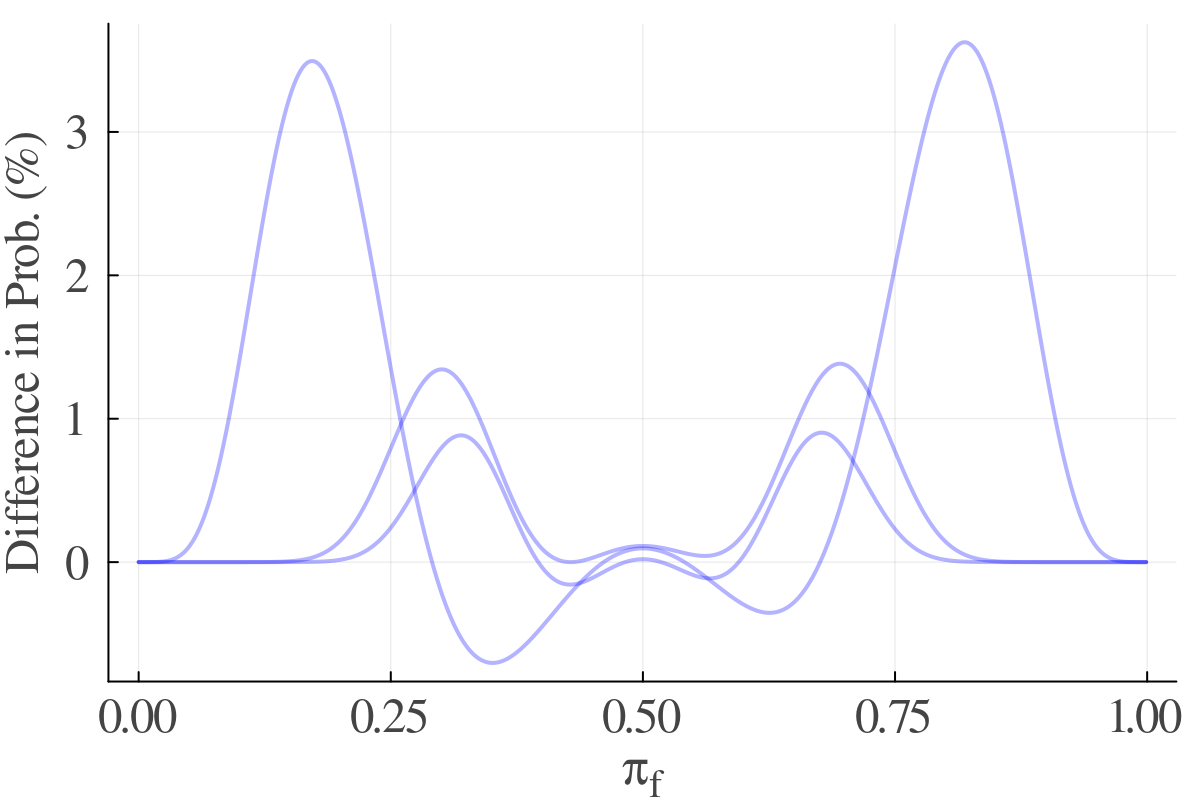}
    \caption{$Q=17,51,67$}
\end{subfigure}
\vspace{0.1in}
\caption{Difference in Limiting Power by $\pi_f$, Spacings minus BC}
\label{fig:diffPowerPi}
\end{figure}

\begin{figure}[h!]
    \centering
    \includegraphics[width=0.45\linewidth]{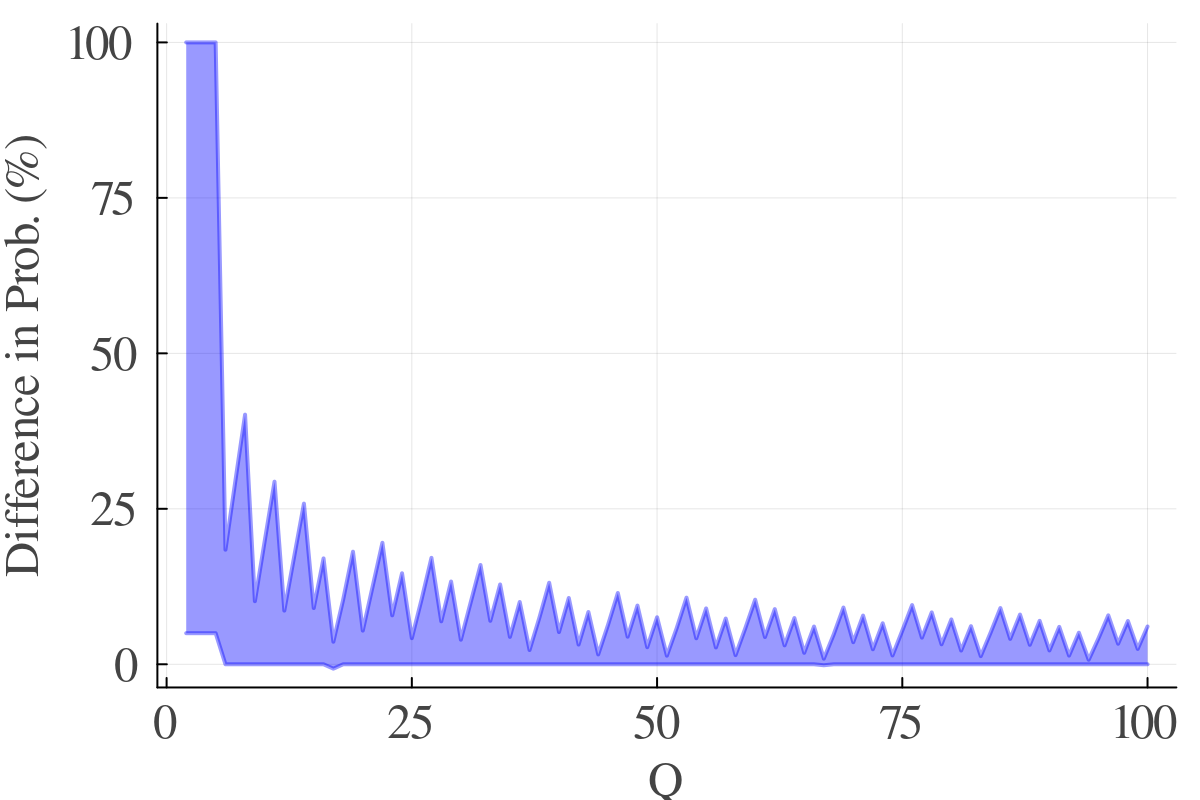}
    \caption{Difference in Limiting Power by $Q$, Spacings minus BC}
    \label{fig:diffPowerQ}
\end{figure}

\begin{table}[!htbp]\centering
\caption{$n=500$, $\alpha=0.05$}
\label{table:500}
\setlength{\tabcolsep}{4pt}
\renewcommand{\arraystretch}{0.95}
\begin{tabular}{ l || c c c c c | c c c c | c c c c | c}
 & \multicolumn{5}{c}{Spacings} & \multicolumn{4}{c}{BC (NR)} & \multicolumn{4}{c}{BC (R)} & CJM\\
 & \multicolumn{5}{c}{$Q$} & \multicolumn{4}{c}{$Q$} & \multicolumn{4}{c}{$Q$}  &\\
 Design & 6 & 10 & 20 & $2q_{rot}$ & $Q_{BC}$ & 6 & 10 & 20 & $Q_{BC}$ & 6 & 10 & 20 & $Q_{BC}$ &\\
 \hline
 D1: $\mu=0$ &4.9&5.0&5.0&5.0&5.1&3.1&2.2&4.1&4.9&5.0&5.0&4.9&5.1&3.6\\
 D1: $\mu=-1$&5.0&5.0&5.5&5.7&5.2&3.2&2.1&4.5&5.2&5.1&5.0&5.1&5.4&5.3\\
 D1: $\mu=-2$&6.4&11.9&49.8&5.4&8.4&4.0&4.8&31.9&6.9&6.0&9.0&34.0&8.2&3.2\\
 \hline
 D2: $\lambda=1$&5.0&5.1&5.4&5.6&5.1&3.1&2.1&4.3&5.0&4.9&5.0&5.2&5.1&3.5\\
 D2: $\lambda=1/3$&4.9&5.0&4.7&4.6&4.6&3.2&2.2&4.2&5.7&5.1&5.0&5.0&5.8&3.8\\
 \hline 
 D3&5.0&5.0&5.5&8.2&12.5&3.1&2.2&4.4&10.6&5.0&5.1&5.3&10.8&6.4\\
 \hline
 D4: $\lambda=0.25$&5.1&5.2&5.4&8.1&12.0&3.2&2.1&4.5&11.2&5.1&5.0&5.4&11.4&3.8\\
 D4: $\lambda=0.10$&5.1&5.5&8.5&28.6&51.6&3.3&2.3&6.4&43.5&5.1&5.3&7.4&43.8&10.0\\
 D4: $\lambda=0.05$&5.3&7.0&20.2&61.6&81.9&3.3&3.0&14.2&77.3&5.2&6.2&15.7&77.5&26.3\\
 \hline
 D5: $\lambda=0.25$&5.0&5.0&5.0&5.0&4.9&3.2&2.1&4.2&4.9&5.1&5.0&5.0&5.1&5.4\\
 D5: $\lambda=0.10$&5.0&4.9&5.0&5.6&9.1&3.1&2.2&4.3&6.2&5.0&5.0&5.0&6.3&12.0\\
 D5: $\lambda=0.05$&5.1&5.1&6.3&24.9&50.7&3.1&2.2&4.3&43.2&4.9&5.0&5.1&43.5&10.6\\ 
 \hline
\end{tabular}
\subcaption{Null Hypothesis}
\label{panel:500_null}
\begin{tabular}{ l || c c c c c | c c c c | c c c c | c}
 & \multicolumn{5}{c}{Spacings} & \multicolumn{4}{c}{BC (NR)} & \multicolumn{4}{c}{BC (R)} & CJM\\
 & \multicolumn{5}{c}{$Q$} & \multicolumn{4}{c}{$Q$} & \multicolumn{4}{c}{$Q$}  &\\
 Design & 6 & 10 & 20 & $2q_{rot}$ & $Q_{BC}$ & 6 & 10 & 20 & $Q_{BC}$ & 6 & 10 & 20 & $Q_{BC}$ &\\
 \hline
 D1: $\mu=0$ &6.7&7.8&9.5&9.5&9.1&4.7&4.1&8.6&8.5&6.9&8.0&9.8&8.6&4.6\\
 D1: $\mu=-1$&6.8&8.1&10.2&10.8&9.3&4.8&4.1&9.2&10.1&6.9&8.1&10.5&10.2&4.4\\
 D1: $\mu=-2$&8.3&15.2&54.4&6.4&10.9&5.3&6.5&36.3&9.0&7.6&11.5&38.5&10.5&3.3\\
 \hline
 D2: $\lambda=1$&7.1&9.0&13.7&15.7&11.7&5.1&4.7&12.4&12.7&7.3&9.0&12.9&13.8&7.2\\
 D2: $\lambda=1/3$&6.4&7.1&7.4&5.9&4.6&4.6&3.6&7.0&5.3&6.7&7.3&8.0&5.4&6.2\\
 \hline 
 D3&6.7&7.6&9.5&13.6&18.9&4.6&4.0&8.1&16.4&6.8&7.8&9.3&16.6&7.5\\
 \hline
 D4: $\lambda=0.25$&7.1&8.8&13.5&22.1&29.1&4.9&4.6&12.1&28.0&7.1&8.9&13.6&28.3&6.9\\
 D4: $\lambda=0.10$&7.5&10.6&21.3&52.8&73.7&5.3&5.4&17.7&67.1&7.5&10.1&19.5&67.4&17.2\\
 D4: $\lambda=0.05$&8.8&14.8&40.3&80.6&92.3&6.0&7.4&30.6&90.5&8.4&12.7&32.7&90.6&38.6\\
 \hline
 D5: $\lambda=0.25$&6.7&7.9&10.1&11.3&10.9&4.8&4.1&9.3&10.6&7.0&8.1&10.6&10.7&6.1\\
 D5: $\lambda=0.10$&6.8&8.2&10.3&10.6&8.7&4.8&4.3&9.4&9.0&7.0&8.3&10.7&9.2&24.9\\
 D5: $\lambda=0.05$&6.7&7.9&8.6&10.7&24.3&4.9&4.2&9.0&18.1&7.1&8.2&10.2&18.3&11.3\\ 
 \hline
\end{tabular}
\subcaption{Alternative Hypothesis}
\label{panel:500_alt}
\end{table}

\begin{table}[!htbp]\centering
\caption{$n=1000$, $\alpha=0.05$}
\label{table:1000}
\setlength{\tabcolsep}{4pt}
\renewcommand{\arraystretch}{0.95}
\begin{tabular}{ l || c c c c c | c c c c | c c c c | c}
 & \multicolumn{5}{c}{Spacings} & \multicolumn{4}{c}{BC (NR)} & \multicolumn{4}{c}{BC (R)} & CJM\\
 & \multicolumn{5}{c}{$Q$} & \multicolumn{4}{c}{$Q$} & \multicolumn{4}{c}{$Q$}  &\\
 Design & 6 & 10 & 20 & $2q_{rot}$ & $Q_{BC}$ & 6 & 10 & 20 & $Q_{BC}$ & 6 & 10 & 20 & $Q_{BC}$ &\\
 \hline
 D1: $\mu=0$ &5.0&5.0&5.0&5.0&5.0&3.1&2.2&4.1&4.9&4.9&5.0&5.0&4.9&3.8\\
 D1: $\mu=-1$&5.0&5.0&5.2&5.5&5.6&3.1&2.2&4.2&5.3&5.0&5.0&5.1&5.5&4.4\\
 D1: $\mu=-2$&5.5&6.5&16.4&5.3&5.9&3.4&2.9&12.1&4.8&5.4&6.1&13.5&5.9&4.5\\
 \hline
 D2: $\lambda=1$&5.0&5.0&5.1&5.4&5.7&3.1&2.1&4.2&5.6&5.0&5.0&5.1&5.7&4.2\\
 D2: $\lambda=1/3$&5.0&5.0&4.9&4.9&4.8&3.1&2.1&4.2&5.2&4.9&4.9&5.0&5.3&4.7\\
 \hline 
 D3&5.1&5.0&5.1&7.1&9.0&3.1&2.1&4.2&8.4&5.0&5.0&5.1&8.4&6.6\\
 \hline
 D4: $\lambda=0.25$&5.0&5.1&5.1&6.9&8.5&3.1&2.1&4.2&8.4&5.0&5.0&5.1&8.5&4.4\\
 D4: $\lambda=0.10$&5.0&5.1&5.9&19.3&32.5&3.2&2.2&4.7&27.6&5.1&5.2&5.6&27.6&14.4\\
 D4: $\lambda=0.05$&5.2&5.5&8.4&58.4&77.2&3.2&2.4&6.4&71.1&5.1&5.3&7.5&71.2&46.6\\
 \hline
 D5: $\lambda=0.25$&5.0&5.1&5.1&5.0&4.9&3.1&2.1&4.1&5.0&5.0&5.0&5.0&5.1&7.0\\
 D5: $\lambda=0.10$&5.0&5.1&5.0&5.0&5.1&3.2&2.1&4.2&5.0&5.1&5.0&5.0&5.0&19.4\\
 D5: $\lambda=0.05$&5.0&5.0&4.9&11.1&24.3&3.1&2.2&4.2&18.3&5.0&5.1&5.0&18.3&15.6\\ 
 \hline
\end{tabular}
\subcaption{Null Hypothesis}
\label{panel:1000_null}
\begin{tabular}{ l || c c c c c | c c c c | c c c c | c}
 & \multicolumn{5}{c}{Spacings} & \multicolumn{4}{c}{BC (NR)} & \multicolumn{4}{c}{BC (R)} & CJM\\
 & \multicolumn{5}{c}{$Q$} & \multicolumn{4}{c}{$Q$} & \multicolumn{4}{c}{$Q$}  &\\
 Design & 6 & 10 & 20 & $2q_{rot}$ & $Q_{BC}$ & 6 & 10 & 20 & $Q_{BC}$ & 6 & 10 & 20 & $Q_{BC}$ &\\
 \hline
 D1: $\mu=0$ &6.7&8.4&11.6&15.5&15.3&4.9&4.4&10.7&15.6&7.1&8.5&12.0&15.6&5.5\\
 D1: $\mu=-1$&6.9&8.4&11.5&14.6&13.4&4.9&4.4&10.7&13.7&7.1&8.5&12.1&13.9&4.8\\
 D1: $\mu=-2$&7.2&9.5&22.1&6.7&8.2&5.0&4.6&16.7&7.2&7.2&8.8&18.4&8.6&4.5\\
 \hline
 D2: $\lambda=1$&7.1&8.9&13.7&20.6&24.7&5.1&4.7&12.4&24.6&7.3&9.0&13.9&24.9&11.5\\
 D2: $\lambda=1/3$&6.8&8.1&10.1&9.8&9.3&4.8&4.2&9.4&9.9&7.0&8.2&10.7&10.1&9.8\\
 \hline 
 D3&6.7&8.2&10.8&16.1&18.6&4.8&4.3&9.9&17.3&7.0&8.3&11.2&17.4&7.5\\
 \hline
 D4: $\lambda=0.25$&7.2&9.0&13.8&29.1&34.7&5.2&5.0&12.6&35.1&7.4&9.2&14.1&35.2&11.7\\
 D4: $\lambda=0.10$&7.3&9.6&16.8&53.2&68.3&5.2&5.1&14.9&63.5&7.5&9.6&16.5&63.5&28.9\\
 D4: $\lambda=0.05$&7.8&11.4&24.7&86.3&94.3&5.5&5.8&20.0&92.0&7.9&10.6&21.9&92.0&65.6\\
 \hline
 D5: $\lambda=0.25$&6.9&8.5&11.8&18.3&18.6&4.8&4.4&10.9&19.0&7.0&8.5&12.3&19.0&8.9\\
 D5: $\lambda=0.10$&6.9&8.5&11.9&18.1&18.6&5.0&4.5&10.8&19.2&7.2&8.6&12.2&19.2&46.6\\
 D5: $\lambda=0.05$&6.9&8.6&12.0&11.2&10.5&4.9&4.5&10.9&7.6&7.1&8.6&12.3&7.6&19.1\\ 
 \hline
\end{tabular}
\subcaption{Alternative Hypothesis}
\label{panel:1000_alt}
\end{table}

\begin{table}[!htbp]\centering
\caption{Effective Sample Size}
\label{table:sample_size}
\begin{tabular}{ l || c c | c c }
 & \multicolumn{2}{c}{$n=500$} & \multicolumn{2}{c}{$n=1000$}\\
 & \multicolumn{2}{c}{$\alpha=0.05$} & \multicolumn{2}{c}{$\alpha=0.05$}\\
 Design & $2q_{rot}$ & $Q_{BC}$ &  $2q_{rot}$ & $Q_{BC}$ \\
 \hline
 D1: $\mu=0$ &44.0&51.0&62.0 & 67.0\\
 &(0.07)&(0.0)&(0.2) & (0.0)\\
 D1: $\mu=-1$&22.9&17.0&31.9 & 30.0\\
 &(1.69)&(0.0)&(1.7) & (13.54)\\
 D1: $\mu=-2$&4.0&9.0&5.2 & 9.0\\
 &(0.63)&(0.0)&(0.99) & (0.04)\\
 \hline
 D2: $\lambda=1$&24.7&17.0&34.6 & 43.5\\
 &(1.86)&(0.09)&(1.86) & (8.01)\\
 D2: $\lambda=1/3$&35.7&46.9&50.2 & 55.5\\
 &(1.53)&(4.44)&(1.52) & (7.2)\\
 \hline 
 D3&37.9&50.8&53.2 & 66.7\\
 &(1.11)&(1.25)&(1.14) & (2.05)\\
 \hline
 D4: $\lambda=0.25$&37.7&50.4&53.0 & 66.2\\
 &(1.34)&(1.94)&(1.34) & (3.59)\\
 D4: $\lambda=0.10$&37.5&50.2&52.7 & 65.5\\
 &(1.37)&(2.28)&(1.37) & (4.61)\\
 D4: $\lambda=0.05$&37.5&50.1&52.6 & 65.4\\
 &(1.37)&(2.34)&(1.38) & (4.81)\\
 \hline
 D5: $\lambda=0.25$&38.3&50.8&53.8 & 66.8\\
 &(1.29)&(1.24)&(1.28) & (1.71)\\
 D5: $\lambda=0.10$&37.6&50.3&52.8 & 65.8\\
 &(1.36)&(2.14)&(1.36) & (4.26)\\
 D5: $\lambda=0.05$&37.5&50.2&52.7 & 65.5\\
 &(1.37)&(2.31)&(1.37) & (4.67)\\
 \hline
\end{tabular}
\end{table}